\documentclass{aa}
\usepackage[T1]{fontenc}

\DeclareRobustCommand{\VAN}[3]{#2}
\let\VANthebibliography\thebibliography
\def\thebibliography{\DeclareRobustCommand{\VAN}[3]{##3}\VANthebibliography}

\usepackage{graphicx}	
\usepackage{txfonts}
\usepackage{amsmath}	
\usepackage{subcaption}
\usepackage{xcolor}
\usepackage{colortbl}
\usepackage{gensymb}
\usepackage{xfrac}
\usepackage{phoenician}
\usepackage{placeins}
\usepackage{hyperref}
\hypersetup{
    colorlinks,
    citecolor=blue,
    linkcolor=blue
    }

\newcommand{\numberOfGRGsCatalogueOei}{3341}
\newcommand{\numberOfGRGsBORG}{281}
\newcommand{\numberOfGRGsBORGzsp}{260}
\newcommand{\numberOfGRGsBORGzspUncertainty}{248}
\newcommand{\numberOfGRGsBORGzspOei}{208}
\newcommand{\numberOfRGsBORG}{1870}
\newcommand{\numberOfRGsBORGzsp}{1443}

\allowdisplaybreaks

\begin{document}
\title{Luminous giants populate the dense Cosmic Web}
\subtitle{The radio luminosity--environmental density relation for radio galaxies in action\thanks{
Table~\ref{tab:GRGsCosmicWeb}'s extension to all selected GRGs, and an analogous table for all selected general RGs, are available in Flexible Image Transport System (FITS) format through the Centre de Donn\'ees astronomiques de Strasbourg (CDS).
One can either use anonymous File Transfer Protocol (FTP) to \url{ftp://cdsarc.cds.unistra.fr} (130.79.128.5) or visit \url{https://cdsarc.cds.unistra.fr/cgi-bin/qcat?J/A+A/}.
}}


\author{Martijn S.\,S.\,L. Oei\inst{1,2} \and
Reinout J. van Weeren\inst{1} \and
Martin J. Hardcastle\inst{3} \and
Aivin R.\,D.\,J.\,G.\,I.\,B. Gast\inst{4} \and
Florent Leclercq\inst{5} \and
Huub J.\,A. R\"ottgering\inst{1} \and
Pratik Dabhade\inst{6,7} \and
Tim W. Shimwell\inst{1,8} \and
Andrea Botteon\inst{9}}

\institute{
Leiden Observatory, Leiden University, Niels Bohrweg 2, 2300 RA Leiden, The Netherlands
\and
Cahill Center for Astronomy and Astrophysics, California Institute of Technology, 1216 E California Blvd, CA 91125 Pasadena, USA
\and
Centre for Astrophysics Research, University of Hertfordshire, College Lane, Hatfield AL10 9AB, UK
\and
Somerville College, University of Oxford, Woodstock Road, Oxford OX2 6HD, UK
\and
CNRS \& Sorbonne Universit\'e, UMR 7095, Institut d'Astrophysique de Paris, 98 bis Boulevard Arago, 75014 Paris, France
\and
Instituto de Astrof\'isica de Canarias, Calle V\'ia L\'actea, s/n, E-38205, La Laguna, Tenerife, Spain
\and
Universidad de La Laguna, Departamento de Astrofisica, E-38206, Tenerife, Spain
\and
ASTRON, the Netherlands Institute for Radio Astronomy, Oude Hoogeveensedijk 4, 7991 PD Dwingeloo, The Netherlands
\and
INAF--IRA, Via P. Gobetti 101, 40129 Bologna, Italy
}

\date{Received 6 June 2023 / Accepted 3 November 2023}




 
  \abstract
   {
   Giant radio galaxies (GRGs, giant RGs, or giants) are megaparsec-scale, jet-driven outflows from accretion disks of supermassive black holes, and represent the most extreme pathway by which galaxies can impact the Cosmic Web around them.
   A long-standing but unresolved question is why giants are so much larger than other radio galaxies.}
   {
   It has been proposed that, in addition to having higher jet powers than most RGs, giants might live in especially low-density Cosmic Web environments.
In this work, we aim to test this hypothesis by pinpointing Local Universe giants and other RGs in physically principled, Bayesian large-scale structure reconstructions.}
   {
   More specifically, we localised a LOFAR Two-metre Sky Survey (LoTSS) DR2--dominated sample of luminous ($l_\nu(\nu = 150\ \mathrm{MHz}) \geq 10^{24}\ \mathrm{W\ Hz^{-1}}$) giants and a control sample of LoTSS DR1 RGs, both with spectroscopic redshifts up to $z_\mathrm{max} = 0.16$, in the BORG SDSS Cosmic Web reconstructions.
   We measured the Cosmic Web density on a smoothing scale of ${\sim}2.9\ \mathrm{Mpc}\ h^{-1}$ for each RG; for the control sample, we then quantified the relation between RG radio luminosity and Cosmic Web density.
   With the BORG SDSS tidal tensor, we also measured for each RG whether the gravitational dynamics of its Cosmic Web environment resemble those of clusters, filaments, sheets, or voids.
   }
   {
   For both luminous giants and general RGs, the Cosmic Web density distribution is gamma-like.
   Luminous giants populate large-scale environments that tend to be denser than those of general RGs.
   This result is corroborated by gravitational dynamics classification and a cluster catalogue crossmatching analysis.
We find that the Cosmic Web density around RGs with 150 MHz radio luminosity $l_\nu$ is distributed as $1 + \Delta_\mathrm{RG}\ \vert\ L_\nu = l_\nu \sim \Gamma(k,\theta)$, where $k = 4.8 + 0.2 \cdot \textphnc{\Alamed}$, $\theta = 1.4 + 0.02 \cdot \textphnc{\Alamed}$, and $\textphnc{\Alamed} \coloneqq \mathrm{log}_{10}(l_\nu\ (10^{23}\ \mathrm{W\ Hz^{-1}})^{-1})$.}
   {This work presents more than a thousand inferred megaparsec-scale densities around radio galaxies, which may be correct up to a factor of order unity --- except in clusters of galaxies, where the densities can be more than an order of magnitude too low.
We pave the way to a future in which megaparsec-scale densities around RGs are common inferred quantities, which help to better understand their dynamics, morphology, and interaction with the enveloping Cosmic Web.
Our data demonstrate that luminous giants inhabit denser environments than general RGs.
This shows that --- at least at high jet powers --- low-density environments are no prerequisite for giant growth.
Using general RGs, we quantified the relation between radio luminosity at 150 MHz and Cosmic Web density on a smoothing scale of ${\sim}2.9\ \mathrm{Mpc}\ h^{-1}$.
This positive relation, combined with the discrepancy in radio luminosity between known giants and general RGs, reproduces the discrepancy in Cosmic Web density between known giants and general RGs.
Our findings are consistent with the view that giants are regular, rather than mechanistically special, members of the radio galaxy population.
}



\maketitle
%


\section{Introduction}
\label{sec:introduction}
Supermassive Kerr black holes are key building blocks of the Universe, on galactic and cosmological scales alike.
During episodes of baryon accretion, they turn into active galactic nuclei (AGN), launching winds and jets that warm and rarefy the interstellar medium \citep[e.g.][]{Fabian12012, King12015}.
These energy flows generally suppress the formation of new stars, although local star formation enhancement can occur within expanding kiloparsec-radius rings \citep[e.g.][]{Dugan12017}.
Meanwhile, jet-mediated AGN outflows --- also known as radio galaxies (RGs) --- can have a vast, megaparsec-scale reach, protruding from both the galaxy and its circumgalactic medium.
The synchrotron radiation from RGs illuminates their jets, lobes, cocoons, and threads \citep[e.g.][]{Ramatsoku12020}, and dominates the known extragalactic radio sky.
Because the behaviour of RGs is linked to the physics of black hole accretion and galactic winds, the pressure field and magnetisation history \citep[e.g.][]{Vazza12017} of the warm--hot intergalactic medium (warm--hot IGM, or WHIM), as well as shocks \citep[e.g.][]{Nolting12019Aligned}, vorticity \citep[e.g.][]{Nolting12019Orthogonal}, and cooling flows \citep[e.g.][]{Fabian11984} in the intracluster medium, a precise understanding of the RG phenomenon is indispensable to modern astrophysics.

A key goal of the study of RGs is to identify the main factors that determine their dynamics and to formulate models \citep[e.g.][]{Scheuer11974, Turner12015, Hardcastle12018} that describe them quantitatively.
One way to investigate the growth of RGs is to search for particularly large examples and to analyse what internal or external traits set them apart from the rest.
Following this logic, \citet{Oei12022Alcyoneus} presented and studied Alcyoneus, a giant radio galaxy (GRG, giant RG, or simply \textit{giant}) whose proper length component in the plane of the sky $l_\mathrm{p} = 4.99 \pm 0.04\ \mathrm{Mpc}$.
In general, giants are members of the RG population for which $l_\mathrm{p} \geq l_\mathrm{p,GRG}$ --- where the latter is some fixed megaparsec-scale threshold --- and that therefore rank among the largest RGs in existence.
However, despite being one of the largest RGs known, Alcyoneus is not particularly luminous, and its host galaxy does not feature a particularly massive central black hole or stellar population --- at least, when compared to other giants and their hosts \citep{Oei12022Alcyoneus}.
The question naturally arises whether external properties, rather than those internal to the host galaxy, are the most important drivers of RG growth.

It is well established that the IGM resists the growth of RGs by forcing jets to convert a part of their kinetic energy into work spent to form lobe cavities \citep[e.g.][]{Hardcastle12020}.
The work needed to free up a cavity is the product of its volume and the local pressure.\footnote{As the pressure field in modern large-scale structure is not constant, as predicted by approximate hydrostatic equilibrium, this expression only holds for sufficiently small volumes.}
Following this line of reasoning, an RG with fixed intrinsic properties should reach a larger end-of-life extent in more tenuous (and colder) Cosmic Web (CW) environments.
It has therefore been proposed that the astonishing growth of giants might be explained by their presumptive tendency to reside in tenuous parts of the Cosmic Web.

Several previous works have investigated the role of the enveloping Cosmic Web density field on GRG growth.
In all cases, the authors traced Cosmic Web environments through three-dimensional galaxy positions estimated with photometric or spectroscopic redshifts.
The pioneering work of \citet{Subrahmanyan12008} presented a case study of GRG MSH 05-2\textit{2} and its Cosmic Web environment as traced by 6dF data \citep{Jones12004}.
Using a sample of 12 giants and environments traced by 2dF/AAOmega data \citep{Sharp12006}, \citet{Malarecki12015} concluded that the lobes of giants grow in directions that avoid denser regions of the Cosmic Web.
In the most comprehensive study yet, \citet{Lan12021} used a sample of 110 giants and environments traced by DESI Legacy Imaging Surveys Data Release (DR) 9 data \citep{Dey12019}.
They did not find evidence that giants occur in more dilute environments than non-giant RGs.
A limitation of this latest study is the use of photometric redshifts to probe the Cosmic Web density field.

In order to determine decisively how GRG growth and the Cosmic Web relate, analyses with both more giants and more accurate Cosmic Web reconstructions seem necessary.
In this work, we present major developments in both regards.
First, the Low-Frequency Array \citep[LOFAR;][]{vanHaarlem12013} Two-metre Sky Survey \citep[LoTSS;][]{Shimwell12017} has made possible the discovery of thousands of previously unknown giants in its Northern Sky imagery at the observing frequency $\nu_\mathrm{obs}=144\ \mathrm{MHz}$ and at resolutions $\theta_\mathrm{FWHM} \in \{6'', 20'', 60'', 90''\}$.
In particular, the joint search efforts of \citet{Dabhade12020March} in the LoTSS DR1 \citep{Shimwell12019}, those of \citet{Simonte12022} in the LoTSS Bo\"otes Deep Field, and those of \citet{Oei12022GiantsSample} in the LoTSS DR2 \citep{Shimwell12022} have tripled the total number of known giants, which now stands at ${\sim}3 \cdot 10^3$.
The number of known giants in the part of the Local Universe covered by the Sloan Digital Sky Survey \citep[SDSS;][]{York12000} DR7 \citep{Abazajian12009} has even quintupled --- a fact whose relevance becomes clear in Sect.~\ref{sec:dataCW}.
Second, the last two decades have seen the development of principled, physics-based Bayesian inference techniques through which the three-dimensional total (i.e. baryonic plus dark) matter density field of the Cosmic Web can be reconstructed \citep[e.g.][]{Wandelt12004, Kitaura12008, Jasche12010July, Jasche12010September, Jasche12013, Jasche12019}.
These techniques make use of the fact that the statistical behaviour of the Early Universe's total matter density field is known, as is the dominant process by which this field evolved over cosmic time: gravity.
By simulating gravity acting on Early Universe density fields and comparing the evolved fields to the observed spatial distribution of galaxies, the late-time density field can be inferred.
These late-time density fields subsequently enable megaparsec-scale density measurements with uncertainties around individual (giant) RGs.
We shall use these measurements as an improved probe of the Cosmic Web density field.

Section~\ref{sec:data} presents the data: radio galaxy observables and late-time density field reconstructions via which we probed the influence of the Cosmic Web on the growth of giants.
In Sect.~\ref{sec:methods}, we explain how we combined these data to determine megaparsec-scale densities and large-scale structure--type probability distributions for giants and general RGs.
Section~\ref{sec:results} presents these results, alongside a quantification of the relation between RG radio luminosity and Cosmic Web density.
We finally evaluate evidence for the claim that the Cosmic Web affects RG growth.
In Sect.~\ref{sec:discussion}, we explain how we tested the reliability of RG Cosmic Web density measurements and dynamical classifications and discuss caveats of and promising future extensions to the current work, before we present conclusions in Sect.~\ref{sec:conclusion}.

\begin{figure*}[t]
    \centering
    \includegraphics[width=\textwidth]{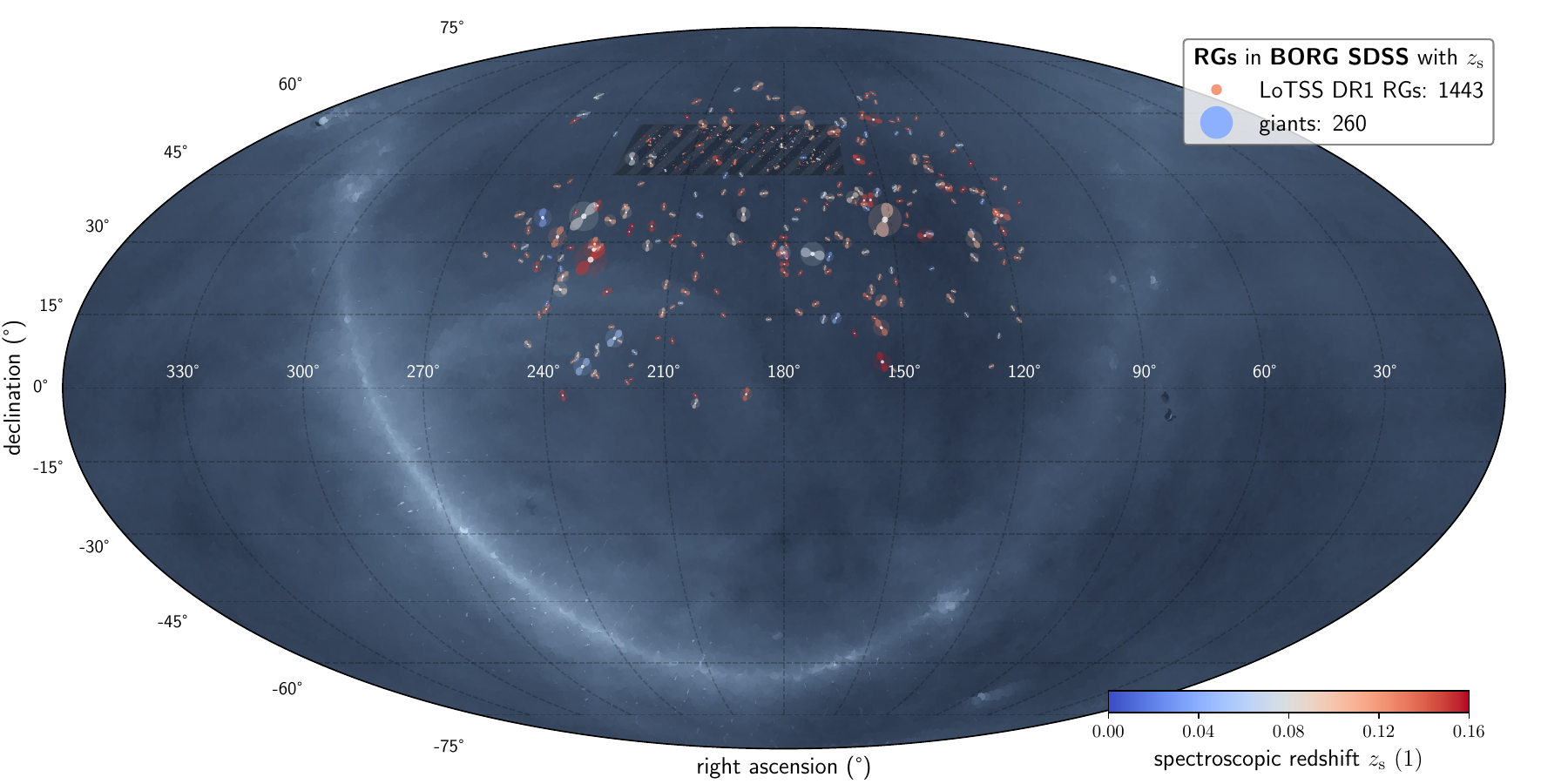}
    \caption{
    Mollweide view of the sky showing the locations of all giants and LoTSS DR1 RGs in the Local Universe for which we inferred Cosmic Web densities and dynamical states.
    The background shows the Milky Way at $150\ \mathrm{MHz}$ \citep{Zheng12017}, on which we overlaid the LoTSS DR1 footprint (hatched dark rectangle).
    All RGs are drawn as lemniscates of Bernoulli; we did not attempt to portray realistic shapes or position angles.
    The colours represent redshifts $z \in (0, z_\mathrm{max} \coloneqq 0.16)$, whilst the diameters are proportional to projected proper lengths $l_\mathrm{p} \in ({\sim}1\ \mathrm{kpc}, 4.6\ \mathrm{Mpc})$.
    Giants are translucent.
    Upon zooming in, the reader can appreciate the wide variety of observed radio galaxy sizes.
    }
    \label{fig:sky}
\end{figure*}

For consistency with \citet{Oei12022Alcyoneus}, \citet{Oei12022GiantsSample}, and \citet{Mostert12023}, we assumed a flat, inflationary $\Lambda$CDM model with parameters from \citet{Planck12020}: $h = 0.6766$, $\Omega_\mathrm{BM,0} = 0.0490$, $\Omega_\mathrm{M,0} = 0.3111$, and $\Omega_{\Lambda,0} = 0.6889$, where $\Omega_\mathrm{DM,0} \coloneqq \Omega_\mathrm{M,0}-\Omega_\mathrm{BM,0} = 0.2621$ and $H_0 \coloneqq h \cdot 100\ \mathrm{km\ s^{-1}\ Mpc^{-1}}$.
With `Local Universe', we refer to the spherical region of space observed to have redshift $z < z_\mathrm{max} \coloneqq 0.16$.
All reported redshifts are heliocentric.
Terminology-wise, we strictly distinguish an RG (a radio-bright structure of plasma and magnetic fields, consisting of a core, jets, hotspots, lobes, a cocoon, and collimated synchrotron threads) from the host galaxy that has generated it.
As in our previous works, we define giants to be RGs with projected proper lengths $l_\mathrm{p} \geq l_\mathrm{p,GRG} \coloneqq 0.7\ \mathrm{Mpc}$.\footnote{In Cosmic Web filament environments, where giants appear most common (Sect.~\ref{sec:results}), lobes may expand along the Hubble flow, rendering their proper and comoving extents different.
To avoid ambiguity, we stress that our projected lengths are proper instead of comoving.
A less precise synonym for `projected proper length' often found in the literature is `largest linear size' (LLS).}
We define the spectral index $\alpha$ so that it relates to flux density $F_\nu$ at frequency $\nu$ as $F_\nu \propto \nu^\alpha$.
Under this convention, and when synchrotron self-absorption is negligible, radio spectral indices are typically negative.

\section{Data}
\label{sec:data}
To compare the Cosmic Web environments of Local Universe giants to those of Local Universe radio galaxies, we combined a GRG catalogue, an RG catalogue, and a Cosmic Web density field reconstruction.
All three data sets are publicly available.

\subsection{Giant radio galaxies}
As our source of giants we used the catalogue aggregated by \citet{Oei12022GiantsSample}, which contains \numberOfGRGsCatalogueOei\ giants with $l_\mathrm{p} \geq l_\mathrm{p,GRG} \coloneqq 0.7\ \mathrm{Mpc}$.\footnote{\citet{Oei12022GiantsSample} provide references to all discovery articles.} This catalogue is intended to be complete up to (and including) September 2022.
Thanks to the steradian-scale Northern Sky coverage of the LoTSS, combined with its arcsecond-scale resolution and sensitivity up to degree scales, LoTSS-discovered giants \citep{Dabhade12020March, Simonte12022, Oei12022GiantsSample} dominate the catalogue.

In their LoTSS DR2 manual search for giants, \citet{Oei12022GiantsSample} used an angular length threshold of $\frac{1}{2}\left(\phi_\mathrm{max} - \phi_\mathrm{min}\right) = 5'$ to limit the duration of their search to a manageable few hundred hours.
At the same time, this $5'$-threshold ensured that the search would yield most giants with sufficient surface brightness in the Local Universe.\footnote{\citet{Oei12022GiantsSample} illustrate this point in their Fig.~9.}
This has been by design: \citet{Oei12022GiantsSample} aimed to build a surface brightness--limited, but otherwise complete census of Local Universe giants in the LoTSS DR2 footprint, with the intent of localising them within the Cosmic Web density field reconstructions of Sect.~\ref{sec:dataCW}.

\begin{figure*}
\centering
    \begin{subfigure}{\columnwidth}
    \centering
    \includegraphics[width=\columnwidth]{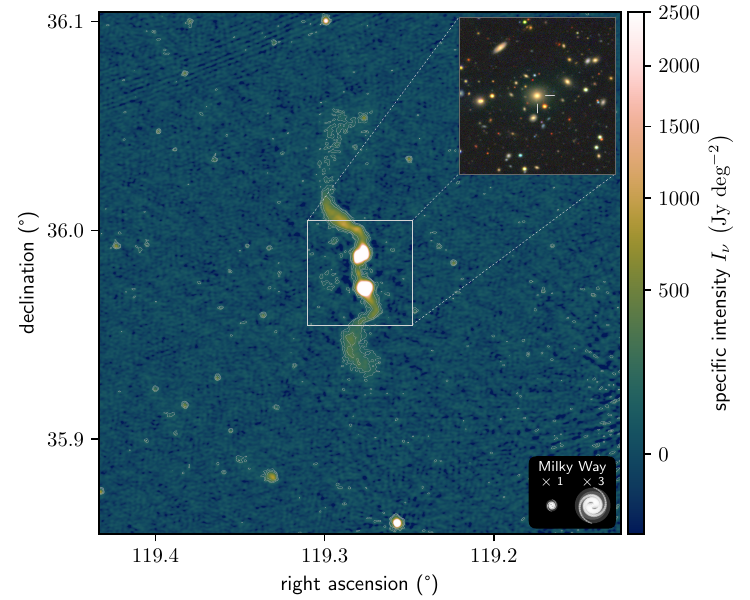}
    \end{subfigure}
    \begin{subfigure}{\columnwidth}
    \centering
    \includegraphics[width=\columnwidth]{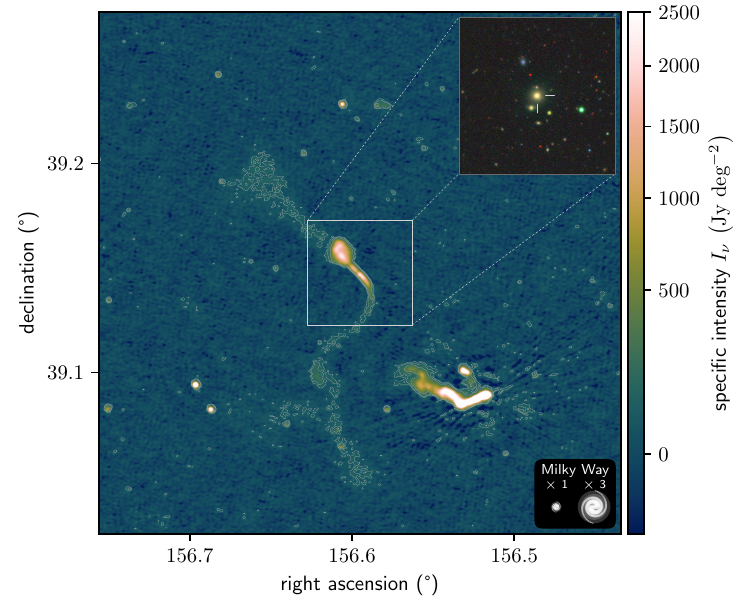}
    \end{subfigure}
    \begin{subfigure}{\columnwidth}
    \centering
    \includegraphics[width=\columnwidth]{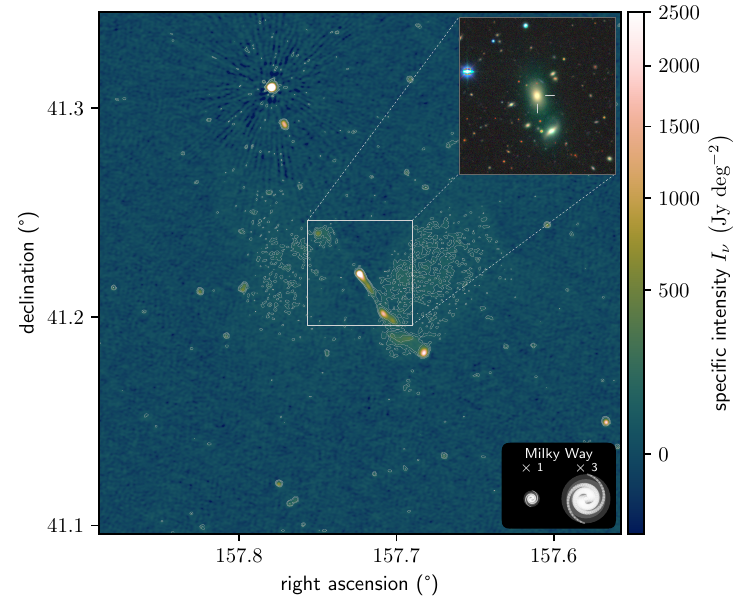}
    \end{subfigure}
    \begin{subfigure}{\columnwidth}
    \centering
    \includegraphics[width=\columnwidth]{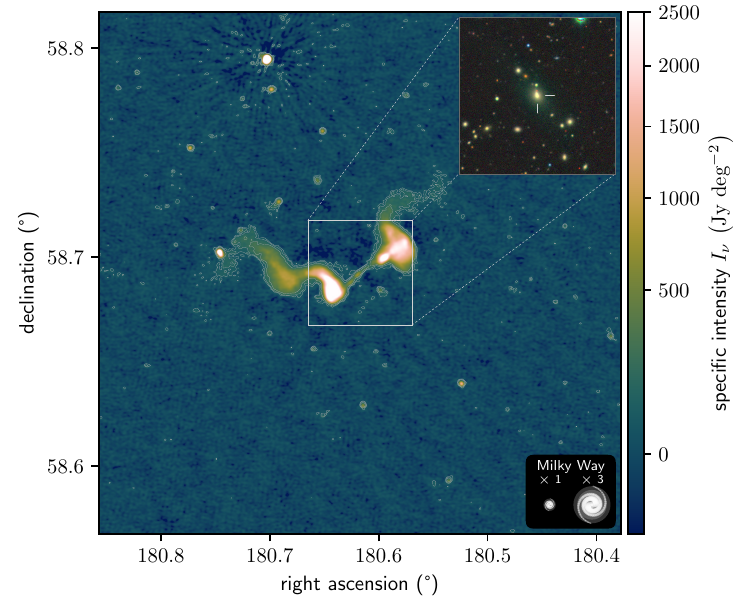}
    \end{subfigure}
    \begin{subfigure}{\columnwidth}
    \centering
    \includegraphics[width=\columnwidth]{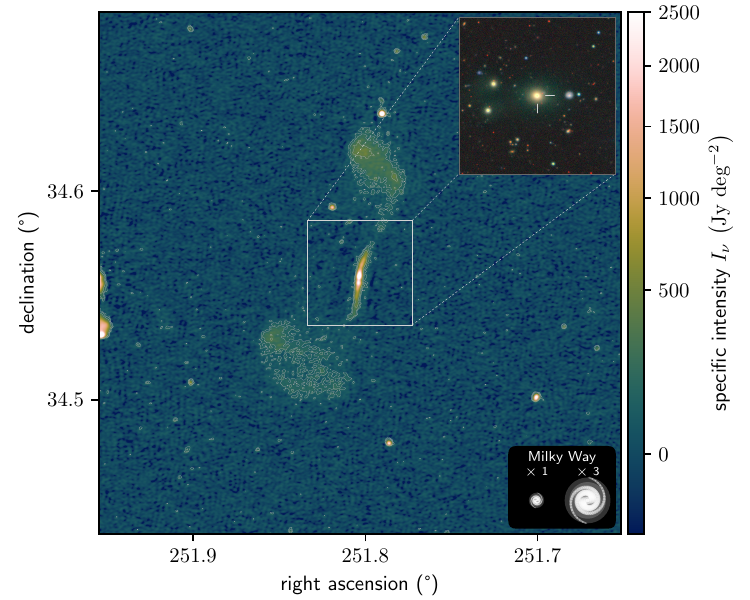}
    \end{subfigure}
    \begin{subfigure}{\columnwidth}
    \centering
    \includegraphics[width=\columnwidth]{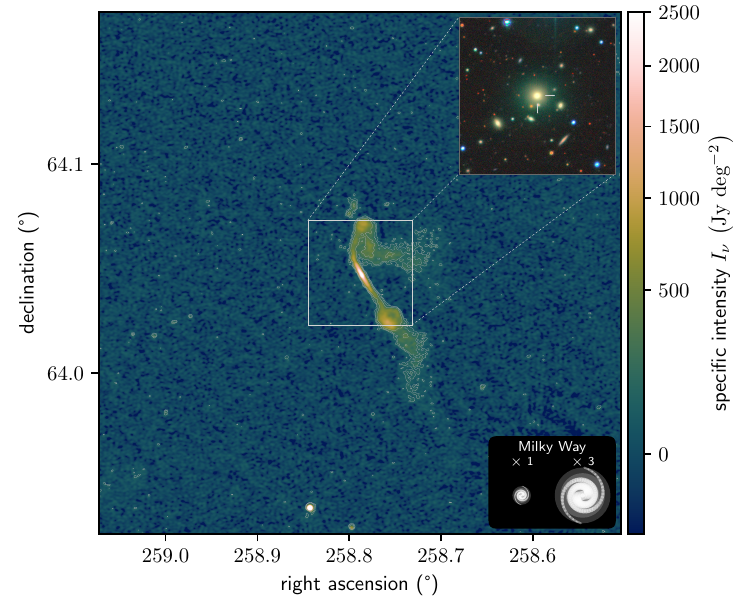}
    \end{subfigure}
    \caption{
    LoTSS DR2 cutouts at central observing frequency $\nu_\mathrm{obs} = 144\ \mathrm{MHz}$ and resolution $\theta_\mathrm{FWHM} = 6''$, centred around giant-generating BCGs of Local Universe clusters.
    All giants shown are discoveries of \citet{Oei12022GiantsSample} for which no previous images have been published.
    Each cutout covers a solid angle of $15' \times 15'$.
    Contours signify 3, 5, and 10 sigma-clipped standard deviations above the sigma-clipped median.
    For scale, we show the stellar Milky Way disk (with a diameter of 50 kpc) generated using the \citet{Ringermacher12009} formula, alongside a 3 times inflated version.
    Each DESI Legacy Imaging Surveys DR9 $(g,r,z)$ inset shows the central $3' \times 3'$ region.
    }
    \label{fig:BCGGiants}
\end{figure*}

As elaborated upon in Sect.~\ref{sec:CWLocalisation}, we retained \numberOfGRGsBORG\ giants in the part of the Local Universe where Cosmic Web analysis is possible.
Of these, \numberOfGRGsBORGzsp\ have spectroscopic redshifts; only these giants could be reliably localised.
We show their sky locations in Fig.~\ref{fig:sky}.
LoTSS DR2 discoveries make up \numberOfGRGsBORGzspOei\ of the final \numberOfGRGsBORGzsp\ giants (80\%). 
Through six example LoTSS DR2 giants, Fig.~\ref{fig:BCGGiants} provides the reader a sense of the quality of the radio imagery underpinning this work, from which angular lengths $\phi$ have been inferred, a sense of the reliability of our host galaxy identification, from which spectroscopic redshifts $z_\mathrm{s}$ have inferred, and a sense of the morphological and surface brightness diversity of the objects under consideration in this study.

\subsection{General radio galaxies}
In order to determine whether the Cosmic Web environments of giants are exceptional, we must create a sample of reference Cosmic Web environments.
For this reason, we also localised within the Cosmic Web a sample of general RGs, selected without regard for their length.

In particular, our starting point for this sample is the radio-bright active galactic nucleus (RLAGN) sub-sample described in \citet{Hardcastle12019}, which contains 23,344 of the 318,520 sources (7\%) in the LoTSS DR1 value-added catalogue of \citet{Williams12019}.
To construct their sub-sample, \citet{Hardcastle12019} combined multiple criteria that separate RLAGN from star-forming galaxies (SFGs).
At the low redshifts considered in this work ($z < z_\mathrm{max} \coloneqq 0.16$), most RLAGN are identified through SDSS spectroscopy, and we therefore expect high RLAGN completeness and minimal contamination from SFGs.

As elaborated upon in Sect.~\ref{sec:CWLocalisation}, we retained \numberOfRGsBORG\ LoTSS DR1 RGs in the part of the Local Universe where Cosmic Web analysis is possible.
Of these, \numberOfRGsBORGzsp\ have spectroscopic redshifts; only these RGs could be reliably localised.
We show also their sky locations in Fig.~\ref{fig:sky}.
The LoTSS DR1 image quality is very similar to the LoTSS DR2 image quality, which Fig.~\ref{fig:BCGGiants} illustrates.

\subsection{Cosmic Web late-time density field}
\label{sec:dataCW}
To localise RGs within the Cosmic Web, we used data products from the Bayesian Origin Reconstruction from Galaxies \citep[BORG;][]{Jasche12013} SDSS run \citep{Jasche12015}.
The BORG SDSS uses second-order Lagrangian perturbation theory \citep[2LPT;][]{Bouchet11995} to forward model structure formation and evaluates the plausibility of a proposed structure formation history by comparing its late-time density field to the three-dimensional positions of galaxies in the SDSS DR7 Main Galaxy Sample \citep[MGS;][]{Abazajian12009}.
As a result, the BORG SDSS provides a late-time total\footnote{The BORG algorithm does not differentiate between baryonic and dark matter, assuming identical behaviour on the multi-megaparsec scale.} matter density field posterior for the part of the Local Universe covered by the SDSS DR7 footprint.
The posterior is represented by a Hamiltonian Monte Carlo \citep[HMC;][]{Duane11987} Markov chain of approximately ten thousand samples after the burn-in phase.
Each sample covers the same volume of $(750\ \mathrm{Mpc}\ h^{-1})^3$ extent with a cubical grid of $256^3$ voxels.
Thus, the side length of a BORG SDSS voxel $L = \frac{1}{256} \cdot 750\ \mathrm{Mpc}\ h^{-1} \approx 2.9\ \mathrm{Mpc}\ h^{-1}$.
The late-time posterior can be compactly summarised by taking the mean and standard deviation (SD) of the samples on a per-voxel basis.
In this work, we use both the individual samples and the summary cubes --- the latter of which we call the BORG SDSS mean and SD.
However, by using these summary cubes, information contained in higher moments of single-voxel posteriors and inter-voxel correlations remains unused.

\citet{Leclercq12015} have extended the BORG SDSS by calculating, for each sample, the Cosmic Web classification as stipulated by the $T$-web definition \citep{Hahn12007}.
The result is a probabilistic classification with a marginal distribution for each voxel.
As one aspect of a broader information theoretic analysis, \citet{Leclercq12016} demonstrated how these classifications can be used to predict galaxy properties, such as $g-r$ colours.
We used these classifications to characterise the dynamical environments of luminous giants and general RGs.

Both the BORG SDSS posterior and a posterior from another BORG run, the BORG 2M++ \citep{Jasche12019}, have been used before to relate properties of active galaxies to the density of the enveloping Cosmic Web \citep{Frank12016, Porqueres12018}.
Our work is the first to relate properties of radio galaxies to BORG (or BORG-like) Cosmic Web reconstructions.

\section{Methods}
\label{sec:methods}
As pointed out in Sect.~\ref{sec:introduction}, the number of known giants has increased substantially in recent years.
For example, within the Local Universe covered by the SDSS DR7, the manual search of \citet{Oei12022GiantsSample} alone has quintupled the number of known giants with spectroscopic redshifts --- from 52 to \numberOfGRGsBORGzsp.
Meanwhile, the BORG SDSS now offers the first physically principled, probabilistic reconstruction of the total matter density field over this volume.
In this section, we explain how we combined both advances by localising giants and general RGs within the BORG SDSS.

\subsection{BORG SDSS localisation procedure}
\label{sec:CWLocalisation}
For each RG in our two samples (be it a giant or a general RG), we first transformed its right ascension, declination, and redshift into a vector with comoving coordinates $\mathbf{r} \coloneqq \left[x, y, z\right]^\top$ following \citet{Jasche12015}'s coordinate system convention.
This transformation is cosmology-dependent.
The BORG SDSS adopts the cosmological parameters of \citet{Jasche12015} to convert SDSS DR7 MGS redshifts into radial comoving distances and subsequently infer the structure formation history of the Local Universe.
In order to obtain valid RG localisations, it was imperative that we used the same conversion between redshift and radial comoving distance.
We therefore adopted the \citet{Jasche12015} cosmology for this particular step.
Afterwards, we simply associated each RG to the BORG SDSS voxel nearest to $\mathbf{r}$.

After associating RGs to voxels, we checked whether the late-time BORG SDSS density field at these locations is sufficiently constrained.
More precisely, we evaluated whether --- at each RG's voxel --- the survey response operator of \citet{Jasche12015}'s lowest $r$-band absolute magnitude bin ($-21.00 < M_r < -20.33$), $R^0$, equals or exceeds some threshold $R^0_\mathrm{min}$.
If indeed $R^0(\mathbf{r}) \geq R^0_\mathrm{min}$, we retained the RG for Cosmic Web analysis; if not, we discarded it.
In this work, we chose $R^0_\mathrm{min} = 0.1$.
Higher choices for $R^0_\mathrm{min}$ rid the sample of comparatively uninformative (that is, more prior-dominated) density and dynamical state measurements, but come at the obvious cost of a reduced sample size.

\begin{figure*}
    \centering
    \includegraphics[width=.99\textwidth]{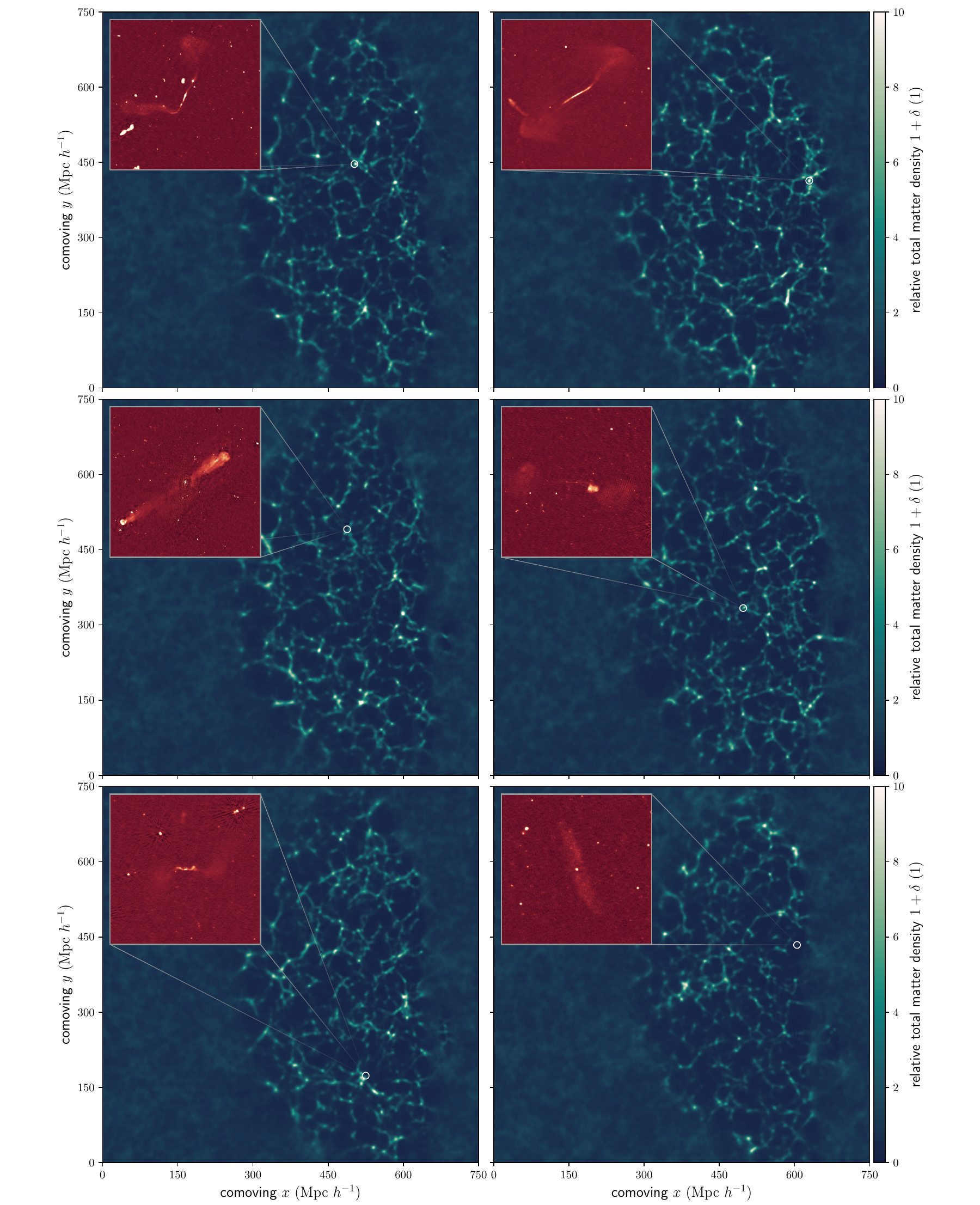}
    \caption{
Example localisations of giant radio galaxies within the large-scale structure of the Local Universe.
The top row shows two cluster giants, the middle row shows two filament giants, and the bottom row shows two sheet giants.
For each giant, we show a slice of constant Cartesian comoving $z$ through the late-time BORG SDSS posterior mean total matter density field and a LoTSS DR2 6$''$ image at $\nu_\mathrm{obs} = 144\ \mathrm{MHz}$ (inset).
Outside of the SDSS DR7--constrained volume, the posterior mean tends to the Universe's late-time mean total matter density $\bar{\rho}_0$.
The locations of the giants are marked by white circles.
}
\label{fig:localisation}
\end{figure*}
In Fig.~\ref{fig:localisation}, we show six example localisations of giants in the Local Universe.
We pinpointed the giants in the upper two panels to galaxy clusters, the giants in the middle two panels to filaments, and the giants in the bottom two panels to sheets.
Next, in order to obtain density distributions for each retained RG, we explored two methods.

\subsubsection{Fixed voxel method}
\label{sec:fixedVoxelMethod}
In the `fixed voxel method', we considered for each RG the marginal posterior density distribution at its voxel --- that is the posterior density distribution for that voxel in isolation --- despite the fact that the BORG SDSS provides reconstructions with complex inter-voxel density correlations.
This method is the simplest of the two methods we have used.

\subsubsection{Flexible voxel method}
\label{sec:flexibleVoxelMethod}
The fixed voxel method has at least two disadvantages.
One issue is the fact that the BORG SDSS has inferred the Cosmic Web with a limited set of bright SDSS DR7 galaxies, which (amongst other factors) causes reconstruction uncertainty.
In practice, this means that a given cluster or filament may morph and wiggle around in different BORG SDSS samples.
If one does not wiggle around the voxel to sample from accordingly (but sticks with the same voxel all the time), one regularly samples outside of the cluster or filament in which the RG of interest resides.
This, of course, biases the inferred densities low.

On top of Cosmic Web reconstruction uncertainty, host galaxy peculiar motion uncertainty leads to additional difficulty in the determination of an RG's Cosmic Web density.
In Appendix~\ref{ap:spectroscopicVsPhotometric}, we show that spectroscopic redshift uncertainties generally lead to sub-voxel localisation uncertainty, even after taking low-mass galaxy cluster--like peculiar motion into account.
For high-mass galaxy clusters however, peculiar motion can cause multi-voxel localisation errors.
We also show that photometric redshift uncertainties cause localisation uncertainties of $10^1$--$10^2\ \mathrm{Mpc}$, or up to tens of voxels, that are unworkably large.
Cosmic Web localisation therefore only seems possible for RGs with spectroscopically detected hosts.

To counteract the fixed voxel method's tendency to sample densities outside of the clusters and filaments in which the RGs truly reside, we propose a flexible voxel method.
In this method, the voxel considered hitherto serves as a reference voxel, around which we search for the most likely correct voxel to sample from.
More precisely, we iterate over $1000$ BORG SDSS samples equally spaced within the MCMC, and consider for each sample all voxels in a sphere of radius $5\ \mathrm{Mpc}\ h^{-1}$ around the reference voxel.
We then simply adopt the voxel with the highest density as the most likely correct voxel for that sample.
This procedure encapsulates our prior belief that the massive ellipticals which give rise to RGs are more likely to occur in a high-density region than in a low-density region of the same volume.

The search radius of $5\ \mathrm{Mpc}\ h^{-1}$ chosen here is arbitrary to some degree.
Clearly, for the flexible voxel method to be any different from the fixed voxel method, this radius must exceed the side length of a single voxel, $2.9\ \mathrm{Mpc}\ h^{-1}$.
For larger radii, we are able to correct for larger peculiar motion errors, and thus provide more accurate densities for RGs located in massive galaxy clusters.
At the same time, for larger radii, we are at risk of straying too far from the reference voxel; for example, this could lead to sampling cluster-like densities for RGs that actually reside in an adjacent filament.
Because the majority of RGs appears to reside in filaments, and because the gravity solver of the BORG SDSS significantly limits the usefulness of cluster densities (see Sect.~\ref{sec:discussion}), we chose a relatively `small' search radius of $5\ \mathrm{Mpc}\ h^{-1}$ --- less than two voxels in each direction.

\begin{figure}
    \centering
    \includegraphics[width=\columnwidth]{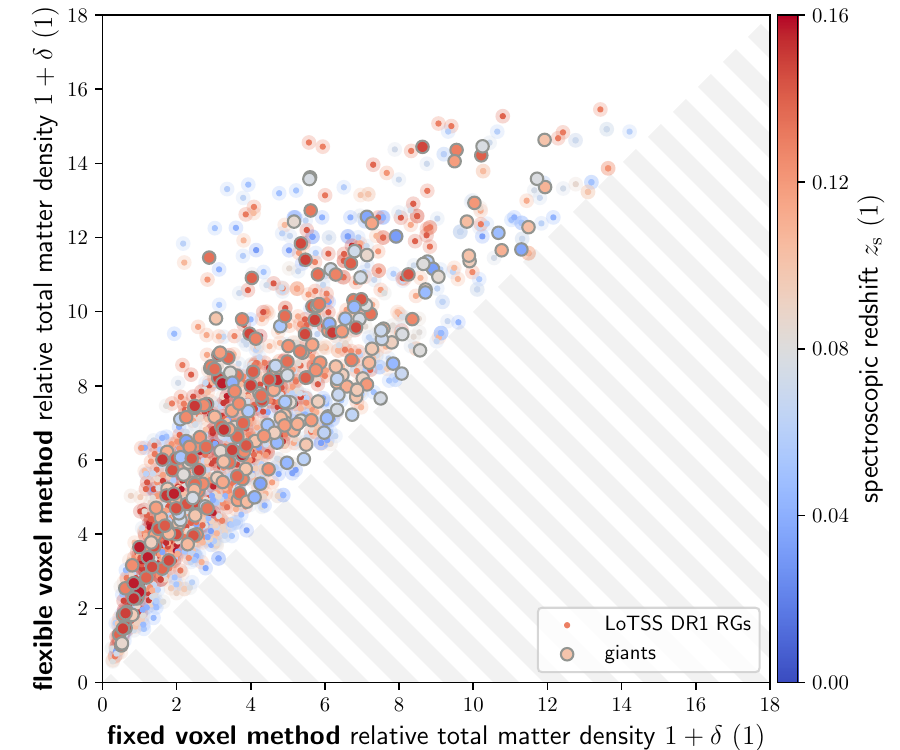}
    \caption{
    Comparison between Cosmic Web densities of giants and LoTSS DR1 RGs in the Local Universe, inferred via two variations of our BORG SDSS--based approach.
    For each RG, we show the mean density measured through Sect.~\ref{sec:fixedVoxelMethod}'s fixed voxel method (horizontal axis) and through Sect.~\ref{sec:flexibleVoxelMethod}'s flexible voxel method (vertical axis). 
    }
    \label{fig:densitiesFixedFlexible}
\end{figure}

In Fig.~\ref{fig:densitiesFixedFlexible}, we compare the mean Cosmic Web densities inferred through the two methods for all RGs considered in this work.
Flexible voxel--based relative densities are typically a factor two higher.
The methods agree most often at the lowest redshifts, where the BORG SDSS Cosmic Web reconstructions are least uncertain.

\subsection{BORG SDSS localisation in practice}
\label{sec:BORGSDSSLocalisationInPractice}
As described in Sect.~\ref{sec:CWLocalisation}, for each giant and general RG in our samples, we measured the marginal posterior density RV.
Of \numberOfGRGsBORG\ giants that lie within the part of the BORG SDSS volume where $R^0(\mathbf{r}) \geq R^0_\mathrm{min}$, there are $\numberOfGRGsBORGzsp$ with a spectroscopic redshift (93\%), which are therefore suitable for Cosmic Web analysis.
Of these, \numberOfGRGsBORGzspOei\ (80\%) are LoTSS DR2 discoveries \citep{Oei12022GiantsSample}.
In exactly the same way, of the \numberOfRGsBORG\ LoTSS DR1 RGs that lie within the constrained BORG SDSS volume, we retained \numberOfRGsBORGzsp\ specimina with spectroscopic redshifts (77\%), which we selected for Cosmic Web analysis.

At the BORG SDSS resolution of $2.9\ \mathrm{Mpc}\ h^{-1}$ per voxel side, the baryonic matter density field approximately equals the dark matter density field scaled down by a factor $\frac{\Omega_\mathrm{BM,0}}{\Omega_\mathrm{DM,0}}$.
The BORG SDSS does not distinguish between these two fields.
Instead, it provides the sum of the baryonic and dark matter density field: the total matter density field.
In this article, we exclusively mention relative total matter densities --- total matter densities divided by today's cosmic mean total matter density $\Omega_\mathrm{M,0} \rho_{\mathrm{c},0}$, where $\rho_{\mathrm{c},0}$ is today's critical density.
To be less verbose, we refer to just `relative densities', and write $1+\delta$, where $\delta$ denotes overdensity in the usual sense.

\begin{figure}
    \centering
    \includegraphics[width=\columnwidth]{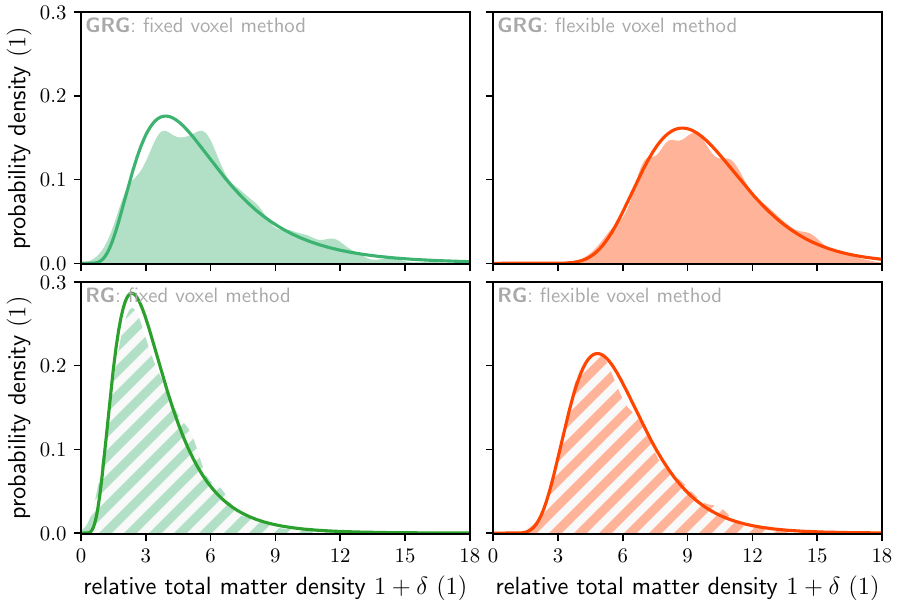}
    \caption{Distributions for the measured relative total matter density RVs $1 + \Delta_\mathrm{GRG,obs}\ \vert\ 1 + \Delta_\mathrm{GRG} = 1 + \delta$ and $1 + \Delta_\mathrm{RG,obs}\ \vert\ 1 + \Delta_\mathrm{RG} = 1 + \delta$ for an individual giant (top row; solid), and for an individual RG (bottom row; hatched).
    Fixed voxel method densities (left column; green) are lower than flexible voxel method densities (right column; orange).
    The MLE-fitted lognormal PDFs (solid curves) demonstrate that the distributions are almost lognormal.
    }
    \label{fig:GRGRGDensityDistributions}
\end{figure}

\begin{figure*}
    \centering
    \includegraphics[width=\textwidth]{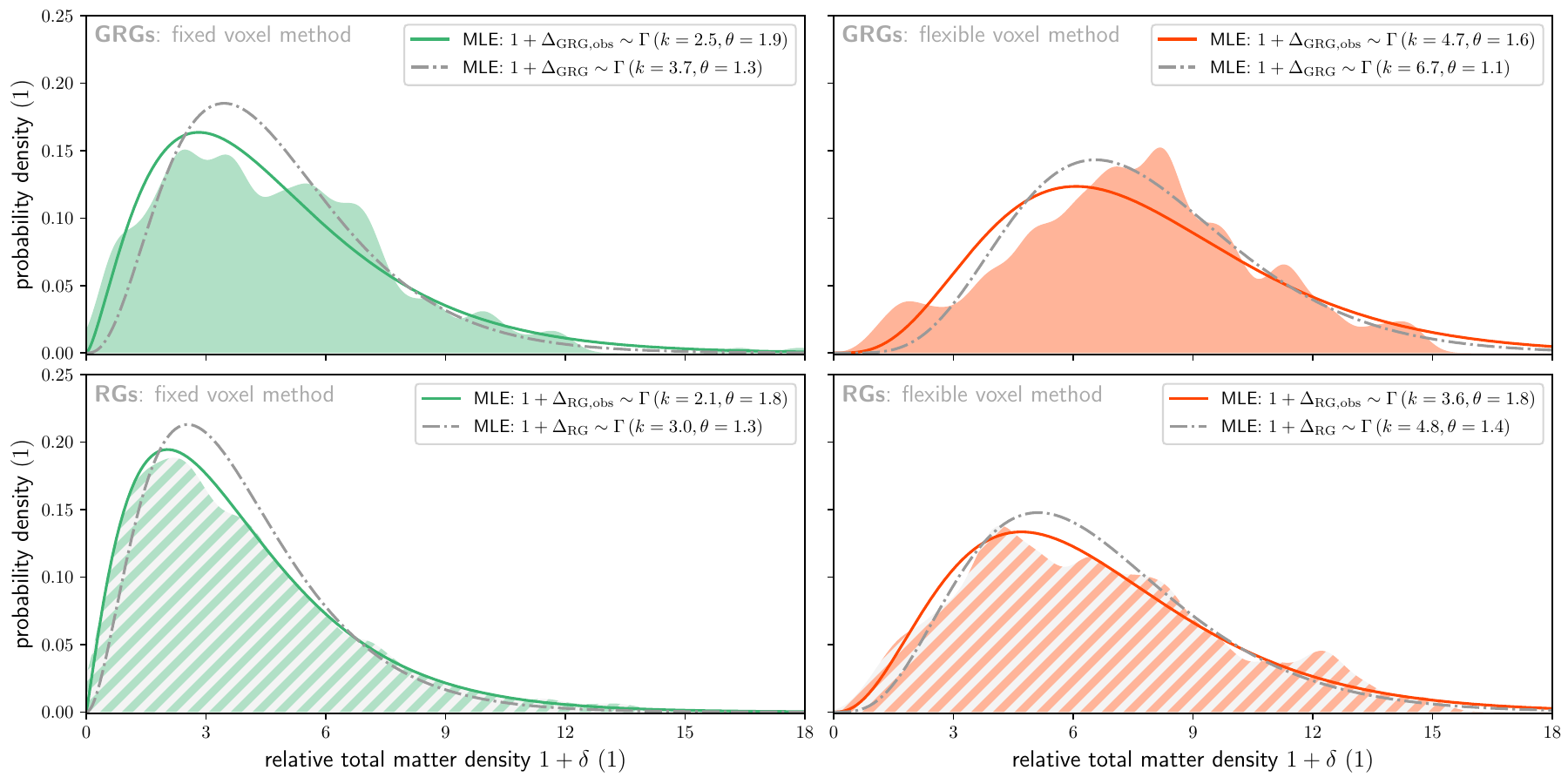}
    \caption{
    Probability density functions (PDFs) of the relative total matter density RVs of \numberOfGRGsBORGzsp\ Local Universe giants (top row), and of \numberOfRGsBORGzsp\ Local Universe RGs (bottom row), determined through the fixed voxel method (left column) and flexible voxel method (right column).
    These RVs correspond to the density field smoothed to a scale of $2.9\ \mathrm{Mpc}\ h^{-1}$.
    We also show PDFs of gamma-distributed RVs with parameters obtained via MLE, both before (solid lines) and after (dash-dotted lines) a heteroskedasticity correction.
    We warn that these distributions are affected by surface brightness selection, and thus represent known populations only.
    }
    \label{fig:distributionsDensity}
\end{figure*}

\section{Results}
\label{sec:results}
In this section, we present the first empirical distributions of the Cosmic Web density and dynamical state around luminous giants and general RGs.
We also determine the RG radio luminosity--Cosmic Web density relation, and test whether it can cause the density distribution discrepancy between luminous giants and general RGs.

\subsection{Cosmic Web density distributions}
\label{sec:CWDensityDistributions}
The relative density posterior distributions of individual RGs resemble lognormal distributions, irrespective of whether the fixed or flexible voxel method was used to determine them.
Figure~\ref{fig:GRGRGDensityDistributions} demonstrates this through example marginals for a typical luminous giant and a typical LoTSS DR1 RG.
Indeed modelling $1 + \Delta_\mathrm{GRG,obs}\ \vert\ 1 + \Delta_\mathrm{GRG} = 1 + \delta \sim \mathrm{Lognormal}(\mu, \sigma^2)$ and $1 + \Delta_\mathrm{RG,obs}\ \vert\ 1 + \Delta_\mathrm{RG} = 1 + \delta \sim \mathrm{Lognormal}(\mu, \sigma^2)$, we can succinctly summarise each RG's measured density distribution with two parameters.
We provide these, for 50 out of \numberOfGRGsBORGzsp\ giants, in Table~\ref{tab:GRGsCosmicWeb}.
(For access to such data for all giants, and for similar data on LoTSS DR1 giants, see the table's footnote.)

\begin{table*}
\centering
\caption{
Cosmic Web properties for 50 out of \numberOfGRGsBORGzsp\ BORG SDSS--constrained giants, sorted by right ascension.\protect\footnotemark
}
\label{tab:GRGsCosmicWeb}
\resizebox{\textwidth}{!}{%
\begin{tabular}{l l l l l l l l l l l}
\hline
rank & host coordinates & spectroscopic & Cosmic Web density & Cosmic Web density & \multicolumn{4}{l}{Cosmic Web $T$-web} & cluster mass & host\\
$\downarrow$ & equatorial J2000 $(\degree)$ & redshift $z_\mathrm{s}\ (1)$ & fixed voxel $\mu, \sigma^2\ (1)$ & flexible voxel $\mu, \sigma^2\ (1)$ & \multicolumn{4}{l}{probabilities $\vec{p}\ (1)$} & $M_{500}\ (10^{14}\ M_\odot)$ & BCG\\
\hline
1 & $111.57678,\ 38.63317$ & $0.15387$ $\pm$ $3 \cdot 10^{-5}$ & $-0.49$, $0.60$ & \textcolor{white}{+}$0.60$, $0.45$ & $(0.0,$ & $0.2,$ & $0.7,$ & $0.1)$ & - & -\\
2 & $113.77185,\ 41.97432$ & $0.08732$ $\pm$ $2 \cdot 10^{-5}$ & \textcolor{white}{+}$1.94$, $0.06$ & \textcolor{white}{+}$2.43$, $0.03$ & $(0.1,$ & $0.9,$ & $0.0,$ & $0.0)$ & - & -\\
3 & $114.98866,\ 43.98326$ & $0.14887$ $\pm$ $2 \cdot 10^{-5}$ & $-0.73$, $0.52$ & \textcolor{white}{+}$0.42$, $0.42$ & $(0.0,$ & $0.2,$ & $0.6,$ & $0.2)$ & - & -\\
4 & $115.96358,\ 28.35779$ & $0.10633$ $\pm$ $2 \cdot 10^{-5}$ & \textcolor{white}{+}$0.88$, $0.53$ & \textcolor{white}{+}$1.87$, $0.20$ & $(0.1,$ & $0.7,$ & $0.2,$ & $0.0)$ & - & -\\
5 & $116.03381,\ 43.99169$ & $0.13484$ $\pm$ $2 \cdot 10^{-5}$ & \textcolor{white}{+}$0.95$, $0.23$ & \textcolor{white}{+}$2.41$, $0.05$ & $(0.0,$ & $0.9,$ & $0.1,$ & $0.0)$ & - & -\\
6 & $117.53962,\ 26.73550$ & $0.13042$ $\pm$ $2 \cdot 10^{-5}$ & \textcolor{white}{+}$1.59$, $0.17$ & \textcolor{white}{+}$2.16$, $0.06$ & $(0.2,$ & $0.8,$ & $0.0,$ & $0.0)$ & - & -\\
7 & $118.14466,\ 35.83983$ & $0.13650$ $\pm$ $3 \cdot 10^{-5}$ & \textcolor{white}{+}$0.95$, $0.25$ & \textcolor{white}{+}$2.09$, $0.10$ & $(0.0,$ & $0.8,$ & $0.2,$ & $0.0)$ & - & -\\
8 & $119.15943,\ 32.46314$ & $0.14620$ $\pm$ $2 \cdot 10^{-5}$ & \textcolor{white}{+}$0.47$, $0.50$ & \textcolor{white}{+}$1.28$, $0.23$ & $(0.1,$ & $0.5,$ & $0.4,$ & $0.0)$ & - & -\\
9 & $119.26227,\ 36.62242$ & $0.13948$ $\pm$ $3 \cdot 10^{-5}$ & $-0.02$, $0.42$ & \textcolor{white}{+}$0.94$, $0.22$ & $(0.0,$ & $0.2,$ & $0.8,$ & $0.0)$ & - & -\\
10 & $119.27937,\ 35.97967$ & $0.13296$ $\pm$ $2 \cdot 10^{-5}$ & \textcolor{white}{+}$1.15$, $0.28$ & \textcolor{white}{+}$2.08$, $0.07$ & $(0.1,$ & $0.8,$ & $0.1,$ & $0.0)$ & $1.3$ & $y$\\
11 & $119.47158,\ 36.67279$ & $0.12844$ $\pm$ $2 \cdot 10^{-5}$ & \textcolor{white}{+}$1.26$, $0.15$ & \textcolor{white}{+}$1.66$, $0.08$ & $(0.1,$ & $0.8,$ & $0.1,$ & $0.0)$ & $0.8$ & $y$\\
12 & $120.25565,\ 13.83117$ & $0.10872$ $\pm$ $2 \cdot 10^{-5}$ & \textcolor{white}{+}$2.46$, $0.03$ & \textcolor{white}{+}$2.58$, $0.02$ & $(0.7,$ & $0.3,$ & $0.0,$ & $0.0)$ & $3.2$ & $y$\\
13 & $120.30535,\ 34.67522$ & $0.08269$ $\pm$ $2 \cdot 10^{-5}$ & \textcolor{white}{+}$1.56$, $0.08$ & \textcolor{white}{+}$1.93$, $0.04$ & $(0.2,$ & $0.8,$ & $0.0,$ & $0.0)$ & - & -\\
14 & $120.38318,\ 47.60446$ & $0.15684$ $\pm$ $3 \cdot 10^{-5}$ & $-0.30$, $0.56$ & \textcolor{white}{+}$1.14$, $0.32$ & $(0.0,$ & $0.4,$ & $0.6,$ & $0.0)$ & - & -\\
15 & $120.80515,\ 51.93263$ & $0.06947$ $\pm$ $2 \cdot 10^{-5}$ & \textcolor{white}{+}$2.00$, $0.04$ & \textcolor{white}{+}$2.21$, $0.03$ & $(0.3,$ & $0.7,$ & $0.0,$ & $0.0)$ & - & -\\
16 & $121.01419,\ 40.80261$ & $0.12617$ $\pm$ $1 \cdot 10^{-5}$ & \textcolor{white}{+}$0.85$, $0.20$ & \textcolor{white}{+}$1.59$, $0.11$ & $(0.1,$ & $0.8,$ & $0.1,$ & $0.0)$ & $2.5$ & $y$\\
17 & $121.32789,\ 28.62417$ & $0.14256$ $\pm$ $2 \cdot 10^{-5}$ & \textcolor{white}{+}$1.18$, $0.39$ & \textcolor{white}{+}$1.78$, $0.17$ & $(0.1,$ & $0.6,$ & $0.3,$ & $0.0)$ & - & -\\
18 & $121.38042,\ 25.80317$ & $0.13698$ $\pm$ $2 \cdot 10^{-5}$ & \textcolor{white}{+}$1.02$, $0.31$ & \textcolor{white}{+}$1.89$, $0.09$ & $(0.0,$ & $0.8,$ & $0.2,$ & $0.0)$ & - & -\\
19 & $121.42954,\ 16.23223$ & $0.10002$ $\pm$ $2 \cdot 10^{-5}$ & \textcolor{white}{+}$2.46$, $0.03$ & \textcolor{white}{+}$2.67$, $0.02$ & $(0.6,$ & $0.4,$ & $0.0,$ & $0.0)$ & $0.9$ & $y$\\
20 & $122.14848,\ 38.91450$ & $0.04081$ $\pm$ $1 \cdot 10^{-5}$ & \textcolor{white}{+}$1.79$, $0.04$ & \textcolor{white}{+}$2.26$, $0.02$ & $(0.1,$ & $0.9,$ & $0.0,$ & $0.0)$ & - & -\\
21 & $122.28630,\ 29.67904$ & $0.12572$ $\pm$ $2 \cdot 10^{-5}$ & \textcolor{white}{+}$1.18$, $0.18$ & \textcolor{white}{+}$2.03$, $0.06$ & $(0.1,$ & $0.8,$ & $0.1,$ & $0.0)$ & - & -\\
22 & $122.31280,\ 41.28897$ & $0.13344$ $\pm$ $3 \cdot 10^{-5}$ & \textcolor{white}{+}$0.86$, $0.38$ & \textcolor{white}{+}$1.48$, $0.14$ & $(0.2,$ & $0.6,$ & $0.2,$ & $0.0)$ & - & -\\
23 & $123.41209,\ 41.36503$ & $0.09984$ $\pm$ $2 \cdot 10^{-5}$ & \textcolor{white}{+}$1.05$, $0.15$ & \textcolor{white}{+}$2.27$, $0.04$ & $(0.0,$ & $0.7,$ & $0.3,$ & $0.0)$ & $0.8$ & $y$\\
24 & $123.91600,\ 50.54063$ & $0.13803$ $\pm$ $3 \cdot 10^{-5}$ & \textcolor{white}{+}$1.16$, $0.29$ & \textcolor{white}{+}$1.85$, $0.08$ & $(0.2,$ & $0.6,$ & $0.2,$ & $0.0)$ & - & -\\
25 & $124.44458,\ 54.70087$ & $0.11867$ $\pm$ $2 \cdot 10^{-5}$ & \textcolor{white}{+}$2.22$, $0.07$ & \textcolor{white}{+}$2.63$, $0.03$ & $(0.2,$ & $0.8,$ & $0.0,$ & $0.0)$ & $4.9$ & $n$\\
26 & $125.78050,\ 10.59828$ & $0.06605$ $\pm$ $1 \cdot 10^{-4}$ & \textcolor{white}{+}$0.95$, $0.14$ & \textcolor{white}{+}$1.53$, $0.05$ & $(0.2,$ & $0.6,$ & $0.2,$ & $0.0)$ & - & -\\
27 & $126.47417,\ 41.60254$ & $0.15359$ $\pm$ $2 \cdot 10^{-5}$ & \textcolor{white}{+}$1.38$, $0.38$ & \textcolor{white}{+}$2.03$, $0.15$ & $(0.2,$ & $0.7,$ & $0.1,$ & $0.0)$ & - & -\\
28 & $127.86456,\ 32.32412$ & $0.05120$ $\pm$ $1 \cdot 10^{-5}$ & \textcolor{white}{+}$0.69$, $0.12$ & \textcolor{white}{+}$1.78$, $0.04$ & $(0.0,$ & $1.0,$ & $0.0,$ & $0.0)$ & - & -\\
29 & $127.99873,\ 30.65853$ & $0.10704$ $\pm$ $2 \cdot 10^{-5}$ & \textcolor{white}{+}$1.80$, $0.08$ & \textcolor{white}{+}$2.07$, $0.04$ & $(0.4,$ & $0.6,$ & $0.0,$ & $0.0)$ & - & -\\
30 & $128.14208,\ 4.41000$ & $0.10600$ $\pm$ $1 \cdot 10^{-5}$ & \textcolor{white}{+}$0.47$, $0.24$ & \textcolor{white}{+}$1.34$, $0.10$ & $(0.0,$ & $0.3,$ & $0.7,$ & $0.0)$ & - & -\\
31 & $129.03256,\ 26.81206$ & $0.08780$ $\pm$ $2 \cdot 10^{-5}$ & \textcolor{white}{+}$1.22$, $0.12$ & \textcolor{white}{+}$1.91$, $0.04$ & $(0.1,$ & $0.7,$ & $0.2,$ & $0.0)$ & - & -\\
32 & $130.18516,\ 58.69709$ & $0.14392$ $\pm$ $3 \cdot 10^{-5}$ & \textcolor{white}{+}$0.15$, $0.40$ & \textcolor{white}{+}$1.21$, $0.19$ & $(0.0,$ & $0.4,$ & $0.6,$ & $0.0)$ & $1.2$ & $y$\\
33 & $130.50142,\ 38.93753$ & $0.11977$ $\pm$ $2 \cdot 10^{-5}$ & \textcolor{white}{+}$1.82$, $0.09$ & \textcolor{white}{+}$2.23$, $0.04$ & $(0.2,$ & $0.8,$ & $0.0,$ & $0.0)$ & $0.9$ & $n$\\
34 & $130.53983,\ 41.94039$ & $0.12565$ $\pm$ $2 \cdot 10^{-5}$ & \textcolor{white}{+}$0.67$, $0.31$ & \textcolor{white}{+}$1.93$, $0.08$ & $(0.0,$ & $0.8,$ & $0.2,$ & $0.0)$ & - & -\\
35 & $131.24330,\ 42.07739$ & $0.14932$ $\pm$ $2 \cdot 10^{-5}$ & \textcolor{white}{+}$1.57$, $0.30$ & \textcolor{white}{+}$2.40$, $0.08$ & $(0.1,$ & $0.8,$ & $0.1,$ & $0.0)$ & - & -\\
36 & $131.36289,\ 44.92399$ & $0.15062$ $\pm$ $3 \cdot 10^{-5}$ & \textcolor{white}{+}$1.70$, $0.27$ & \textcolor{white}{+}$2.19$, $0.10$ & $(0.3,$ & $0.6,$ & $0.1,$ & $0.0)$ & $0.8$ & $n$\\
37 & $132.73665,\ 42.80464$ & $0.09238$ $\pm$ $1 \cdot 10^{-5}$ & \textcolor{white}{+}$1.05$, $0.17$ & \textcolor{white}{+}$1.68$, $0.05$ & $(0.2,$ & $0.8,$ & $0.0,$ & $0.0)$ & $1.3$ & $y$\\
38 & $133.45742,\ 14.87390$ & $0.06933$ $\pm$ $2 \cdot 10^{-5}$ & \textcolor{white}{+}$2.00$, $0.04$ & \textcolor{white}{+}$2.24$, $0.02$ & $(0.8,$ & $0.2,$ & $0.0,$ & $0.0)$ & $0.7$ & $y$\\
39 & $133.80972,\ 49.19334$ & $0.11773$ $\pm$ $1 \cdot 10^{-5}$ & \textcolor{white}{+}$0.99$, $0.36$ & \textcolor{white}{+}$2.14$, $0.09$ & $(0.0,$ & $0.8,$ & $0.2,$ & $0.0)$ & - & -\\
40 & $134.23524,\ 47.95594$ & $0.14927$ $\pm$ $2 \cdot 10^{-5}$ & \textcolor{white}{+}$1.20$, $0.44$ & \textcolor{white}{+}$2.00$, $0.17$ & $(0.1,$ & $0.7,$ & $0.2,$ & $0.0)$ & - & -\\
41 & $134.58152,\ 46.37086$ & $0.11711$ $\pm$ $1 \cdot 10^{-5}$ & \textcolor{white}{+}$1.19$, $0.24$ & \textcolor{white}{+}$1.54$, $0.11$ & $(0.2,$ & $0.6,$ & $0.2,$ & $0.0)$ & - & -\\
42 & $135.36742,\ 55.04455$ & $0.04602$ & \textcolor{white}{+}$2.15$, $0.03$ & \textcolor{white}{+}$2.35$, $0.01$ & $(0.4,$ & $0.6,$ & $0.0,$ & $0.0)$ & - & -\\
43 & $135.45964,\ 55.92429$ & $0.14090$ $\pm$ $2 \cdot 10^{-5}$ & \textcolor{white}{+}$1.60$, $0.21$ & \textcolor{white}{+}$2.06$, $0.09$ & $(0.4,$ & $0.6,$ & $0.0,$ & $0.0)$ & - & -\\
44 & $136.09149,\ 19.72455$ & $0.09953$ $\pm$ $2 \cdot 10^{-5}$ & \textcolor{white}{+}$1.30$, $0.10$ & \textcolor{white}{+}$1.72$, $0.06$ & $(0.1,$ & $0.9,$ & $0.0,$ & $0.0)$ & - & -\\
45 & $137.13529,\ 18.28034$ & $0.11554$ $\pm$ $3 \cdot 10^{-5}$ & \textcolor{white}{+}$1.17$, $0.17$ & \textcolor{white}{+}$1.96$, $0.06$ & $(0.1,$ & $0.8,$ & $0.1,$ & $0.0)$ & - & -\\
46 & $137.23745,\ 58.38413$ & $0.14364$ $\pm$ $3 \cdot 10^{-5}$ & \textcolor{white}{+}$0.17$, $0.52$ & \textcolor{white}{+}$1.33$, $0.20$ & $(0.0,$ & $0.6,$ & $0.4,$ & $0.0)$ & - & -\\
47 & $137.87830,\ 48.54753$ & $0.10865$ $\pm$ $2 \cdot 10^{-5}$ & \textcolor{white}{+}$0.32$, $0.30$ & \textcolor{white}{+}$1.45$, $0.09$ & $(0.0,$ & $0.6,$ & $0.4,$ & $0.0)$ & - & -\\
48 & $139.74751,\ 31.86128$ & $0.06194$ $\pm$ $1 \cdot 10^{-5}$ & \textcolor{white}{+}$1.65$, $0.08$ & \textcolor{white}{+}$1.78$, $0.04$ & $(0.6,$ & $0.4,$ & $0.0,$ & $0.0)$ & - & -\\
49 & $139.95191,\ 57.84889$ & $0.13695$ $\pm$ $3 \cdot 10^{-5}$ & \textcolor{white}{+}$1.21$, $0.17$ & \textcolor{white}{+}$1.67$, $0.09$ & $(0.1,$ & $0.8,$ & $0.1,$ & $0.0)$ & $0.8$ & $n$\\
50 & $140.34212,\ 54.86499$ & $0.04469$ $\pm$ $1 \cdot 10^{-5}$ & \textcolor{white}{+}$1.90$, $0.02$ & \textcolor{white}{+}$2.31$, $0.02$ & $(0.0,$ & $0.6,$ & $0.4,$ & $0.0)$ & - & -\\
\end{tabular}%
}
\end{table*}
\footnotetext{
We provide MLE parameters of lognormal fits to total matter (i.e. both baryonic and dark matter) relative density distributions.
We report parameters for both the fixed and flexible voxel methods.
Densities span the BORG SDSS comoving voxel volume of $(2.9\ \mathrm{Mpc}\ h^{-1})^3$ and are relative to today's cosmic mean total matter density.
The mean and variance of a density distribution are $\mathbb{E}[1+\Delta_\mathrm{GRG,obs}\ \vert\ 1+\Delta_\mathrm{GRG} = 1 + \delta_i] = \exp{(\mu + \frac{1}{2}\sigma^2)}$ and $\mathbb{V}[1+\Delta_\mathrm{GRG,obs}\ \vert\ 1+\Delta_\mathrm{GRG} = 1 + \delta_i] = (\exp{(\sigma^2)}-1)\exp{(2\mu + \sigma^2)}$.
The components of the $T$-web probability vector $\vec{p} = (p_1, p_2, p_3, p_4)$ correspond to clusters, filaments, sheets, and voids, respectively.
Cluster masses stem from crossmatching with \citet{Wen12015}.
We share a full table (with all \numberOfGRGsBORGzsp\ entries), alongside an analogous table for our \numberOfRGsBORGzsp\ selected LoTSS DR1 RGs, via the CDS.
}

By aggregating just the means of these marginal relative density distributions, we could analyse the distributions for our observed populations as a whole.
In green, Fig.~\ref{fig:distributionsDensity} shows kernel density estimated (KDE) relative density distributions for both luminous giants (top panels) and the broader population of RGs (bottom panels) in the Local Universe.
The panels in the left column represent the fixed voxel method, whilst the panels in the right column represent the flexible voxel method.
More precisely, these KDE distributions approximate the distributions of the observed GRG relative density RV $1+\Delta_\mathrm{GRG,obs}$ and the observed RG relative density RV $1+\Delta_\mathrm{RG,obs}$.

We sought to summarise these distributions parametrically.
After testing various two-parameter distributions for continuous, non-negative RVs (such as the gamma distribution and the lognormal distribution), the KDE distributions of $1+\Delta_\mathrm{RG,obs}$ in the bottom panels --- which are based on all \numberOfRGsBORGzsp\ selected LoTSS DR1 RGs --- appeared best approximated by a gamma distribution.\footnote{In Appendix~\ref{ap:gammaAnsatz}, we provide an astrophysical--statistical argument that could explain the gamma distribution's emergence.}
Maximum likelihood estimation (MLE) suggested $1 + \Delta_\mathrm{RG,obs}\sim \Gamma(k = 2.1, \theta = 1.8)$ for the fixed voxel method and $1 + \Delta_\mathrm{RG,obs}\sim \Gamma(k = 3.6, \theta = 1.8)$ for the flexible voxel method.
Similarly, fitting a gamma distribution to $1+\Delta_\mathrm{GRG,obs}$ through MLE gave $1 + \Delta_\mathrm{GRG,obs} \sim \Gamma(k = 2.5, \theta = 1.9)$ for the fixed voxel method and $1 + \Delta_\mathrm{GRG,obs} \sim \Gamma(k = 4.7, \theta = 1.6)$ for the flexible voxel method, although the latter does not provide a tight fit.
These gamma distributions constitute practical, two-parameter representations of the underlying data and are drawn as solid lines in Fig.~\ref{fig:distributionsDensity}.

The mean of a gamma-distributed RV $1 + \Delta \sim \Gamma(k, \theta)$ is $\mathbb{E}[1+\Delta] = k\theta$.
Thus, for the fixed voxel method, $\mathbb{E}[1+\Delta_\mathrm{GRG,obs}] = 4.8$ and $\mathbb{E}[1+\Delta_\mathrm{RG,obs}] = 3.8$.
Similarly, for the flexible voxel method, $\mathbb{E}[1+\Delta_\mathrm{GRG,obs}] = 7.5$ and $\mathbb{E}[1+\Delta_\mathrm{RG,obs}] = 6.5$.
Naively, it appears that we can conclude that, in a statistical sense, giants occupy denser regions of the Cosmic Web than radio galaxies in general.
Using a two-sample Kolmogorov--Smirnov (KS) test, we formally tested the null hypothesis that the observed giant and general RG relative density distributions of Fig.~\ref{fig:distributionsDensity} share a common underlying distribution.
For both the fixed and flexible voxel methods, the $p$-value $p \lesssim 10^{-6}$.
The null hypothesis is thus rejected (for typical significance levels).
These distributions come with at least two major caveats, however.

\begin{figure*}[t]
    \centering
    \includegraphics[width=\textwidth]{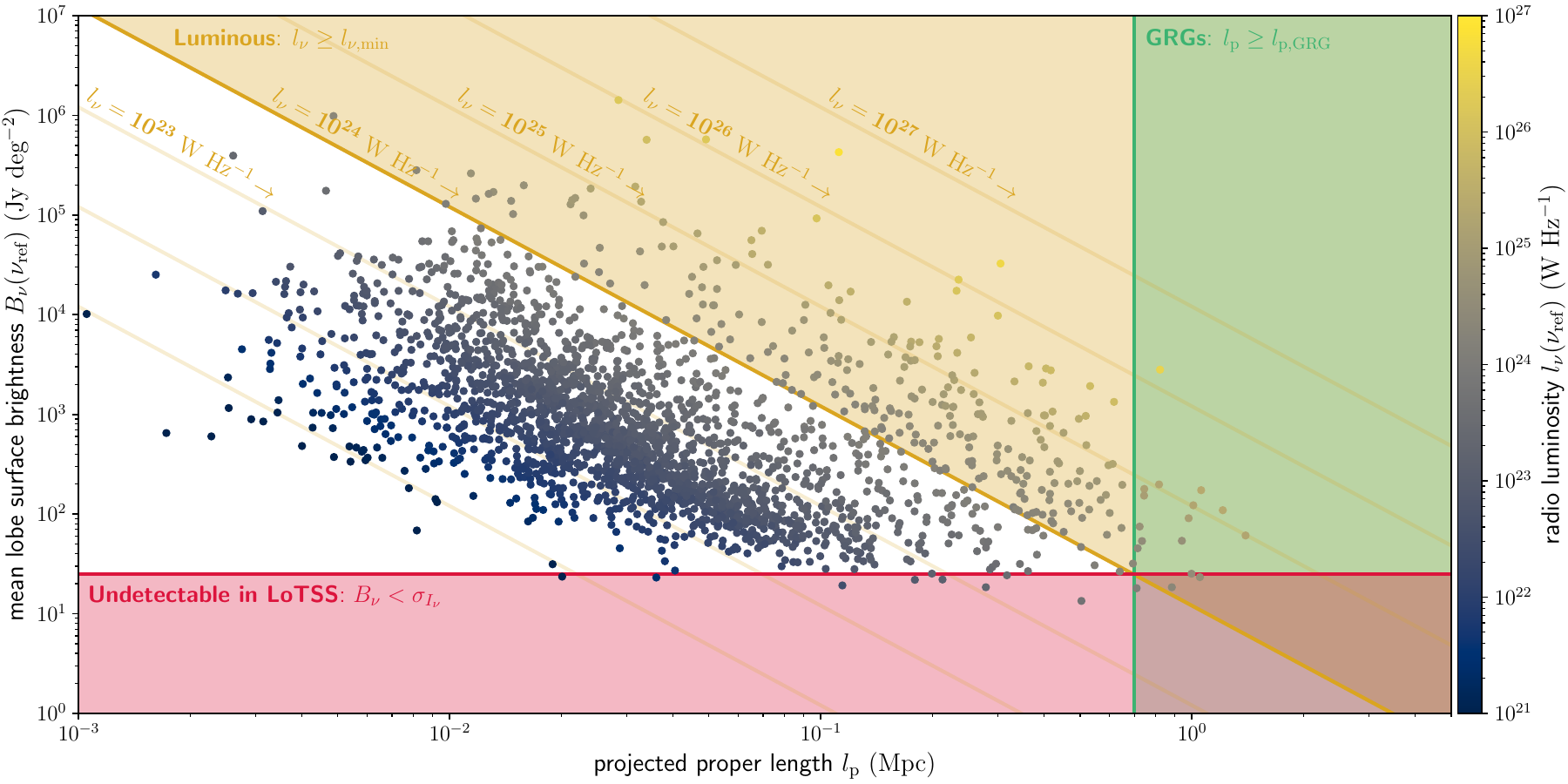}
    \caption{
    Mean lobe surface brightness estimates $B_\nu$ at $\nu_\mathrm{ref} = 150\ \mathrm{MHz}$ for LoTSS DR1 RGs at $z < 0.2$, shown as a function of projected proper length $l_\mathrm{p}$.
    During the majority (${\sim}90\%$) of their lifetime, RGs grow while maintaining a roughly constant radio luminosity.
    If, additionally, growth is shape-preserving, surface brightness decreases quadratically with $l_\mathrm{p}$.
    The seven golden lines therefore denote approximate evolutionary tracks of RGs at $z = 0$ with end-of-life radio luminosities $l_\nu \in \{10^{21}\ \mathrm{W\ Hz^{-1}}, ..., 10^{27}\ \mathrm{W\ Hz^{-1}}\}$.
    The region of parameter space where RGs can become detectable giants, $l_\nu \geq l_{\nu,\mathrm{min}}\coloneqq 10^{24}\ \mathrm{W\ Hz^{-1}}$, is shaded gold.
    The region of parameter space occupied by giants, $l_\mathrm{p} \geq l_\mathrm{p,GRG} \coloneqq 0.7\ \mathrm{Mpc}$, is shaded green.
    The region of parameter space inaccessible to the LoTSS, $B_\nu < \sigma_{I_\nu} = 25\ \mathrm{Jy\ deg^{-2}}$, is shaded red.
    }
    \label{fig:SBSelection}
\end{figure*}

\subsubsection{Heteroskedasticity}
First, BORG SDSS density measurements are heteroskedastic: higher densities have higher measurement errors than lower densities \citep[see e.g. Fig. 6 of][]{Jasche12015}.
Heteroskedasticity causes the observed distributions to differ from the intrinsic distributions; not only by widening them (as also occurs in the more familiar homoskedastic setting), but also by systematically shifting the distributions towards lower densities.
Appendix~\ref{ap:heteroskedasticity} presents a simple method to infer distributions corrected for this effect.
We thus distinguish between $1 + \Delta_\mathrm{GRG,obs}$ and $1 + \Delta_\mathrm{GRG}$, and between $1 + \Delta_\mathrm{RG,obs}$ and $1 + \Delta_\mathrm{RG}$; in both cases, the latter RV is heteroskedasticity-corrected.
As the effect induces minor shifts only, we assumed that the heteroskedasticity-free distributions are also approximately gamma.
We then applied the method of Appendix~\ref{ap:heteroskedasticity} to both our GRG and RG data.
Having no closed-form MLE expressions at our disposal, we obtained the MLE parameters of the heteroskedasticity-free gamma distributions by evaluating the likelihood function over an exhaustive grid of $(k, \theta)$ values.
For the fixed voxel method, this yielded $1 + \Delta_\mathrm{GRG} \sim \Gamma(k = 3.7, \theta = 1.3)$ and $1 + \Delta_\mathrm{RG} \sim \Gamma(k = 3.0, \theta = 1.3)$; $\mathbb{E}[1+\Delta_\mathrm{GRG}] = 4.8$ and $\mathbb{E}[1+\Delta_\mathrm{RG}] = 3.9$.
For the flexible voxel method, this yielded $1 + \Delta_\mathrm{GRG} \sim \Gamma(k = 6.7, \theta = 1.1)$ and $1 + \Delta_\mathrm{RG} \sim \Gamma(k = 4.8, \theta = 1.4)$; $\mathbb{E}[1+\Delta_\mathrm{GRG}] = 7.4$ and $\mathbb{E}[1+\Delta_\mathrm{RG}] = 6.7$.
We draw the corresponding PDFs as grey, dash-dotted lines in Fig.~\ref{fig:distributionsDensity}.

\begin{figure*}
    \centering
    \includegraphics[width=\textwidth]{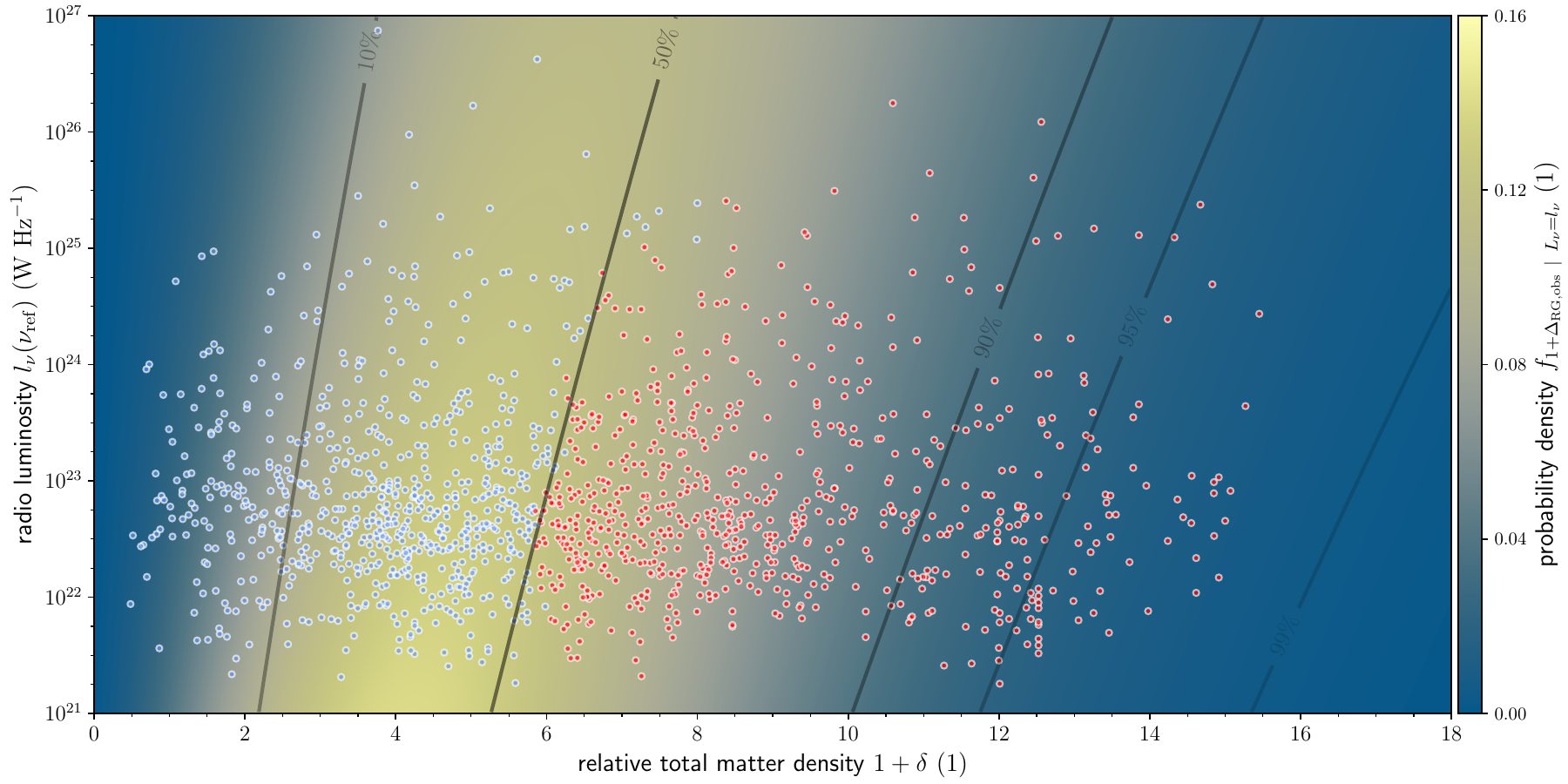}
    \caption{
    PDFs of the observed RG relative total matter density RV given a radio luminosity at $\nu_\mathrm{ref} = 150\ \mathrm{MHz}$, using MLE parameter values for the model described in Sect.~\ref{sec:radioLuminosityCosmicWebDensity}.
    The black contours denote CDF values.
    We overplot all \numberOfRGsBORGzsp\ selected LoTSS DR1 RGs (dots), with those above the empirical median density coloured red, and those below coloured blue.
    We used flexible voxel method densities.
    For fixed voxel method densities, see Fig.~\ref{fig:modelDensityLuminosityScatterFixed}.
    }
    \label{fig:modelDensityLuminosityScatterFlexible}
\end{figure*}

\subsubsection{Surface brightness selection}
The second --- and plausibly most important --- caveat is the imprint of selection effects, and in particular the surface brightness selection effect induced by the LoTSS noise level $\sigma_{I_\nu} = 25\ \mathrm{Jy\ deg^{-2}}$ at $\theta_\mathrm{FWHM} = 6''$.
This effect causes RGs whose lobe surface brightnesses are below the noise level to evade sample inclusion.
Lobe surface brightness depends on the RG's projected proper length $l_\mathrm{p}$, morphological parameters $\textphnc{\Af}_l$ and $\textphnc{\Af}_{l_\nu}$, radio luminosity $l_\nu$, redshift $z$, and lobe spectral index $\alpha$.
Equation~40 of \citet{Oei12022GiantsSample} presents an approximate formula for the mean lobe surface brightness $B_\nu$:
\begin{align}
    B_\nu(\nu_\mathrm{ref}) = \frac{2 \cdot \textphnc{\Af}_{l_\nu} \cdot l_\nu(\nu_\mathrm{ref})}{\pi^2 \cdot \mathbb{E}[D](\eta(\textphnc{\Af}_l))\cdot \textphnc{\Af}^2_l \cdot l_\mathrm{p}^2 \cdot (1+z)^{3-\alpha}},
\label{eq:surfaceBrightness}
\end{align}
where $\textphnc{\Af}_{l_\nu}$ is the fraction of the total radio luminosity that belongs to the lobes, $\textphnc{\Af}_l$ is the fraction of the RG's axis that lies inside the lobes, $\eta(\textphnc{\Af}_l) = \frac{\textphnc{\Af}_l}{2-\textphnc{\Af}_l}$, and $\mathbb{E}[D]$ is the mean deprojection factor as given by Eq.~A.29 of \citet{Oei12022GiantsSample}:
\begin{align}
    \mathbb{E}[D](\eta) = (1+\eta)\left(\frac{\pi}{2}-\frac{\eta}{\sqrt{1-\eta^2}}\ln{\left(\frac{\sqrt{1-\eta^2}}{\eta}+\frac{1}{\eta}\right)}\right).
\end{align}
To explicitly demonstrate that surface brightness selection biases our Local Universe GRG and RG samples towards short lengths and high radio luminosities, we approximated $B_\nu(\nu_\mathrm{ref} \coloneqq 150\ \mathrm{MHz})$ for all LoTSS DR1 RGs with $z < 0.2$, assuming $\textphnc{\Af}_l = \textphnc{\Af}_{l_\nu} = 0.3$ \citep{Oei12022Alcyoneus} and $\alpha = -0.7$.
Figure~\ref{fig:SBSelection} shows the results.
As mature RGs grow, they trace out a curve approximately parallel to the golden lines (which represent constant radio luminosities), directed towards the bottom-right of the plot.
If their lives last long enough, they therefore reach a length --- which depends on their radio luminosity --- beyond which they disappear into the noise of the LoTSS.
In particular, our calculations suggest that the only giants that can be detected through the LoTSS are those with $l_\nu(\nu_\mathrm{ref}) \gtrsim 10^{24}\ \mathrm{W\ Hz^{-1}}$.\footnote{In Fig.~\ref{fig:SBSelection}, the approximate evolutionary track for $l_\nu = 10^{24}\ \mathrm{W\ Hz^{-1}}$ shows that when RGs with this radio luminosity turn into giants (i.e. when the fourth golden line intersects the green line), they disappear in the LoTSS noise (this golden line intersects the red line).}
Indeed, the lowest--radio luminosity giant in the LoTSS DR1 GRG sample of \citet{Dabhade12020March} --- which consists of 239 giants --- has $l_\nu = 1 \cdot 10^{24}\ \mathrm{W\ Hz^{-1}}$.
Clearly, if RGs with $l_\nu \sim 10^{23}\ \mathrm{W\ Hz^{-1}}$ can also form giants, then this subpopulation will be missing in LoTSS GRG samples.
Equivalently, the giants that \emph{are} included in our LoTSS GRG sample necessarily have radio luminosities $l_\nu \gtrsim 10^{24}\ \mathrm{W\ Hz^{-1}}$.
It is overwhelmingly likely that the GRG sample's radio luminosities are biased high.
Concerningly, \citet{Croston12019} have shown that, for RGs at $z < 0.4$, radio luminosity and environmental richness correlate positively.
This gives rise to the possibility that the trend shown in Fig.~\ref{fig:distributionsDensity}, where giants appear to reside in denser Cosmic Web environments than RGs in general, is a result of surface brightness selection: a GRG sample biased high in radio luminosity will be biased high in Cosmic Web density.
\begin{figure*}
    \centering
    \includegraphics[width=.9\textwidth]{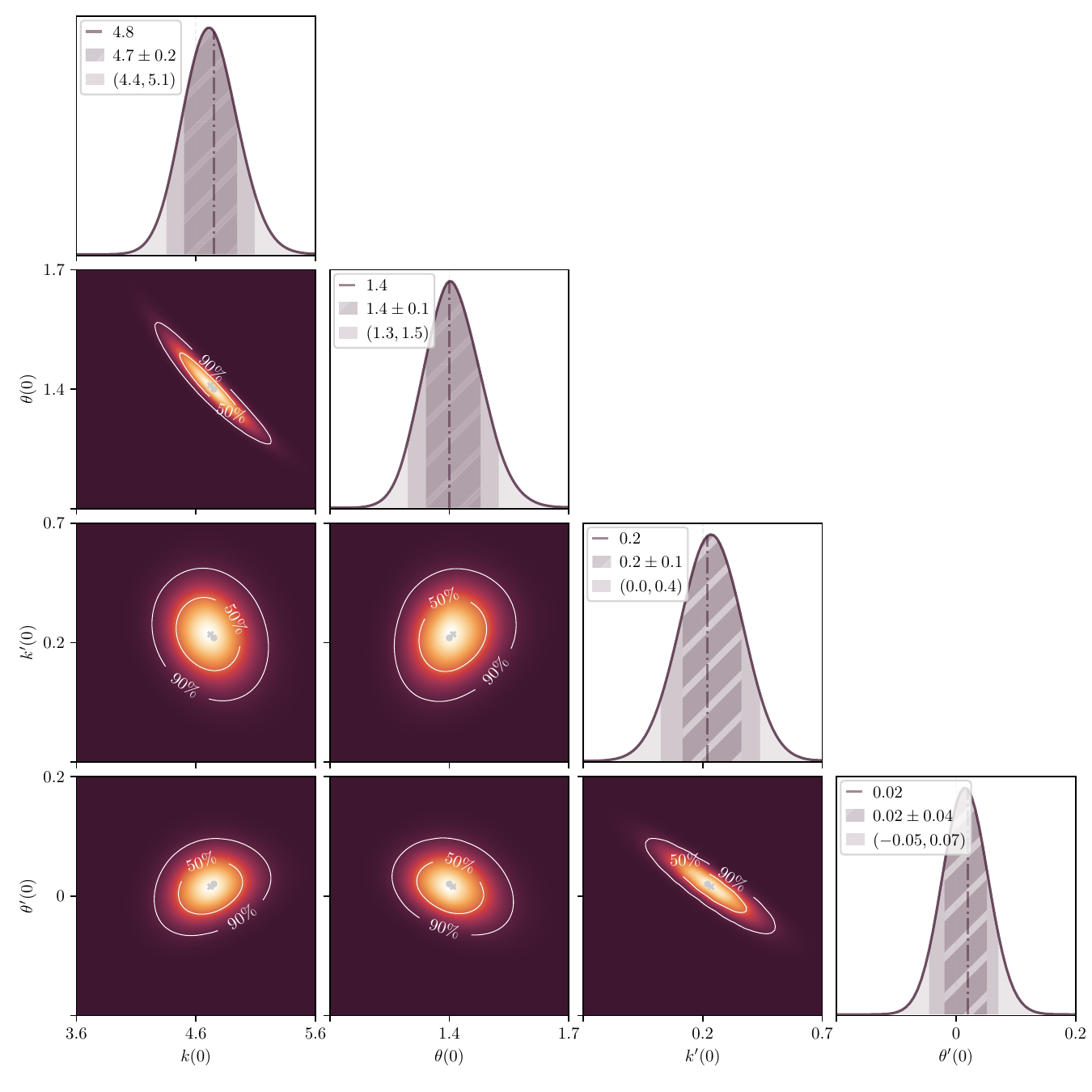}
    \caption{
    Posterior distribution over $k(0)$, $\theta(0)$, $k'(0)$, and $\theta'(0)$, based on selected Local Universe LoTSS DR1 RGs.
    We show all two-parameter marginals of the likelihood function, with contours enclosing $50\%$ and $90\%$ of total probability.
    We mark the MLE (grey dot) and the maximum a posteriori (MAP; grey cross).
    The one-parameter marginals again show the MLE (dash-dotted line), a mean-centred interval of standard deviation--sized half-width (hashed region), and a median-centred $90\%$ credible interval (shaded region).
    We used flexible voxel method densities.
    For fixed voxel method densities, see Fig.~\ref{fig:modelDensityLuminosityCornerFixed}.
    }
    \label{fig:modelDensityLuminosityCornerFlexible}
\end{figure*}

\subsection{Radio luminosity--Cosmic Web density relation}
\label{sec:radioLuminosityCosmicWebDensity}
Interestingly, our data allow us to revisit the radio luminosity--environmental richness correlation that \citet{Croston12019} have found.
In Fig.~\ref{fig:modelDensityLuminosityScatterFlexible}, we plot all 1443 selected LoTSS DR1 RGs in Cosmic Web density--radio luminosity parameter space.
By eye, it appears that for higher radio luminosities, the Cosmic Web density distribution shifts upwards.
To quantify the apparent tendency that more luminous radio galaxies occupy denser regions of the Cosmic Web, we extended the gamma distribution fitting discussed in Sect.~\ref{sec:CWDensityDistributions}.
In particular, we assumed that the RG relative density RV at some fixed radio luminosity, $1+\Delta_\mathrm{RG}\ \vert\ L_\nu = l_\nu$, is gamma distributed:
\begin{align}
1 + \Delta_\mathrm{RG}\ \vert\ L_\nu = l_\nu \sim \Gamma(k(\textphnc{\Alamed}), \theta(\textphnc{\Alamed})).
\end{align}
Instead of assuming constant parameters $k$ and $\theta$, we assumed that $k$ and $\theta$ depend on radio luminosity:
\begin{align}
    \textphnc{\Alamed} = \textphnc{\Alamed}(l_\nu) \coloneqq \mathrm{log}_\mathrm{10}\left(\frac{l_\nu}{l_{\nu,\mathrm{ref}}}\right).
\end{align}
The natural minimal extension to assuming that $k$ and $\theta$ remain constant as $\textphnc{\Alamed}$ changes, is to consider Taylor polynomial approximations of degree one (around $\textphnc{\Alamed} = 0$ --- i.e. Maclaurin polynomials):
\begin{align}
    k(\textphnc{\Alamed}) \approx k(0) + k'\left(0\right) \cdot \textphnc{\Alamed},\ \ \ \ \ \ \ \ \theta(\textphnc{\Alamed}) \approx \theta(0) + \theta'\left(0\right) \cdot \textphnc{\Alamed}.
\end{align}
However, this model does not allow for a direct comparison to our data, which are samples from $1 + \Delta_\mathrm{RG,obs}\ \vert\ L_\nu = l_\nu$ rather than from $1 + \Delta_\mathrm{RG}\ \vert\ L_\nu = l_\nu$.
Therefore, we had to extend the model by applying heteroskedastic measurement errors.
In line with Appendix~\ref{ap:heteroskedasticity}, we assumed
\begin{align}
    1 + \Delta_\mathrm{RG,obs}\ \vert\ 1 + \Delta_\mathrm{RG} = 1 + \delta \sim \mathrm{Lognormal}(\mu, \sigma^2),
\end{align}
where the parameters $\mu = \mu(a,b)$ and $\sigma^2 = \sigma^2(a, b)$ are given by
\begin{align}
    \mu = \ln{\left(1+\delta\right)} - \frac{1}{2}\ln{\left(1 + a\left(1+\delta\right)^{b-2}\right)};\\
    \sigma^2 = \ln{\left(1 + a\left(1+\delta\right)^{b-2}\right)}.
\end{align}
The parameters $a$ and $b$ characterise the heteroskedasticity.
By considering the means and SDs of the relative densities of all selected LoTSS DR1 RGs, we found $a = 0.4$ and $b = 1.1$ for the fixed voxel method and $a = 0.4$ and $b = 0.9$ for the flexible voxel method.

In order to obtain a fit that is valid over many orders of magnitude in radio luminosity, rather than only over the order of magnitude in which most data are concentrated, we calculated the likelihood in such a way that all decades in radio luminosity that contain at least $20$ RGs receive equal weight.
We simply determined the likelihood function, and thus the posterior distribution for a flat prior, through a grid search over the four parameters $k(0)$, $\theta(0)$, $k'(0)$, and $\theta'(0)$.
We defined $l_{\nu,\mathrm{ref}} \coloneqq 10^{23}\ \mathrm{W\ Hz^{-1}}$.
Figure~\ref{fig:modelDensityLuminosityCornerFlexible} visualises this flat-prior posterior.
In addition, Fig.~\ref{fig:modelDensityLuminosityScatterFlexible} shows percentile scores of $10\%$, $50\%$, $90\%$, $95\%$, and $99\%$ corresponding to the MAP model.
Figures~\ref{fig:modelDensityLuminosityScatterFixed} and \ref{fig:modelDensityLuminosityCornerFixed} demonstrate that, although the densities are lower, the results remain qualitatively the same upon switching to the fixed voxel method.
We summarise the posterior for both methods in Table~\ref{tab:posterior}.

\begin{center}
\captionof{table}{MAP and posterior mean and SD of the free parameters in our Cosmic Web density--RG radio luminosity inference.\protect\footnotemark
}
\vspace{1mm}
\textbf{fixed voxel method:}
\vspace{1mm}\\
\begin{tabular}{c c c}
\hline
parameter & MAP & posterior mean and SD\\
 [3pt] \hline\arrayrulecolor{lightgray}
$k(0)$ & $2.9$ & $2.9 \pm 0.1$\\
\hline
$\theta(0)$ & $1.4$ & $1.4 \pm 0.1$\\
\hline
$k'(0)$ & $0.04$ & $0.04 \pm 0.08$\\
\hline
$\theta'(0)$ & $0.08$ & $0.08 \pm 0.04$
\end{tabular}
\label{tab:posterior}
\end{center}
\footnotetext{As we assume a flat prior, the MAP parameters are also the MLE parameters.}

\begin{center}
\textbf{flexible voxel method:}
\vspace{1mm}\\
\begin{tabular}{c c c}
\hline
parameter & MAP & posterior mean and SD\\
 [3pt] \hline\arrayrulecolor{lightgray}
$k(0)$ & $4.8$ & $4.7 \pm 0.2$\\
\hline
$\theta(0)$ & $1.4$ & $1.4 \pm 0.1$\\
\hline
$k'(0)$ & $0.2$ & $0.2 \pm 0.1$\\
\hline
$\theta'(0)$ & $0.02$ & $0.02 \pm 0.04$
\end{tabular}
\end{center}
As values of $k'(0)$ and especially $\theta'(0)$ tend to be positive, we find evidence for a positive radio luminosity--environmental richness relation, in line with the results of \citet{Croston12019}.
More quantitatively, we call the relation positive (at $l_\nu = l_{\nu,\mathrm{ref}}$) when
\begin{align}
    \left.\frac{\mathrm{d}\mathbb{E}[1+\Delta_\mathrm{RG}\ \vert\ L_\nu = l_\nu]}{\mathrm{d}\textphnc{\Alamed}}\right\vert_{\textphnc{\Alamed} =0} = k(0)\theta'(0) + k'(0)\theta(0) > 0.
\label{eq:relationPositive}
\end{align}
We combined the posterior with Eq.~\ref{eq:relationPositive} to obtain the posterior probability that the radio luminosity--Cosmic Web density relation is positive --- that is to say that more luminous RGs tend to live in denser regions of the Cosmic Web.
For both the fixed and flexible voxel methods, this probability exceeds $99\%$.

\subsubsection{Surface brightness selection}
However, we must again be mindful of a possible surface brightness selection effect.
RGs of a given radio luminosity may be able to grow larger in more tenuous Cosmic Web environments, as radio luminosity traces jet power \citep[in close-to-linear fashion; e.g.][]{Willott11999, Shabala12013, Hardcastle12018}.
If this is indeed the case, then the lobe surface brightnesses of RGs in tenuous environments are likely lower than those of RGs with the same radio luminosity in dense environments.
As a result, of all RGs with the same radio luminosity, those in tenuous environments are more likely to be missing from observational samples.
This selection effect, that favours higher relative densities, holds for all radio luminosities, and could only explain the trend of Fig.~\ref{fig:modelDensityLuminosityScatterFlexible} if the effect becomes more severe for higher radio luminosities.
On the one hand, RGs with higher radio luminosities at a given relative density are expected to grow longer.
On the other hand, as is clear from Fig.~\ref{fig:SBSelection}, the projected proper length range over which RGs retain detectable lobe surface brightnesses increases with radio luminosity.
In other words, at higher radio luminosities, larger RGs remain detectable.

Whether the net result of these counteracting effects is less or more bias towards higher relative densities as radio luminosity increases, depends on the scaling between radio luminosity and growth.
Self-similar growth predicts \citep{Kaiser11997} that an RG's projected proper length scales with jet power $Q_\mathrm{jet}$, Cosmic Web density at the host galaxy $1+\delta_\mathrm{g}$, and time since birth $t$ as
\begin{align}
    l_\mathrm{p} \propto Q_\mathrm{jet}^{\frac{1}{5-\beta}} \cdot (1+\delta_\mathrm{g})^\frac{-1}{5-\beta} \cdot t^{\frac{3}{5-\beta}},
    \label{eq:scalingLengthJetPowerTime}
\end{align}
where $-\beta$ is the exponent of the local Cosmic Web density profile:
\begin{align}
    (1+\delta)(r) \propto (1+\delta_\mathrm{g}) \cdot r^{-\beta}.
\end{align}
Under the additional assumption of equipartition between the relativistic electron--positron kinetic energy density and the magnetic field energy density, and upon neglecting electron--positron energy losses (such as adiabatic, synchrotron, and inverse Compton losses), the RG's radio luminosity obeys \citep{Shabala12013}
\begin{align}
    l_\nu \propto Q_\mathrm{jet}^\frac{5+p}{6} l_\mathrm{p}^{3 - \frac{(4+\beta)(5+p)}{12}}.
\end{align}
Here, $-p$ is the exponent of the initial post-acceleration electron--positron kinetic energy distribution $n(\gamma) \propto \gamma^{-p}$, valid for Lorentz factors between some minimum and maximum values $\gamma_\mathrm{min}$ and $\gamma_\mathrm{max}$.
Solving for $Q_\mathrm{jet}$, substituting the result in Eq.~\ref{eq:scalingLengthJetPowerTime}, and solving for $l_\mathrm{p}$, yields --- at fixed $1+\delta_\mathrm{g}$ and $t$ ---
\begin{align}
    l_\mathrm{p} \propto l_\nu^\frac{12}{36 + (5+p)(6-3\beta)}.
    \label{eq:scalingLengthRadioLuminosity}
\end{align}
For a constant-density environment (i.e. $\beta = 0$) and a strong-shock spectral index $p = 2$, one finds $l_\mathrm{p} \propto l_\nu^{\sfrac{2}{13}}$.
For an environment where density falls off quadratically with distance, $\beta = 2$, and therefore $l_\mathrm{p} \propto l_\nu^{\sfrac{1}{3}}$.
Meanwhile, Eq.~\ref{eq:surfaceBrightness} predicts that the maximum detectable projected proper length for RGs with a given morphology, at a given redshift, and with given lobe spectral index, scales as $l_\mathrm{p,max} \propto \sqrt{l_\nu}$.
These arguments suggest that $l_\mathrm{p,max}$ increases more rapidly with $l_\nu$ than $l_\mathrm{p}$ does.
Thus, amongst a family of RGs with varying radio luminosity but seen at the same age (i.e. time after birth) and with the same morphology and Cosmic Web density, the most luminous is most easily detectable.
This provides some reassurance that the trend seen in Fig.~\ref{fig:modelDensityLuminosityScatterFlexible} is real: surface brightness selection at a fixed relative density seems to become less severe as radio luminosity increases.

\begin{figure}
    \centering
    \includegraphics[width=\linewidth]{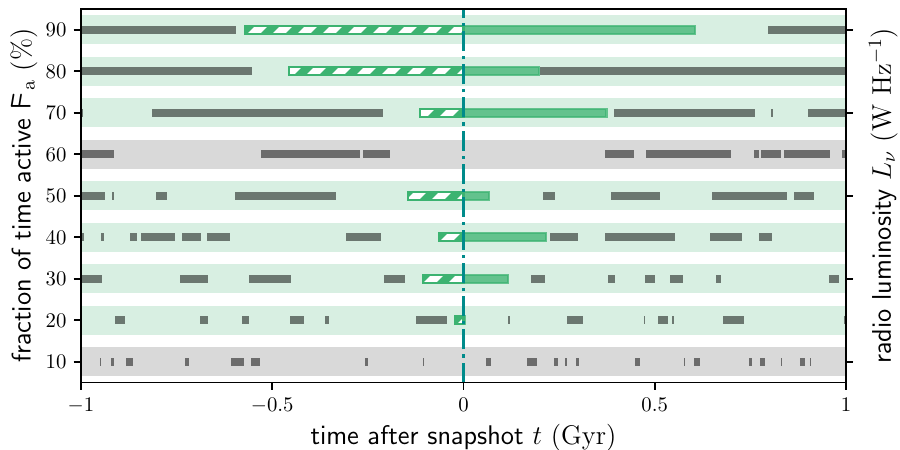}
    \caption{
    Monte Carlo simulation of a population of nine galaxies (thick stripes) with successive active and quiescent phases.
    We mark the active phases (thin stripes).
    Each galaxy spends a different fraction $\textphnc{\Af}_\mathrm{a}$ of the time in the active phase.
    We model the time spent in both phases as draws from exponential distributions: $t_\mathrm{a} \sim \mathrm{Exp}(\lambda_\mathrm{a})$ and $t_\mathrm{q} \sim \mathrm{Exp}(\lambda_\mathrm{q})$.
    We assume that all galaxies share the same quiescent phase rate parameter $\lambda_\mathrm{q} = 10\ \mathrm{Gyr}^{-1}$, although this may not be entirely correct \citep{Turner12015}.
    This assumption fixes the active phase rate parameter: $\lambda_\mathrm{a} = (\textphnc{\Af}_\mathrm{a}^{-1}-1)\lambda_\mathrm{q}$.\protect\footnotemark
    \ On human time scales, we can access only a snapshot view of this population.
    During the present snapshot (vertical line), seven galaxies generate RGs, and could be included in an RG sample.
    RG ages at the snapshot (hatched parts of the thin green stripes) generally increase with $\textphnc{\Af}_\mathrm{a}$.
    }
    \label{fig:RGLifeSnapshot}
\end{figure}

However, there are at least two caveats here.
First, more luminous RGs may live longer than less luminous RGs, as jointly suggested by the radio luminosity--stellar mass relation and the stellar mass--radio AGN occurrence relation of \citet{Sabater12019}.
As a result, in a snapshot of the RG population at a given instant, more luminous RGs may typically be older than less luminous RGs.
Figure~\ref{fig:RGLifeSnapshot} illustrates this point.
If more luminous RGs indeed tend to be older than less luminous RGs, then the relation between projected length and radio luminosity at a fixed Cosmic Web density for a snapshot population will be steeper than the one of Eq.~\ref{eq:scalingLengthRadioLuminosity}, which assumes equal ages.
Secondly, we warn for the possibility --- which remains subject of debate \citep{Mingo12019} --- that the RG shape distribution at a fixed Cosmic Web density may change with radio luminosity.

\footnotetext{
We define
\begin{align}
    \textphnc{\Af}_\mathrm{a} \coloneqq \frac{\mathbb{E}[t_\mathrm{a}]}{\mathbb{E}[t_\mathrm{a} + t_\mathrm{q}]} = \frac{\mathbb{E}[t_\mathrm{a}]}{\mathbb{E}[t_\mathrm{a}] + \mathbb{E}[t_\mathrm{q}]} = \frac{\lambda_\mathrm{a}^{-1}}{\lambda_\mathrm{a}^{-1} + \lambda_\mathrm{q}^{-1}} =
    \left(1+\frac{\lambda_\mathrm{a}}{\lambda_\mathrm{q}}\right)^{-1}.
    \label{eq:fractionActiveSimple}
\end{align}
We note that the more natural definition
\begin{align}
    \textphnc{\Af}_\mathrm{a} \coloneqq \mathbb{E}\left[\frac{t_\mathrm{a}}{t_\mathrm{a}+t_\mathrm{q}}\right] = \mathbb{E}\left[\left(1 + \frac{t_\mathrm{q}}{t_\mathrm{a}}\right)^{-1}\right]
\end{align}
gives, for $r \coloneqq \frac{\lambda_\mathrm{a}}{\lambda_\mathrm{q}}$,
\begin{align}
    \textphnc{\Af}_\mathrm{a} = \begin{cases}
    \frac{1}{2} & \text{if } r = 1;\\
    \frac{r \ln{r} - r + 1}{(r-1)^2} & \text{if } r > 0, r \neq 1.
    \end{cases}
\label{eq:fractionActiveNatural}
\end{align}
To arrive at Eq.~\ref{eq:fractionActiveNatural}, we used standard identities for the PDFs of the ratio distribution and the inverse distribution, and applied l'H\^{o}pital's rule twice.
When $r = 1$, that is to say when $\lambda_\mathrm{a} = \lambda_\mathrm{q}$, both definitions predict the same $\textphnc{\Af}_\mathrm{a} = \frac{1}{2}$.
For $r < 1$, our definition predicts higher $\textphnc{\Af}_\mathrm{a}$ than the natural definition does; for $r > 1$, our definition predicts lower $\textphnc{\Af}_\mathrm{a}$.
To find $\lambda_\mathrm{a}$ given $\lambda_\mathrm{q}$ and $\textphnc{\Af}_\mathrm{a}$ under the natural definition, one must solve Eq.~\ref{eq:fractionActiveNatural} numerically.
For simplicity, we generated Fig.~\ref{fig:RGLifeSnapshot} using Eq.~\ref{eq:fractionActiveSimple}.
}

\subsubsection{Predicting the density distribution of observed giants by treating them as luminous general RGs}
\begin{figure}[t]
    \centering
    \includegraphics[width=\columnwidth]{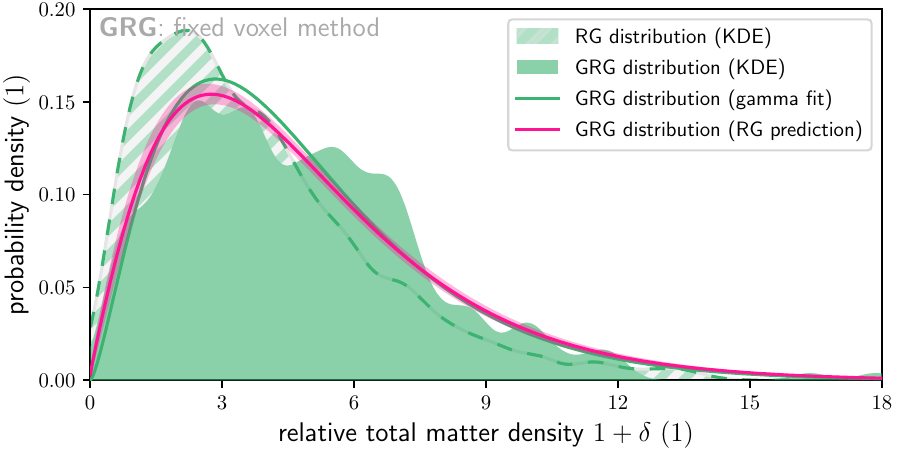}
    \caption{
    Density distribution predicted for observed giants combining Sect.~\ref{sec:radioLuminosityCosmicWebDensity}'s RG radio luminosity--Cosmic Web density relation and a radio luminosity cut-off $l_{\nu,\mathrm{min}} = 10^{25}\ \mathrm{W\ Hz^{-1}}$ (pink); in addition, we show the KDE density distributions for giants with gamma distribution MLE fit (solid green) and for general RGs (hatched green) --- all as in the left column of Fig.~\ref{fig:distributionsDensity}.
    }
    \label{fig:distributionDensityGRGExplained}
\end{figure}

We then tested whether Fig.~\ref{fig:distributionsDensity}'s higher densities for observed giants than for observed general RGs could be explained by these giants' higher radio luminosities and Fig.~\ref{fig:modelDensityLuminosityScatterFlexible}'s radio luminosity--Cosmic Web density relation.
We first calculated the RG relative total matter density RV for radio luminosities of $l_{\nu,\mathrm{min}}$ and higher.
By applying standard PDF identities, we found
\begin{align}
    &f_{1+\Delta_\mathrm{RG}\ \vert\ L_\nu \geq l_{\nu,\mathrm{min}}}(1+\delta) =\nonumber\\
    &\frac{\int_{l_{\nu,\mathrm{min}}}^\infty f_{1+\Delta_\mathrm{RG}\ \vert\ L_\nu = l_\nu}(1+\delta) \cdot f_{L_\nu}(l_\nu)\ \mathrm{d}l_\nu}{\int_{l_{\nu,\mathrm{min}}}^\infty f_{L_\nu}(l_\nu)\ \mathrm{d}l_\nu} =\nonumber\\
    &\frac{\int_{l_{\nu,\mathrm{min}}}^\infty f_{1+\Delta_\mathrm{RG}\ \vert\ L_\nu = l_\nu}(1+\delta) \cdot f_{L_\nu\ \vert\ L_\nu \geq l_{\nu,\mathrm{min}}}(l_\nu)\ \mathrm{d}l_\nu}{\int_{l_{\nu,\mathrm{min}}}^\infty f_{L_\nu\ \vert\ L_\nu \geq l_{\nu,\mathrm{min}}}(l_\nu)\ \mathrm{d}l_\nu},
\label{eq:PDFRGDensityAboveRadioLuminosity}
\end{align}
To proceed, one must specify $f_{L_\nu\ \vert\ L_\nu \geq l_{\nu,\mathrm{min}}}(l_\nu)$ --- up to a constant, at least.
We modelled $L_\nu\ \vert\ L_\nu \geq l_{\nu,\mathrm{min}} \sim \mathrm{Pareto}(l_{\nu,\mathrm{min}}, \lambda)$, where $\lambda$ is the tail index or rate parameter of the associated exponential distribution.\footnote{Equivalently, one could model
\begin{align}
\ln{\left(\frac{L_\nu\ \vert\ L_\nu \geq l_{\nu,\mathrm{min}}}{l_{\nu,\mathrm{min}}}\right)} \sim \mathrm{Exp}(\lambda),
\end{align}
or
\begin{align}
\mathrm{log}_{10}\left(\frac{L_\nu\ \vert\ L_\nu \geq l_{\nu,\mathrm{min}}}{l_{\nu,\mathrm{min}}}\right) \sim \mathrm{Exp}\left(\lambda \cdot \ln{10}\right).
\end{align}}
For both $l_{\nu,\mathrm{min}} = 10^{23}\ \mathrm{W\ Hz^{-1}}$ and $l_{\nu,\mathrm{min}} = 10^{24}\ \mathrm{W\ Hz^{-1}}$, we found $\lambda_\mathrm{MLE} = 0.6$ (based on $1 \cdot 10^3$ and $3 \cdot 10^2$ RGs, respectively).
For higher $l_{\nu,\mathrm{min}}$, $\lambda_\mathrm{MLE}$ increases somewhat, but becomes less reliable because of decreasing sample sizes.
We therefore stuck with $\lambda_\mathrm{MLE} = 0.6$.
We then evaluated Eq.~\ref{eq:PDFRGDensityAboveRadioLuminosity} for $l_{\nu,\mathrm{min}} = 10^{25}\ \mathrm{W\ Hz^{-1}}$ and added heteroskedastistic noise in order to compare our prediction to observations.
We propagated the correlated uncertainties on the parameters $k(0)$, $\theta(0)$, $k'(0)$, and $\theta'(0)$, which together determine $f_{1+\Delta_\mathrm{RG}\ \vert\ L_\nu = l_\nu}$, by rejection sampling from the joint posterior distribution.
We show the resulting mean prediction with standard deviation half-width (pink) in Fig.~\ref{fig:distributionDensityGRGExplained}.
\noindent
Although the prediction is based solely on observations of general RGs --- and so is built without knowledge of the observed GRG distribution (solid green KDE) --- it appears to fit it better than a direct MLE gamma fit to the GRG data (solid green curve) does.
However, a better fit is not necessarily also a good fit.
We therefore tested whether the prediction is consistent with our sample of Cosmic Web densities for observed giants.
A KS test suggested an answer in the affirmative, as it yielded a $p$-value of $0.7$.

Section~\ref{sec:CWDensityDistributions} presented a paradox: known giant radio galaxies reside in denser Cosmic Web environments than their smaller kin, in apparent violation of the long-standing hypothesis that giants should emerge predominantly in the dilute Cosmic Web.
The modelling in this section suggests that the higher densities found for known giants than for general RGs can be explained by a combination of these giants' higher radio luminosities and the general relation between RG radio luminosity and Cosmic Web density.
Fainter giants --- which might inhabit the more dilute Cosmic Web --- might exist, but are currently not observationally accessible.

\subsection{Radio galaxy number densities}
After accreting baryons, supermassive black holes (SMBHs) can launch jets that give rise to radio galaxies.
Although it is clear that SMBH jets and RGs play an important role in galaxy evolution and cosmology, the physics of jet launching is not yet fully understood.
Moreover, because SMBHs are astronomical unit--sized, galaxies are kiloparsec-sized (ratio $10^8$--$10^9$), and the Cosmic Web is megaparsec-sized (ratio $10^{11}$--$10^{12}$), it is not yet possible to build cosmological simulations in which a realistic interplay between SMBHs, their host galaxies, and the enveloping Cosmic Web naturally arises.
Thus, for now, major advances in our understanding of RGs on the cosmological scale must come from observations instead of from simulations.
It is therefore of considerable interest to observationally constrain the occurrence of RGs as a function of Cosmic Web density.

Enticingly, our relative total matter density measurements --- in combination with the BORG SDSS HMC Markov chain samples in their entirety --- allow for a determination of the RG number density as a function of Cosmic Web density: $n_\mathrm{RG}(1+\delta)$.
To see why, we first emphasise that the PDF $f_{1+\Delta_\mathrm{RG}}$, which we have approximated through observations and visualised in the bottom row of Fig.~\ref{fig:distributionsDensity}, provides a key ingredient for $n_\mathrm{RG}(1+\delta)$.
It shows that if a volume of cosmological extent is surveyed, one will obtain more RGs with moderate (filament-like) densities than with low (void-like) or high (cluster-like) densities.
We remark that in such a survey sample the number of RGs whose densities fall within a given interval depends not only on the RG number density at the interval's densities, but also on the prevalence of environments with such densities.
As a result, if a sample contains more filament-inhabiting RGs than, say, cluster-inhabiting RGs, one need not necessarily conclude that the RG number density in filaments is higher than in clusters; in fact, the RG number density in the latter environment could well be higher, provided that clusters are sufficiently rare.

Conveniently, the density fields of the BORG SDSS HMC Markov chain samples enable one to quantify the rarity of each Cosmic Web environment through a PDF $f_{1+\Delta_\mathrm{CW}}$, which corresponds to the Cosmic Web relative density RV $1 + \Delta_\mathrm{CW}$.
This RV represents the relative total matter density at a randomly chosen point in the Local Universe (and at the BORG SDSS resolution of $2.9\ \mathrm{Mpc}\ h^{-1}$).
We determine the distribution of $1 + \Delta_\mathrm{CW}$ simply by binning the density fields of a few hundred Markov chain samples.
As a result, $f_{1+\Delta_\mathrm{CW}}$ is best determined at low densities, which correspond to the most common environments, and least so at high densities, which are rare.\footnote{At first sight, it is tempting to approximate $f_{1+\Delta_\mathrm{CW}}$ by binning the BORG SDSS mean rather than individual BORG SDSS samples.
This, however, leads to incorrect results.
The mean tends to the cosmic mean wherever the structure of the Cosmic Web is not well constrained by the SDSS DR7 MGS, causing low-density environments to be overrepresented in it and high-density environments underrepresented.
By contrast, the samples have low-density and high-density environments statistically correctly proportioned --- that is, insofar as the gravity solver allows.
When using the BORG SDSS mean, the probability densities $f_{1+\Delta_\mathrm{CW}}(1+\delta)$ for $1 + \delta \gtrsim 10$ are up to an order of magnitude lower than predicted by BORG SDSS samples.}
As we expect $f_{1+\Delta_\mathrm{CW}}$ to be smooth, we applied a Savitzky--Golay filter (of polynomial order 1) to the binned data; however, this step is not essential.

Appendix~\ref{ap:relativeNumberDensity} shows that $n_\mathrm{RG}(1+\delta)$, the RG number density as would arise in an environment of constant density $1+\delta$, obeys
\begin{align}
\frac{n_\mathrm{RG}(1+\delta)}{\bar{n}_\mathrm{RG}} = \frac{f_{1 + \Delta_\mathrm{RG}}(1 + \delta)}{f_{1 + \Delta_\mathrm{CW}}(1 + \delta)}.
    \label{eq:RGNumberDensityAtRelativeDensity}
\end{align}
Here, $\bar{n}_\mathrm{RG}$ is the cosmic mean RG number density.
Thus, a simple point-wise division of the PDFs of $1+\Delta_\mathrm{RG}$ and $1 + \Delta_\mathrm{CW}$ yields the RG number density at a given Cosmic Web density relative to the cosmic mean value.

\begin{figure}[t]
    \centering
    \includegraphics[width=\columnwidth]{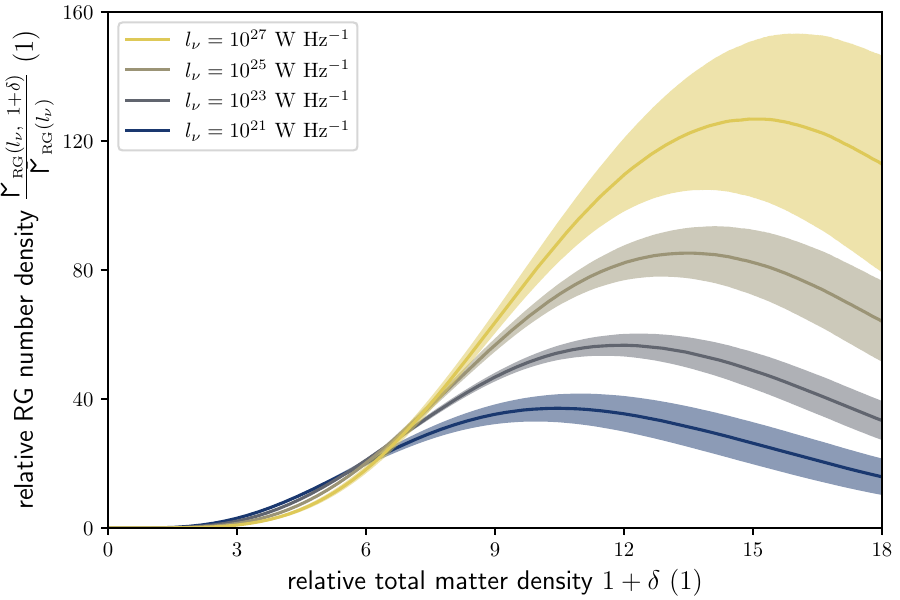}
    \caption{
    Number densities of RGs with a given radio luminosity $l_\nu$, as a function of Cosmic Web density at the $2.9\ \mathrm{Mpc}\ h^{-1}$ scale.
    The number density functions reflect conditions in the Local Universe, and are given relative to the cosmic mean number density at $l_\nu$.
    The solid curves reflect the posterior mean, whilst the shaded areas around the mean denote $-1$ to $+1$ posterior standard deviation ranges.
    We used flexible voxel method densities.
    }
    \label{fig:RGNumberDensityRelativeFlexible}
\end{figure}

Unfortunately, as discussed in Sect.~\ref{sec:CWDensityDistributions}, our measurements of $f_{1 + \Delta_\mathrm{RG}}$ suffer from surface brightness selection.
To avoid selection effects, we formulated a more specific version of Eq.~\ref{eq:RGNumberDensityAtRelativeDensity}.
In particular, we considered the number density of RGs per unit of radio luminosity as a function of radio luminosity and relative density: $\textphnc{\Anun}_\mathrm{RG}(l_\nu, 1+\delta)$ --- again, relative to a cosmic mean value: $\bar{\textphnc{\Anun}}_\mathrm{RG}(l_\nu)$.
Appendix~\ref{ap:relativeNumberDensity} shows that this ratio equals
\begin{align}
\frac{\textphnc{\Anun}_\mathrm{RG}(l_\nu, 1 + \delta)}{\bar{\textphnc{\Anun}}_\mathrm{RG}(l_\nu)} = \frac{f_{1 + \Delta_\mathrm{RG} \vert L_\nu = l_\nu}(1 + \delta)}{f_{1 + \Delta_\mathrm{CW}}(1 + \delta)}.
\label{eq:RGNumberDensityAtRadioLuminosityAndRelativeDensity}
\end{align}
This radio luminosity--dependent relative number density is, as before, equal to a point-wise division of PDFs.
With respect to the right-hand side (RHS) of Eq.~\ref{eq:RGNumberDensityAtRelativeDensity}, only the PDF in the numerator has changed.
Section~\ref{sec:radioLuminosityCosmicWebDensity} explains how we modelled this PDF.
It depends on the four parameters $k(0)$, $k'(0)$, $\theta(0)$, and $\theta'(0)$, for which the data shown in Fig.~\ref{fig:modelDensityLuminosityScatterFlexible} induce the posterior distribution shown in Fig.~\ref{fig:modelDensityLuminosityCornerFlexible}.
We evaluated the RHS of Eq.~\ref{eq:RGNumberDensityAtRadioLuminosityAndRelativeDensity} by rejection sampling from the posterior, calculating $f_{1 + \Delta_\mathrm{RG} \vert L_\nu = l_\nu}$ for each sampled quartet $(k(0), k'(0), \theta(0), \theta'(0))$, and dividing the resulting PDFs point-wise by $f_{1+\Delta_\mathrm{CW}}$.
In Fig.~\ref{fig:RGNumberDensityRelativeFlexible}, for four observationally accessible radio luminosities, we visualise the posterior mean alongside a range that is two standard deviations wide.

As expected, RG number densities generally increase with Cosmic Web density.
The more luminous RGs become, the more pronounced the contrast between their number density in clusters and their cosmic mean number density is.
Our results indicate that high-luminosity RGs, with $l_\nu \in 10^{25}$--$10^{27}\ \mathrm{W\ Hz^{-1}}$, attain number densities in clusters that are ${\sim}10^2$ times higher than average.
Interestingly, for radio luminosities $l_\nu \in 10^{21}$--$10^{23}\ \mathrm{W\ Hz^{-1}}$, we obtain statistically significant, preliminary evidence that the RG number density peaks at a cluster-like density that depends positively on $l_\nu$.\footnote{For radio luminosities $l_\nu \in 10^{25}$--$10^{27}\ \mathrm{W\ Hz^{-1}}$, the uncertainty margins are too large to claim evidence for a peak.}
Beyond this density, these low-luminosity RGs become less prevalent per unit of volume, possibly because the underlying galaxy population starts generating more luminous RGs instead.

We stress that these results are tentative, as both $f_{1 + \Delta_\mathrm{RG}\vert L_\nu = l_\nu}$ and $f_{1 + \Delta_\mathrm{CW}}$ suffer from systematic errors at higher densities.
These systematic errors are not reflected in the uncertainties of Fig.~\ref{fig:RGNumberDensityRelativeFlexible}.
In the case of $f_{1 + \Delta_\mathrm{RG}\vert L_\nu = l_\nu}$, the gamma distribution approximation loses validity in the tail: beyond some point, it will consistently under- or overestimate the probability density.
Our LoTSS DR1 sample contains only a modest number (${\sim}10^2$) of RGs at relative densities $1 + \delta \gtrsim 10$ --- especially when selecting those around a given value of $l_\nu$ only (in which case just ${\sim}10^1$ RGs remain).
As a result, in its tail the gamma PDF takes on the character of an extrapolation function rather than of a fitting function, and should consequently be treated with caution.
In the future, larger RG samples can resolve this situation.
In the case of $f_{1 + \Delta_\mathrm{CW}}$, the 2LPT model of structure formation used by the BORG SDSS loses validity at cluster densities.
As a result, this work's estimate of $f_{1 + \Delta_\mathrm{CW}}$ underpredicts the prevalence of cluster-like densities.
Newer BORG runs, such as the BORG 2M++, employ an enhanced gravity solver \citep{Jasche12019}, relieving this limitation.

\begin{figure}[t]
    \centering
    \includegraphics[width=\columnwidth]{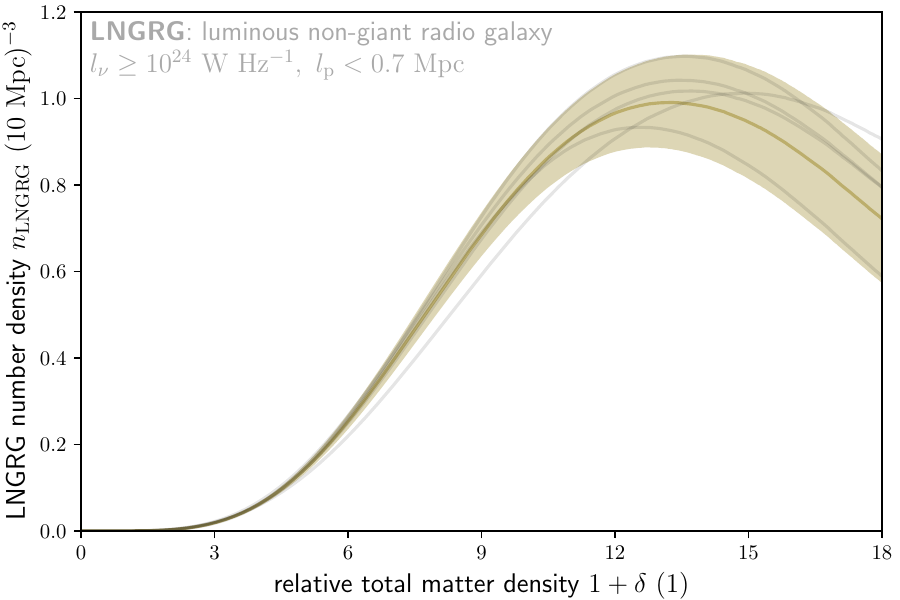}
    \caption{Number density of luminous non-giant radio galaxies in the Local Universe, as a function of Cosmic Web density.
    We define these as RGs with $150\ \mathrm{MHz}$ radio luminosities $l_\nu \geq 10^{24}\ \mathrm{W\ Hz^{-1}}$, projected proper lengths $l_\mathrm{p} < l_\mathrm{p,GRG} \coloneqq 0.7\ \mathrm{Mpc}$, and redshifts $z < z_\mathrm{max} \coloneqq 0.16$.
    The Cosmic Web densities encompass baryonic and dark matter, are defined on a $2.9\ \mathrm{Mpc}\ h^{-1}$ scale, and are given relative to the cosmic mean.
    We used flexible voxel method densities here.
    We show the posterior mean with a shaded $-1$ to $+1$ posterior standard deviation range.
    We also show five individual realisations (grey).
    }
    \label{fig:LNGRGNumberDensityFlexible}
\end{figure}

Although preliminary in nature, the inferences shown in Fig.~\ref{fig:RGNumberDensityRelativeFlexible} could in principle be used to calibrate cosmological simulations that feature RG feedback.
The fact that our number densities are given with respect to a cosmic mean value is not problematic, because such a cosmic mean number density is straightforward to calculate from simulation snapshots.
However, apart from making sure that the RG number densities in simulated environments of various densities are correctly proportioned, it is also important to calibrate the absolute level of RG feedback in simulations.
Interestingly, there are variations on Eq.~\ref{eq:RGNumberDensityAtRelativeDensity} for which an absolute number density can be explicitly calculated with current-day data.
One can consider, for example, non-giant radio galaxies (i.e. $l_\mathrm{p} < l_\mathrm{p,GRG} \coloneqq 0.7\ \mathrm{Mpc}$) in the Local Universe with radio luminosities $l_\nu \geq 10^{24}\ \mathrm{W\ Hz^{-1}}$.
Figure~\ref{fig:SBSelection} suggests that this RG subpopulation is already fully accessible by the LoTSS.
Calling these RGs luminous non-giant radio galaxies (LNGRGs), we have
\begin{align}
    n_\mathrm{LNGRG}(1+\delta) = \bar{n}_\mathrm{LNGRG} \cdot \frac{f_{1 + \Delta_\mathrm{LNGRG}}(1 + \delta)}{f_{1 + \Delta_\mathrm{CW}}(1 + \delta)}.
    \label{eq:LNGRGNumberDensityAtRelativeDensity}
\end{align}
We calculated $\bar{n}_\mathrm{LNGRG}$ by counting all LNGRGs in the LoTSS DR1 footprint up to $z_\mathrm{max}$, finding $N_\mathrm{LNGRG} = 172$.
Because of calibration problems and higher noise levels towards the edges of the footprint, a fraction of the footprint is --- artificially --- devoid of LNGRGs.
We thus applied a correction factor to $N_\mathrm{LNGRG}$ of 1.2, yielding $N_\mathrm{LNGRG} = 200$.
Assuming that $N_\mathrm{LNGRG} \sim \mathrm{Poisson}(\lambda_\mathrm{LNGRG})$, maximum likelihood estimation simply suggests $\hat{\lambda}_\mathrm{LNGRG,MLE} = 200$.
We combined this MLE Poisson distribution for $N_\mathrm{LNGRG}$ with the comoving volume $V = 16 \cdot 10^6\ \mathrm{Mpc}^3$ to obtain a probability distribution for $\bar{n}_\mathrm{LNGRG}$.
We find $\bar{n}_\mathrm{LNGRG} = 12 \pm 1\ (100\ \mathrm{Mpc})^{-3}$.
Next, we approximated $f_{1 + \Delta_\mathrm{LNGRG}} \approx f_{1+\Delta_{\mathrm{RG}}\ \vert\ L_\nu \geq 10^{24}\ \mathrm{W\ Hz^{-1}}}$, and calculated the latter through Eq.~\ref{eq:PDFRGDensityAboveRadioLuminosity} and Sect.~\ref{sec:radioLuminosityCosmicWebDensity}'s subsequent procedure.
Taking $f_{1+\Delta_\mathrm{CW}}$ as before, we obtained a probability distribution over LNGRG number density functions, visualised in Fig.~\ref{fig:LNGRGNumberDensityFlexible}.

As it should, the LNGRG number density function follows the same trend as the functions of Fig.~\ref{fig:RGNumberDensityRelativeFlexible}; the novelty here is the prediction of absolute number densities.
We reiterate that systematic errors at cluster-like densities render only the left side of the plot reliable.
It predicts that, in filaments, $n_\mathrm{LNGRG} \sim 10^{-1}\ (10\ \mathrm{Mpc})^{-3}$, or $n_\mathrm{LNGRG} \sim 10^{-4}\ \mathrm{Mpc}^{-3}$.
As major filaments have volumes $V \in 10^2$--$10^3\ \mathrm{Mpc}^3$, only one in ten major filaments harbours an LNGRG.

\subsection{Cosmic Web dynamical state distributions}
\label{sec:CWDynamicalStateDistributions}
\begin{figure}
\centering
\includegraphics[width=\linewidth]{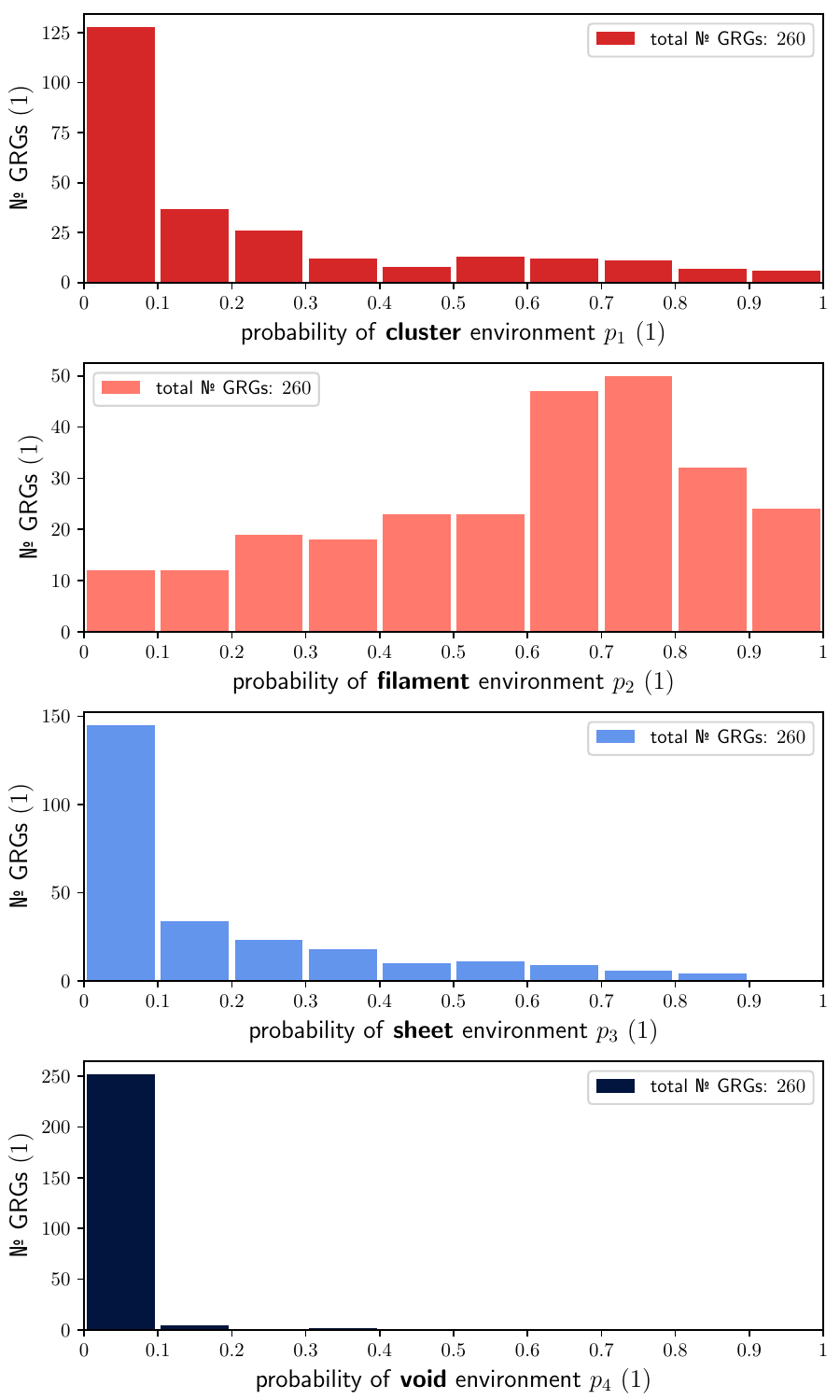}
\caption{
Bayesian Cosmic Web reconstructions allow one to determine, for each giant, a probability distribution $\vec{p}$ over the four $T$-web environment classes (clusters, filaments, sheets, and voids).
Here we summarise these distributions: each panel shows how a component of $\vec{p} \coloneqq (p_1, p_2, p_3, p_4)$ is itself distributed.
}
\label{fig:distributionsProbabilityTWeb}
\end{figure}

\begin{figure*}
    \centering
    \begin{subfigure}{.33\textwidth}
    \includegraphics[width=\textwidth]{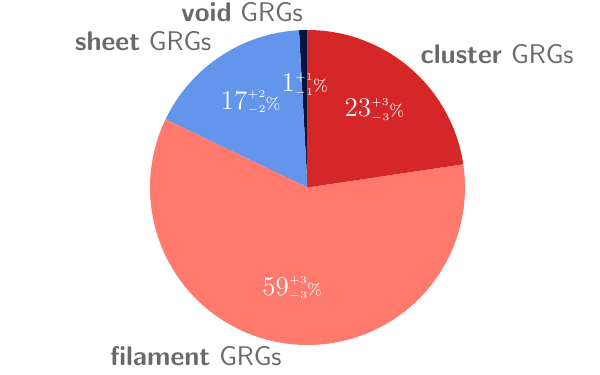}
    \end{subfigure}
    \begin{subfigure}{.33\textwidth}
    \includegraphics[width=\textwidth]{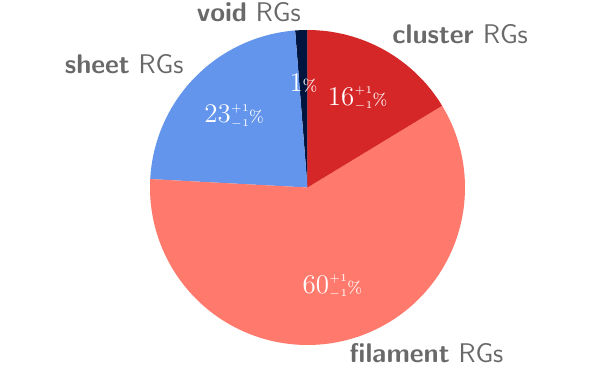}
    \end{subfigure}
    \begin{subfigure}{.33\textwidth}
    \includegraphics[width=\textwidth]{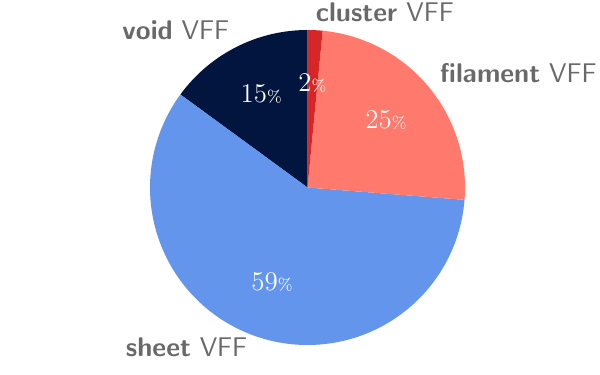}
    \end{subfigure}
    \caption{
    Dynamical environment classification of giant radio galaxies (left), radio galaxies in general (centre), and the Local Universe in its entirety \citep[right; from Table 3 of][]{Leclercq12015}.
    We define clusters, filaments, sheets, and voids in the $T$-web sense.
    These distributions depend on the scale to which the Cosmic Web density field is smoothed; in this case, the smoothing scale is $2.9\ \mathrm{Mpc}\ h^{-1}$.
    If giants and RGs were scattered uniformly throughout the Cosmic Web, their distributions would be similar to that of the volume-filling fractions (VFFs).
    Instead, observed RGs --- and observed giants in particular --- favour cluster and filament environments.
    }
    \label{fig:piesTWeb}
\end{figure*}

For each RG in the two samples, we obtained a probability distribution over $T$-web environment classes from the BORG SDSS extensions of \citet{Leclercq12015}.
The $T$-web scheme classifies the Cosmic Web at each point on the basis of its local gravitational dynamics.
On an intuitive level, the scheme counts the number of dimensions along which the orbit of a test particle released at rest with respect to the environment is stable.
More formally, at any location the Hessian of the gravitational potential, or tidal tensor $T$, has either three, two, one, or zero positive eigenvalues; the environment is interpreted correspondingly as a cluster, filament, sheet, or void.
\citet{Leclercq12015} have determined the $T$-web environment for each voxel of each BORG SDSS sample.
By iterating over samples and counting the frequencies of the four environment classes on a per-voxel basis, we obtained a marginal environment class distribution $\vec{p} = (p_1, p_2, p_3, p_4)$ for each voxel.
In Fig.~\ref{fig:distributionsProbabilityTWeb}, we show how each of $\vec{p}$'s components is distributed for luminous giants.
We note the following.

For 99\% of our giants, the most likely environment class is more than 50\% probable.
For 20\% of giants, it is more likely than not that they inhabit a cluster (i.e. $p_1 > 50\%$); for 68\% of giants, it is more likely than not that they inhabit a filament (i.e. $p_2 > 50\%$); and for 11\% of giants, it is more likely than not that they inhabit a sheet (i.e. $p_3 > 50\%$).
Finally, for 0\% of giants, it is more likely than not that they inhabit a void (i.e. $p_4 > 50\%$).

Some giants inhabit sheets (e.g. those in the bottom row of Fig.~\ref{fig:localisation}), but concluding this in individual cases rarely has a high degree of certainty: for just 0.5\% of giants, it is very likely that they inhabit a sheet (i.e. $p_3 > 80\%$).

A cluster environment is very unlikely (i.e. $p_1 < 20\%$) for a clear majority of 63\% of giants, whilst a filament environment is very unlikely (i.e. $p_2 < 20\%$) for only 9\% of giants; the filament environment hypothesis can thus seldom be excluded.
A sheet environment is very unlikely (i.e. $p_3 < 20\%$) for 71\% of giants; finally, a void environment is nearly always very unlikely (i.e. $p_4 < 20\%$) --- this holds for 99\% of giants.

The corresponding distributions for LoTSS DR1 RGs (not shown) are largely similar.
This is also clear from Fig.~\ref{fig:piesTWeb}, in which we compare the environment class distributions of luminous giants, LoTSS DR1 RGs, and the Local Universe in its entirety.
Evidently, luminous giants and general RGs are not scattered uniformly throughout the Cosmic Web, but trace cluster and filament environments.
Remarkably, nearly a quarter of all luminous giants occurs in clusters, which have a volume-filling fraction (VFF) of just 1--2\%.
It is easy to derive that the ratio between a GRG environment class probability (left pie) and the corresponding VFF (right pie) yields the GRG number density of that class relative to the cosmic mean GRG number density.
Of course, the analogous statement for RGs also holds.
In Table~\ref{tab:relativeNumberDensities}, we present relative number densities for luminous giants and general RGs in the Local Universe.
\begin{table}
    \centering
    \caption{
    Number densities of observed Local Universe giants and general RGs relative to their cosmic means.
    These numbers depend on environment class definitions and the scale to which the Cosmic Web density field is smoothed; in this case, we use the $T$-web classification scheme and a smoothing scale of $2.9\ \mathrm{Mpc}\ h^{-1}$.
    Clusters and filaments are (giant) radio galaxy overdense; sheets and voids are underdense.
    }
    \label{tab:relativeNumberDensities}
    \resizebox{\columnwidth}{!}{%
    \begin{tabular}{l l l l l}
    \hline
    & \textbf{cluster} & \textbf{filament} & \textbf{sheet} & \textbf{void}\\
    & environment & environment & environment & environment\\
    \hline
GRGs & $15 \pm 2$ & $2.4 \pm 0.1$ & $0.29 \pm 0.04$ & $0.06 \pm 0.04$\\
RGs & $10.6 \pm 0.6$ & $2.41 \pm 0.05$ & $0.39 \pm 0.02$ & $0.08 \pm 0.02$\\
    \end{tabular}%
    }
\end{table}
Clusters are strongly (G)RG overdense; voids are strongly (G)RG underdense.

The left and central pie charts of Fig.~\ref{fig:piesTWeb} can be regarded as prior distributions over Cosmic Web dynamical states for luminous giants and RGs in or near the Local Universe.
This fact can be useful whenever direct BORG SDSS environment characterisation is not possible.
For a randomly picked luminous giant, Fig.~\ref{fig:piesTWeb}'s left pie chart specifies the prior probability distribution over dynamical states to be $\vec{p} = (23\%, 59\%, 17\%, 1\%)$.
Interestingly, there is often additional information available to constrain the dynamical state.
For example, multiwavelength imagery demonstrates that Alcyoneus \citep{Oei12022Alcyoneus}, one of the longest known RGs, does not inhabit a galaxy cluster; in addition, the number of luminous galaxies in its vicinity excludes the possibility of a void environment.\footnote{Characterising Alcyoneus's Cosmic Web environment is of scientific interest, as it could help understand how the largest RGs form.
However, at $z = 0.25$, it is too far away to probe directly with the BORG SDSS.}
Combining this information with the prior, we obtain --- after rounding --- the posterior probability distribution $\vec{p} = (0\%, 80\%, 20\%, 0\%)$.
We thus estimate the probability that Alcyoneus inhabits a filament to be 80\%.
Of course, this line of reasoning generalises to any luminous giant outside of galaxy clusters that has massive galactic neighbours in its megaparsec-scale vicinity.

\section{Discussion}
\label{sec:discussion}
\subsection{Cosmic Web density accuracy}
\label{sec:CosmicWebDensityAccuracy}
In this work, we have determined megaparsec-scale total matter densities around RGs in the Local Universe, and used these to compare the environments of luminous giants to those of general RGs.
A natural question to ask is how accurate these densities are, and to which degree they can be used in other analyses, such as in the fitting of dynamical models to individual RGs \citep[e.g.][]{Hardcastle12018}.
To quantify accuracy, one would ideally know the ground-truth densities that the inferred densities are meant to approach.
Problematically, however, no ground-truth densities are known --- especially for filaments, the environment type which most RGs seem to inhabit.
For Local Universe clusters though, approximate masses are known.
We exploited this fact to test the accuracy of our densities for cluster RGs.
We immediately remark that our density accuracy is likely to vary with density, so that the accuracy of cluster RG densities might not be informative of the accuracy of filament RG densities.
This is because the 2LPT gravity solver of the BORG SDSS has particularly limited validity in the strongly non-linear cluster regime.

To test the cluster RG density accuracy, we cross-matched our samples with \citet{Wen12015}'s SDSS DR12--based galaxy cluster catalogue.
To specify each cluster's location, this catalogue provides the right ascension, declination, and redshift of the brightest cluster galaxy (BCG).
First, we selected all clusters in the Local Universe ($z < z_\mathrm{max}$).
Next, we cross-matched in an angular sense, and considered an RG matched with a BCG if their angular separation is less than $3''$.
We thus retained 60 BCG giants and 113 BCG LoTSS DR1 RGs.
Figure~\ref{fig:BCGGiants} shows six members (10\%) of the BCG giants sample.
The perturbed jet and lobe shapes seen in the radio, and the massive ellipticals and crowded environments seen in the optical, together visually demonstrate the adequacy of our BCG giant identification.
We could identify LoTSS DR1 BCG RGs with equal confidence.

As a quality check of the cluster masses, we tested whether they are accurate enough to contain fingerprints of known physical effects.
In particular, we tested whether clusters with RG-generating BCGs are more massive than other Local Universe clusters.
The AGN feedback paradigm suggests that this is the case: in low-mass clusters, AGN feedback converts cool cores to non-cool cores within a gigayear, shutting down further AGN feedback for many gigayears to come; by contrast, high-mass clusters maintain stable AGN feedback cycles with gigayear-scale periods \citep{Nobels12022}, and thus their BCGs have a higher probability of being observed while generating RGs.
The median mass of clusters with RG-generating BCGs is $M_{500} = 1.1 \cdot 10^{14}\ M_\odot$, while for other clusters it is $M_{500} = 1.0 \cdot 10^{14}\ M_\odot$.
A Kolmogorov--Smirnov test yielded a $p$-value $p = 7\%$, suggesting to reject the null hypothesis of equality in distribution only for high significance levels (e.g. $\alpha = 10\%$).
We also tested whether clusters with giant-generating BCGs are more massive than other clusters.
The median mass of clusters with giant-generating BCGs is $M_{500} = 1.2 \cdot 10^{14}\ M_\odot$, while for other clusters it is $M_{500} = 1.0 \cdot 10^{14}\ M_\odot$.
A KS test yielded $p = 5\%$, suggesting to reject equality in distribution only for rather high significance levels (e.g. $\alpha = 5\%$ or $\alpha = 10\%$).

As a first quality check of our densities, we tested whether our BCG giants have higher densities than our non-BCG giants.
The median relative densities for BCG giants and non-BCG giants are $1 + \delta = 8.5$ and $1+\delta = 7.2$, respectively.
A KS test yielded $p = 0.2\%$, indeed suggesting to reject equality in distribution.
Similarly, we tested whether our LoTSS DR1 BCG RGs have higher densities than our LoTSS DR1 non-BCG RGs.
The median relative densities for BCG RGs and non-BCG RGs are $1+\delta = 8.4$ and $1+\delta = 5.9$, respectively.
A KS test yielded $p \ll 0.1\%$, again suggesting to reject equality in distribution.

\begin{figure}
    \centering
    \includegraphics[width=\columnwidth]{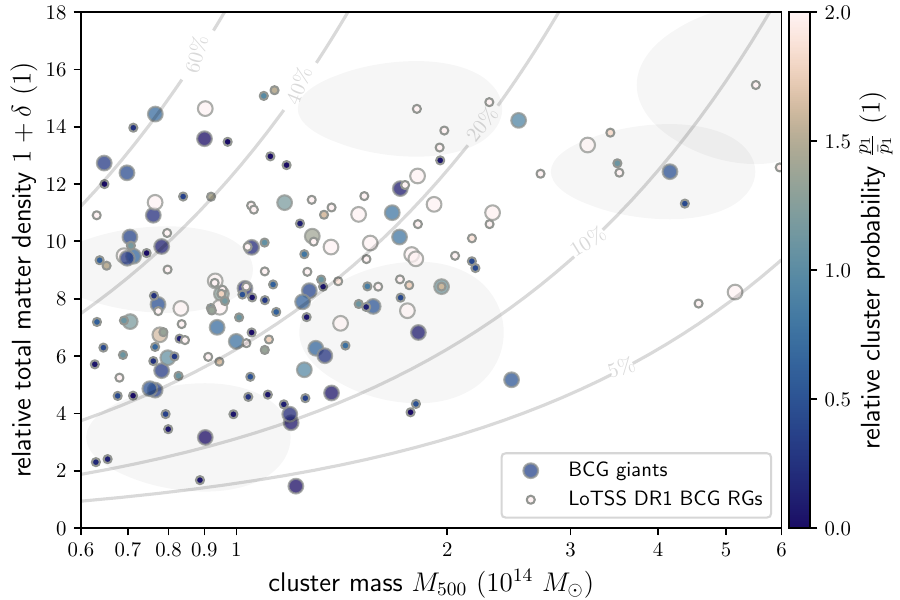}
    \caption{
    Comparison between environmental estimates for luminous giants (large dots) and LoTSS DR1 RGs (small dots) generated by BCGs in the Local Universe.
    We compare cluster masses $M_{500}$ from \citet{Wen12015} with relative densities $1+\delta$ from the BORG SDSS, and colour the dots based on their cluster probability $p_1$ (relative to the average cluster probability $\bar{p}_1$ for luminous giants and LoTSS DR1 RGs).
    For three randomly chosen giants and three randomly chosen RGs, we visualise uncertainties.
    Generally, as expected, RG-generating BCGs claimed to be in more massive clusters by \citet{Wen12015} have higher BORG SDSS densities and $T$-web cluster probabilities.
    The contours indicate fractions of the density expected through $M_{500}$, and show that our relative densities fall short significantly in the cluster regime --- by a factor that ranges from less than two to more than an order of magnitude.
    Densities of low-mass clusters are less biased than densities of high-mass clusters.
    We used flexible voxel method densities here.
    }
    \label{fig:GRGRGClusters}
\end{figure}

Figure~\ref{fig:GRGRGClusters} shows the relation between cluster mass and relative density.
Although it is qualitatively clear that BCG RGs in more massive clusters have higher inferred densities, the densities are all lower than one would expect from spreading out the cluster mass over a BORG SDSS voxel.
(For reference, a cluster of mass $M \sim 10^{14}\ M_\odot$ should have a $2.9\ \mathrm{Mpc}\ h^{-1}$--scale relative density $1+\delta \sim 40$.)
This systematic discrepancy cannot be attributed to errors in the cluster masses, as \citet{Wen12015} have made sure that their estimates are unbiased.
The alternative is to conclude that our densities are biased low in the cluster regime.
This is unsurprising, as the assumptions of the BORG SDSS's 2LPT gravity solver are violated in clusters.
Our densities are least accurate for high-mass clusters, where they can be off by an order of magnitude, and most accurate for low-mass clusters, where they can be less than a factor two too low.
It is unclear whether, and if so how, these accuracy characterisations can be extrapolated to the filament density regime.
The increased applicability of 2LPT to the filament density regime suggests that the accuracy will increase.
We therefore expect our filament densities to be correct up to a small factor of order unity.

Until Bayesian Cosmic Web reconstructions are further developed, we urge caution in using our inferred densities as precision estimates: in filament environments, they should be trusted up to a factor of order unity only; in cluster environments, for the moment, other sources (such as cluster catalogues) appear to provide more accurate density estimates.
If one would insist in using the cluster densities, then Fig.~\ref{fig:GRGRGClusters} suggests that one should correct them by multiplying by a factor of ${\sim}5$.

\subsection{Cosmic Web dynamical state accuracy}
Similar questions of accuracy can be raised about the Cosmic Web dynamical state distributions for luminous giants and RGs presented in the left and central pie charts of Fig.~\ref{fig:piesTWeb}.
To test the cluster GRG and cluster RG percentages, we again used the cluster catalogue of \citet{Wen12015}.

Cluster giants can be subdivided into BCG giants and non-BCG cluster giants.
In Sect.~\ref{sec:CosmicWebDensityAccuracy}, we determined that 60 out of \numberOfGRGsBORGzsp\ giants (23\%) are BCG giants.\footnote{In comparison, \citet{Dabhade12020March} found that only 20 out of 128 LoTSS DR1 giants with $z \leq 0.55$ (16\%) were BCG giants.
We hypothesise that we find a higher occurrence of BCG giants because our study is restricted to the Local Universe ($z < z_\mathrm{max} \coloneqq 0.16$), where cluster catalogues are generally more complete.}
We defined non-BCG cluster giants as giants generated by non-BCG galaxies for which the comoving distance to the BCG is less than some megaparsec-scale threshold.
We obtained approximate comoving coordinates from right ascensions, declinations, and spectroscopic redshifts, assuming vanishing peculiar motion.
In order to allow for a fair comparison to the BORG SDSS, we set the megaparsec-scale threshold to a BORG SDSS voxel side length: $2.9\ \mathrm{Mpc}\ h^{-1}$.
Although much larger than a galaxy cluster virial radius, this threshold does ensure that cluster galaxies with significant peculiar motion --- a standard deviation $\sigma_{v_\mathrm{p}} = 10^2$--$10^3\ \mathrm{km\ s^{-1}}$ is typical --- have a reasonable chance to be linked to their cluster.
For this choice of threshold, we found 21 non-BCG cluster giants.
We thus find a total of 81 cluster giants (31\%), which is more than expected based on the BORG SDSS measurement ($23\substack{+3\\-3}\%$) shown in Fig.~\ref{fig:piesTWeb}'s GRG pie chart.
This suggests that the BORG SDSS underpredicts the occurrence of cluster giants.
Indeed, the strongly underpredicted densities of galaxy clusters discussed in Sect.~\ref{sec:CosmicWebDensityAccuracy} also leave their mark on the accuracy of our dynamical states.
For example, the giant shown in the top-left panel of Fig.~\ref{fig:BCGGiants} and tabulated on the tenth row of Table~\ref{tab:GRGsCosmicWeb}, clearly occupies a cluster-like environment with an estimated $M_{500} = 1.3 \cdot 10^{14}\ M_\odot$, but is erroneously assigned an 80\% chance to inhabit a filament.

Similarly, cluster RGs can be subdivided into BCG RGs and non-BCG cluster RGs.
In Sect.~\ref{sec:CosmicWebDensityAccuracy}, we determined that 113 out of \numberOfRGsBORGzsp\ selected LoTSS DR1 RGs (8\%) are BCG RGs.
By defining non-BCG cluster RGs in a way analogous to non-BCG cluster giants, we found 109 non-BCG cluster RGs.\footnote{Whereas three-quarters of all luminous cluster giants are BCG giants, only half of all general cluster RGs are BCG RGs.
Thus, whenever a luminous giant is found in a cluster, it is most likely generated by the BCG; by contrast, whenever a general RG is found in a cluster, a BCG origin is just as likely as a non-BCG origin.}
We thus find a total of 222 cluster RGs (15\%), consistent with the BORG SDSS measurement ($16\substack{+1\\-1}\%$) shown in Fig.~\ref{fig:piesTWeb}'s RG pie chart.

To assess the accuracy of the filament GRG and RG fractions shown in Fig.~\ref{fig:piesTWeb}, we cross-matched our giants and selected LoTSS DR1 RGs with the galaxy group and cluster catalogue of \citet{Tempel12017}, which provides estimates of $M_{200}$.
\citet{Wen12015} define clusters through $M_{500} \geq 0.6 \cdot 10^{14}\ M_\odot$, which translates to $M_{200} \geq 0.9 \cdot 10^{14}\ M_\odot$: a typical conversion is $M_{500} = 0.7\ M_{200}$ \citep[e.g.][]{Pierpaoli12003}.
To separate groups from clusters, we therefore defined groups through $M_{200} < 0.9 \cdot 10^{14}\ M_\odot$.
As with our cluster tests, we associated RGs to groups on the basis of estimated comoving coordinates, assuming vanishing peculiar motion.
In the case of groups, this assumption is more reasonable.
We again used a $2.9\ \mathrm{Mpc}\ h^{-1}$ association threshold.
We associated 155 out of \numberOfGRGsBORGzsp\ luminous giants (60\%) and 903 out of \numberOfRGsBORGzsp\ general RGs (63\%) with galaxy groups.
These percentages provide lower bounds to the occurrence of filament giants and RGs, as there are also filament giants and RGs that do not reside in galaxy groups.
It is therefore likely that the filament GRG and RG fractions of Fig.~\ref{fig:piesTWeb} ($59\substack{+3\\-3}\%$ and $60\substack{+1\\-1}\%$, respectively) are underestimates.

Finally, as a probe of sheet and void environments, we explored the occurrence of giants and RGs for which the closest galaxy group or cluster is more than $10\ \mathrm{Mpc}$ away.
We identified 18 out of \numberOfGRGsBORGzsp\ luminous giants (7\%) and 115 out of \numberOfRGsBORGzsp\ general RGs (8\%) as occupying sheets and voids.
This suggests that the combined sheet and void GRG and RG fractions of Fig.~\ref{fig:piesTWeb} ($18\substack{+2\\-2}\%$ and $24\substack{+1\\-1}\%$, respectively) are significant overestimates.

Rounded to an accuracy of $10\%$, our ancillary environment analysis suggests an environment class probability distribution for luminous giants of $\vec{p} = (30\%, 60\%, 10\%)$, representing clusters, filaments, and a combination of sheets and voids, respectively.
For general RGs, it suggests $\vec{p} = (20\%, 70\%, 10\%)$.
Both distributions are in fair agreement with the BORG SDSS measurement $\vec{p} = (20\%, 60\%, 20\%)$.
The ancillary analysis reinforces the idea that, in the Local Universe, luminous giants detectable at LoTSS-like image qualities occupy denser environments than general RGs.

\subsection{Cosmic Web density resolution}
\label{sec:CosmicWebDensityResolution}
All Cosmic Web density and dynamical state measurements presented in this work correspond to a density field with a resolution, or smoothing scale, of $2.9\ \mathrm{Mpc}\ h^{-1}$.
In fact, given the granular nature of matter, one must always adopt a smoothing scale in order to define an informative notion of density.\footnote{As there are many situations in which there exists a natural smoothing scale, the smoothing scale is often left unmentioned.}
Ideally, this scale is chosen such that the resulting density field varies smoothly over the physical scales of interest --- avoiding both unwanted stochastic variations over space and time (as would occur when the smoothing scale is too small), and excessive blurring of the physical phenomena under study (as would occur when the smoothing scale is too large).
In the case of this work, the smoothing scale of the Cosmic Web density field could be considered too large.
It is set by the limitations of the BORG SDSS 2LPT gravity solver, the ambition to cover the entire Local Universe in the direction of the SDSS MGS footprint, and --- most practically --- the relation between the number of voxels and the numerical cost of generating the HMC Markov chain.
As galaxy clusters have typical radii of ${\sim}1\ \mathrm{Mpc}\ h^{-1}$, they are not resolved by the BORG SDSS; the same therefore certainly holds for galaxy groups.
Thus, even if the gravitational dynamics of the reconstructions were exact, the utility of the BORG SDSS in improving dynamical model fits to individual RGs is limited: the smoothing scale of the density field is larger than the scale required to resolve (beta model--like) density profiles within filaments and clusters ($0.1$--$1\ \mathrm{Mpc}\ h^{-1}$).
At best, our densities reflect the correct multi-megaparsec-scale averages, but these are still much lower than the, say, central $1\ \mathrm{Mpc}$--scale densities.
This is because the massive galaxies that usually host RGs tend to occupy the bottom of their local gravitational potential wells, where the large-scale density is typically higher than everywhere in the vicinity.
Relying on simulations, \citet{Oei12023} have estimated the relation between the $2.9\ \mathrm{Mpc}\ h^{-1}$--scale relative total matter density $1+\delta$ (as used throughout this work), and the $1\ \mathrm{Mpc}$--scale relative IGM density $1 + \delta_\mathrm{IGM}$.
For galaxies with a stellar mass $M_\star = 10^{11}\ M_\odot$, their Fig.~11 suggests $1 + \delta_\mathrm{IGM} = 10 \cdot (1+\delta)^{0.75}$.
For luminous giants, the mode of the $2.9\ \mathrm{Mpc}\ h^{-1}$--scale relative total matter density RV is $1 + \delta = 3.5$ for the fixed voxel method and $1 + \delta = 6.3$ for the flexible voxel method (see Fig.~\ref{fig:distributionsDensity}); the corresponding host-centred $1\ \mathrm{Mpc}$--scale relative IGM density is $1 + \delta_\mathrm{IGM} = 23$ for the fixed voxel method and $1 + \delta_\mathrm{IGM} = 40$ for the flexible voxel method.
This relation in principle allows one to perform Bayesian inference of the distributions of the $1\ \mathrm{Mpc}$--scale Cosmic Web density around RG hosts --- possibly even conditioned on RG radio luminosity, as in Sect.~\ref{sec:radioLuminosityCosmicWebDensity} --- by converting parametrised high-resolution distributions to the lower BORG SDSS resolution, applying heteroskedastic measurement errors, and finally comparing the resulting distribution to data.
We leave this to future work.

Concerning resolution, we finally warn that care must be taken when calculating (AGN feedback--related) quantities that relate quadratically to density, such as the local bremsstrahlung emissivity: these tend to be underestimated if computed from low-resolution densities.
To see why, we considered $\rho$, the density at some natural smoothing scale, alongside $\langle \rho \rangle$, the density averaged over some larger-than-natural smoothing scale.
We assumed that only the latter is known.
Next, we picked a quantity $y = a\rho^2$, which we aimed to compute at the larger smoothing scale: $\langle y \rangle = \langle a\rho^2 \rangle = a \langle \rho^2 \rangle$.
Naively, one would be tempted to proceed in calculating $\langle y \rangle$ by assuming $\langle \rho^2 \rangle \approx \langle \rho \rangle^2$, as to use $\langle y \rangle \approx a\langle \rho \rangle ^2$.
However, by doing so, one risks underestimating $\langle y \rangle$, because $\langle \rho^2 \rangle \geq \langle \rho \rangle^2$ by Jensen's inequality.
In Appendix~\ref{ap:densitySquared}, we explicitly calculated the ratio between $\langle \rho^2 \rangle$ and $\langle \rho \rangle^2$ for sheets and voids, filaments, and clusters.
For sheets and voids, this ratio is at most $\frac{4}{3}$; for filaments, it is at most a number of order unity; for clusters, it can exceed a thousand.


\subsection{Spectroscopic redshift selection}
\label{sec:selectionEffects}
\begin{figure}
    \centering
    \includegraphics[width=\columnwidth]{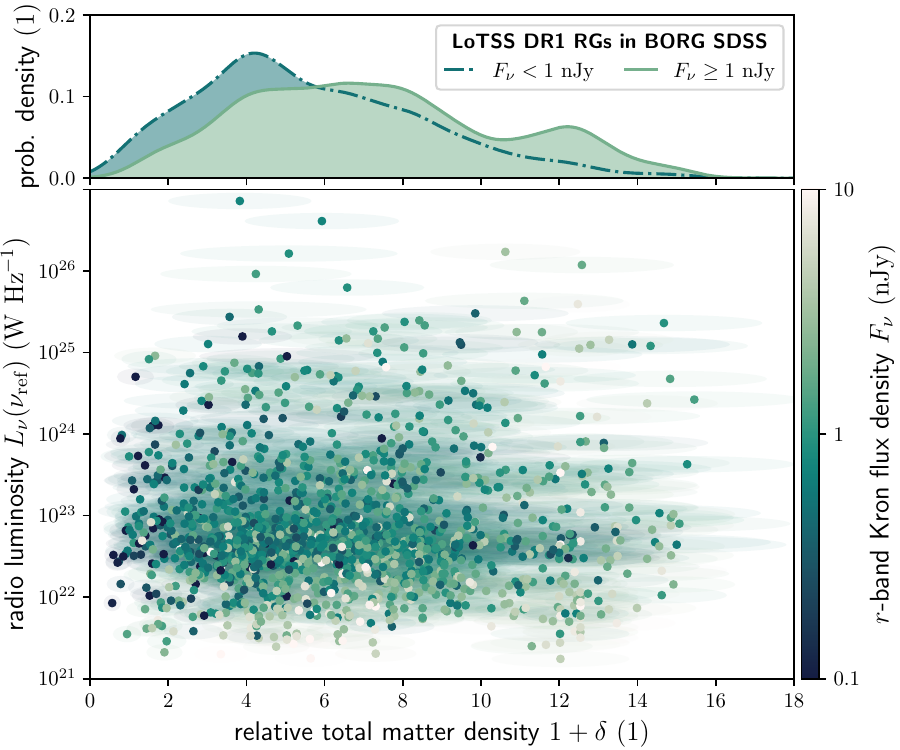}
    \caption{Observations underpinning the RG radio luminosity--Cosmic Web density relation of Fig.~\ref{fig:modelDensityLuminosityScatterFlexible}, but with Pan-STARRS $r$-band Kron flux densities $F_\nu$ of the host galaxies indicated.
    We show uncertainty ellipses whose semi-major and semi-minor axes represent one standard deviation along each dimension.
    In the top panel, we show KDE Cosmic Web density distributions for $F_\nu < 1\ \mathrm{nJy}$, and for $F_\nu \geq 1\ \mathrm{nJy}$.
    These suggest that spectroscopic redshift selection biases the observed Cosmic Web density distributions high.
    }
    \label{fig:RGDensitiesRadioLuminositiesFluxesKronR}
\end{figure}

\begin{figure}
    \centering
    \includegraphics[width=\columnwidth]{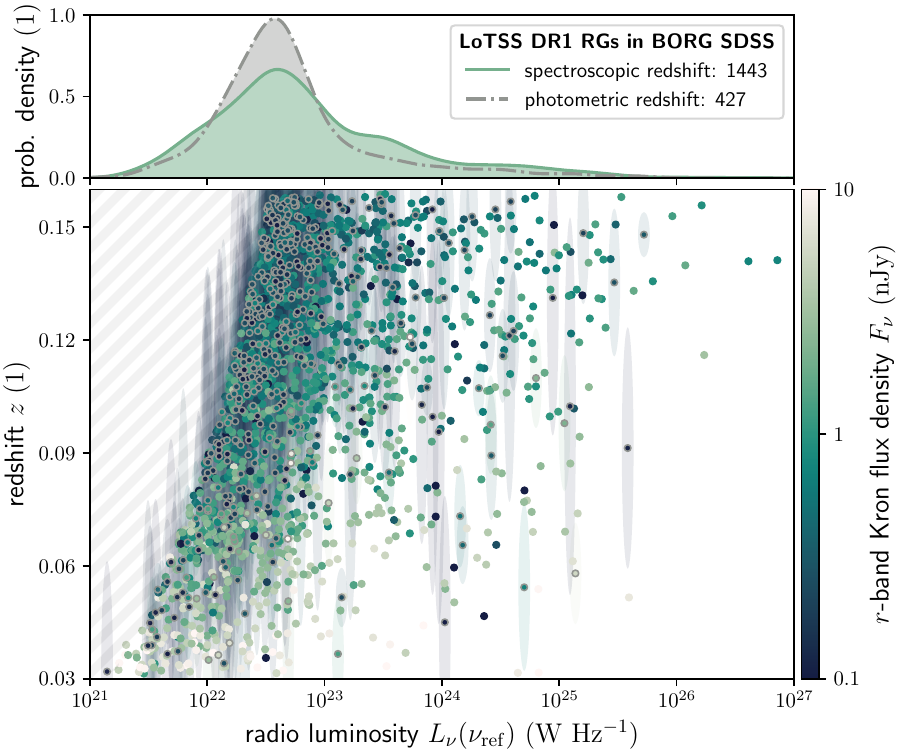}
    \caption{
    Overview of radio luminosity selection resulting from spectroscopic redshift selection, radio surface brightness limitations, and the correlation between radio and optical luminosity.
    The top panel compares radio luminosity distributions at $\nu_\mathrm{ref} = 150\ \mathrm{MHz}$ of LoTSS DR1 RGs in the constrained part of the BORG SDSS, distinguishing between those with spectroscopic redshifts (green) and those without (grey).\protect\footnotemark
    \ Radio galaxies with spectroscopic redshifts dominate (77\%) the population.
    The bottom panel shows the two subpopulations in radio luminosity--redshift space, with RGs without spectroscopic redshifts marked (grey circles).
    We show uncertainty ellipses whose semi-major and semi-minor axes represent half a standard deviation along each dimension.
    One retrieves the top panel by collapsing the data on the horizontal axis.
    }
    \label{fig:RGRadioLuminositiesSpectroscopicPhotometric}
\end{figure}
\footnotetext{We perform KDE on samples of $\mathrm{log}_{10}(L_\nu(\nu_\mathrm{ref}) \cdot (\mathrm{W\ Hz^{-1}})^{-1})$ rather than of $L_\nu(\nu_\mathrm{ref})$, and thus the probability densities on the vertical axis are dimensionless.}\noindent
To localise RGs in the Cosmic Web and subsequently measure their relative total matter densities, we required them to have a spectroscopic redshift.
By selecting on spectroscopic redshift availability, we implicitly selected on both Cosmic Web density and radio luminosity.
The key reason is that spectroscopic redshifts are preferentially available for galaxies with high optical flux densities --- so galaxies that are nearby,\footnote{There is an exception to this rule: physically large galaxies at low redshifts, such as ellipticals in nearby galaxy clusters, can be angularly too extended for SDSS fibres and are therefore sometimes excluded from SDSS's spectroscopic survey.
Excluding the RGs associated to these galaxies from our samples introduces a bias against high density environments.
This artificially vacates the right sides of Figs.~\ref{fig:GRGRGDensityDistributions} and \ref{fig:modelDensityLuminosityScatterFlexible}, and leads towards an underestimation of the occurrence of cluster RGs in our environment analysis.} and galaxies that have high optical luminosities.
Galaxies with high optical luminosities tend to inhabit dense environments and, when active, tend to generate RGs with high radio luminosities.

To demonstrate that spectroscopic redshift selection indeed affects our RG radio luminosity--Cosmic Web density relation, Fig.~\ref{fig:RGDensitiesRadioLuminositiesFluxesKronR} revisits the same Local Universe LoTSS DR1 RG data as shown in Fig.~\ref{fig:modelDensityLuminosityScatterFlexible}, but now with the Pan-STARRS $r$-band Kron flux densities $F_\nu$ of the host galaxies indicated.
The top panel of Fig.~\ref{fig:RGDensitiesRadioLuminositiesFluxesKronR} shows KDEs of $1+\Delta_\mathrm{RG,obs}$ for hosts with $F_\nu < 1\ \mathrm{nJy}$, and for hosts with $F_\nu \geq 1\ \mathrm{nJy}$.
Hosts with low optical flux densities tend to inhabit more tenuous environments than hosts with high optical flux densities.
By requiring spectroscopic redshifts, we have discarded RGs with only photometric redshifts, whose hosts almost exclusively have $F_\nu < 1\ \mathrm{nJy}$.
As a result, our estimates of $1+\Delta_\mathrm{RG,obs}$, $1+\Delta_\mathrm{RG}$, and $1+\Delta_\mathrm{RG}\ \vert\ L_\nu = l_\nu$ are biased high.
The severity of the bias depends on the percentage of RGs that have been selected out.
As discussed in Sect.~\ref{sec:BORGSDSSLocalisationInPractice}, 23\% of all LoTSS DR1 RGs within the constrained BORG SDSS volume lack a spectroscopic redshift.
Our sample of luminous giants is much less affected: only 7\% of the population within the constrained BORG SDSS volume lack spectroscopy.
Thus, correcting for spectroscopic redshift selection would shift the Cosmic Web density distribution for general RGs to somewhat lower densities, while it would not appreciably shift the corresponding distribution for luminous giants.
The result would be an even bigger discrepancy between the Cosmic Web density distributions of luminous giants and general RGs than established in this work, strengthening our results.

Spectroscopic redshift selection also leaves an imprint on the sample's radio luminosities.
The top panel of Fig.~\ref{fig:RGRadioLuminositiesSpectroscopicPhotometric} shows that the radio luminosities of the \numberOfRGsBORG\ LoTSS DR1 RGs in the constrained BORG SDSS volume are distributed differently between the subpopulations with and without spectroscopic redshifts.
In particular, the subpopulation with spectroscopic redshifts is more spread out over radio luminosity compared to the subpopulation without: both very low and very high values are more probable.\footnote{The distribution is not only more dispersed, but is also shifted rightwards: the median and mean radio luminosities of the spectroscopic subpopulation are 1.3 and 3.2 times larger, respectively.}
To understand why, we turn to the bottom panel of Fig.~\ref{fig:RGRadioLuminositiesSpectroscopicPhotometric}.
The LoTSS $6''$ noise level, the inverse square law, and cosmological surface brightness dimming together determine an RG radio luminosity detection threshold that increases with redshift.
This threshold induces the hatched region, which is inaccessible to LoTSS-like radio observations.
As a straightforward consequence, RGs with low radio luminosities can be detected only at low redshifts.
It is not immediately clear whether the host galaxies of low--radio luminosity RGs at low redshifts have low or high optical flux densities.
On the one hand, low radio luminosities suggest low optical luminosities: both forms of spectral luminosity correlate positively with stellar mass \citep[as evinced by, for example, Figs. 2 of][]{Mahajan12018, Sabater12019} and therefore\footnote{There exist cases in which this argument does not hold: correlation is not necessarily transitive.} positively with eachother.
However, one must again be cautious of selection effects, which can inject strong but spurious luminosity--luminosity correlations in observations \citep{Singal12019}.
On the other hand, observing low-redshift galaxies is easier, again because of the inverse square law and cosmological surface brightness dimming.
The bottom panel of Fig.~\ref{fig:RGRadioLuminositiesSpectroscopicPhotometric}, and in particular its lower left corner, shows that the correlation between radio and optical luminosity is not strong enough to deny the observational advantages of low redshifts: low--radio luminosity RGs at low redshifts have high optical flux densities.
As a result, spectroscopic redshifts are more common for such RGs.
Similarly, the correlation between radio and optical luminosity explains that spectroscopic redshifts are more common for high--radio luminosity RGs.

This indirect radio luminosity selection effect is not expected to change the radio luminosity--Cosmic Web density relation inferred in this work and shown in Figs.~\ref{fig:modelDensityLuminosityScatterFlexible} and \ref{fig:modelDensityLuminosityCornerFlexible} (as well as in Figs.~\ref{fig:modelDensityLuminosityScatterFixed} and \ref{fig:modelDensityLuminosityCornerFixed}).
This is because the inferred relation concerns the Cosmic Web density distribution at given radio luminosity.
Changes in the number of RGs available in each decade of radio luminosity --- and in particular an increase in the number of RGs with $10^{22}\ \mathrm{W\ Hz^{-1}} < L_\nu(\nu = 150\ \mathrm{MHz}) < 10^{23}\ \mathrm{W\ Hz^{-1}}$ --- do not materially affect our analysis, as we already performed data rebalancing that downweights our sample's most populated decades.\footnote{The hatched region of Fig.~\ref{fig:RGRadioLuminositiesSpectroscopicPhotometric} shows that radio surface brightness selection causes our Local Universe RG sample to be strongly incomplete in the radio luminosity decades $10^{21}$--$10^{22}$ and $10^{22}$--$10^{23}\ \mathrm{W\ Hz^{-1}}$.
Extending our sample in the lowest of these decades upweights this decade's importance, likely revealing that the positive scaling between RG radio luminosity and Cosmic Web density is stronger than we suggest in Sect.~\ref{sec:radioLuminosityCosmicWebDensity}.}


\subsection{Future outlook}
Excitingly, there appear to be many opportuntities to improve our understanding of the large-scale environments of radio galaxies through Cosmic Web reconstructions, both with current and near-future data.

In this work, we have used LoTSS DR1 Local Universe RLAGN with spectroscopic redshifts from \citet{Hardcastle12019}.
With the release of the LoTSS DR2 radio data \citep{Shimwell12022} and the associated optical crossmatching results \citep{Hardcastle12023}, it will become possible to redo our analysis on a footprint that is more than 13 times larger.
Doing so would increase the size of the general RG sample by an order of magnitude, from ${\sim}10^3$ to ${\sim}10^4$.
In turn, this would allow us to revisit Sect.~\ref{sec:radioLuminosityCosmicWebDensity}'s RG radio luminosity--Cosmic Web density relation, but with all of Fig.~\ref{fig:modelDensityLuminosityScatterFlexible}'s six radio luminosity decades fully evenly weighted.
If the positive correlation between radio luminosity and Cosmic Web density is genuine, it will emerge more clearly upon weighting evenly.
Such a LoTSS DR2 sample might even be large enough to meaningfully investigate possible differences in the radio luminosity--Cosmic Web density relation for different types of RGs, such as those generated by quasars, and those generated by other AGN.
In the near future, the WEAVE--LOFAR Survey \citep{Smith12016} will generate ${\sim}10^6$ spectra of LOFAR-selected galaxies, which will help alleviate the effects of spectroscopic redshift selection described in Sect.~\ref{sec:selectionEffects}.
Finally, the availability of optical spectra for all AGN in the current analysis and for those in similar follow-ups means that a comprehensive classification into HERGs and LERGs appears possible.
In turn, this classification would allow for studies into the relationships between AGN accretion mode, RG growth, and the surrounding Cosmic Web.
As these AGN reside in the Local Universe, many could be further characterised by crossmatching with X-ray catalogues.

Another promising extension of our work would be to use Cosmic Web reconstructions that are complementary to or improve upon the BORG SDSS.
Notably, the BORG 2M++ \citep{Jasche12019} offers one such possibility: it provides density reconstructions for the entire sky and at a higher spatial resolution, and features more accurate selection functions, bias modelling, and gravitational dynamics than the BORG SDSS.
However, its reconstructions are limited to $z_\mathrm{max} = 0.1$, whereas those of the BORG SDSS extend to $z_\mathrm{max} = 0.16$.
Using the BORG SDSS and the BORG 2M++ in conjunction would allow for an extension of the number of giants for which a density can be determined: the catalogue of giants by \citet{Oei12022GiantsSample}, introduced in Sect.~\ref{sec:data}, contains 83 giants with spectroscopic redshifts $z_\mathrm{s} < 0.1$ that lie outside of the constrained BORG SDSS volume.
Adding these to the \numberOfGRGsBORGzsp\ giants for which we have already determined densities would increase the number of BORG giants by 32\%.
Perhaps more importantly, there are 89 giants for which both BORG SDSS and BORG 2M++ density estimates are possible.
Comparing these densities allows for a quality assessment of the BORG SDSS density measurements at large.
In particular, given its better gravity solver, we expect BORG 2M++ cluster densities to perform much better in Sect.~\ref{sec:CosmicWebDensityAccuracy}'s Cosmic Web density--cluster mass test.
In addition to the BORG 2M++, additional BORG runs are on their way.
Besides, there exist comparable Cosmic Web reconstruction frameworks, such as COSMIC BIRTH \citep[e.g.][]{Kitaura12021}.
In any case, thanks to the combination of new spectroscopic galaxy surveys and advances in computing power, high-fidelity Bayesian Cosmic Web reconstructions will become available for an increasingly large fraction of the Local Universe and beyond.
We expect that Cosmic Web density $1+\delta$ will therefore become a standard RG observable, alongside properties such as projected proper length $l_\mathrm{p}$ and radio luminosity $l_\nu$.
Through the work of \citet{Leclercq12015}, we have derived probabilistic Cosmic Web environment classifications, based on the $T$-web definition, for \numberOfGRGsBORGzsp\ luminous giants and \numberOfRGsBORGzsp\ LoTSS DR1 RGs.
However, \citet{Leclercq12015} provides additional environmental classifications for the BORG SDSS volume that allow us to measure and understand the Cosmic Web environments of giants and other RGs in complementary ways.
Future work could explore and compare RG environments in the $T$-web, DIVA, ORIGAMI, and LICH sense \citep{Leclercq12016, Leclercq12017}.

\section{Conclusion}
\label{sec:conclusion}
Using Bayesian Cosmic Web reconstructions of the Local Universe, we compared the large-scale environments of giant radio galaxies with those of the radio galaxy population in general.
In particular, we measured multi-megaparsec-scale Cosmic Web densities around the hosts of ${\sim}10^2$ luminous giants and ${\sim}10^3$ general RGs.
This reveals that the currently observable population of giants inhabits denser regions of the Cosmic Web than general RGs --- contradicting the popular hypothesis that giants primarily form in the dilute Cosmic Web.
Our interpretation is that high jet powers, as implied by the high radio luminosities of the known population of giants, enable these RGs to overcome the IGM's resistance to their megaparsec-scale growth.

Next, we quantified, for the first time, the relation between radio luminosity, a proxy for jet power, and Cosmic Web density.
Our radio luminosity--conditioned Cosmic Web density distributions reinforce the idea that AGN that generate weak jets primarily reside in the dilute Cosmic Web, while AGN that generate powerful jets primarily reside in the dense Cosmic Web.
We subsequently showed that the radio luminosity--Cosmic Web density relation, which we have inferred using our sample of general RGs only, can accurately explain the density distribution observed for luminous giants.
Additional evidence corroborates that luminous giants form more often in the dense Cosmic Web than RGs in general.
Classifying Cosmic Web environments as clusters, filaments, sheets, or voids on the basis of the local gravitational field, we find that luminous giants occur more often in clusters than RGs in general.
We also find evidence that clusters with giant-generating BCGs are more massive than other clusters; the evidence for the analogous claim for clusters with RG-generating BCGs is weaker.
We now summarise our main findings in more detail.

\begin{enumerate}
    \item Using the BORG SDSS, we have performed the most physically principled measurements yet of the Cosmic Web environments of giant radio galaxies and radio galaxies in general.
    Our characterisations require spectroscopic redshifts, and are currently confined to the Local Universe ($z < z_\mathrm{max} \coloneqq 0.16$).
    In particular, we have determined $2.9\ \mathrm{Mpc}\ h^{-1}$--scale total matter densities and gravitational environment classifications for \numberOfGRGsBORGzsp\ luminous giants, of which \numberOfGRGsBORGzspOei\ (80\%) have recently been discovered through the LoTSS DR2, and for \numberOfRGsBORGzsp\ LoTSS DR1 RGs.
    \item While the marginal probability distributions of the Cosmic Web density around individual RGs are approximately lognormal, the probability distribution of the Cosmic Web density for the RG population as a whole is approximately gamma.
    We provide a physical argument that motivates why the gamma distribution could have arisen.
    The probability distribution of the Cosmic Web density for luminous giants also appears roughly gamma, but favours higher densities.
    Heteroskedastic measurement errors, which are a general feature of our densities, spuriously shift the population distributions towards lower densities.
    We demonstrate a forward modelling method to correct for heteroskedasticity.
    \item We quantified the radio luminosity--Cosmic Web density relation for general RGs.
    We find that the $2.9\ \mathrm{Mpc}\ h^{-1}$--scale Cosmic Web density distribution at a given $150\ \mathrm{MHz}$ radio luminosity is well described by $1 + \Delta_\mathrm{RG}\ \vert\ L_\nu = l_\nu \sim \Gamma(k,\theta)$, where $k = 4.8 + 0.2 \cdot \textphnc{\Alamed}$, $\theta = 1.4 + 0.02 \cdot \textphnc{\Alamed}$, and $\textphnc{\Alamed} \coloneqq \mathrm{log}_{10}(l_\nu \cdot (10^{23}\ \mathrm{W\ Hz^{-1}})^{-1})$.
    This shows that more luminous RGs tend to live in denser regions of the Cosmic Web.
    Treating giants as ordinary RGs with $l_\nu \geq 10^{25}\ \mathrm{W\ Hz^{-1}}$, we used this relation to predict the Cosmic Web density distribution for luminous giants.
    The prediction is consistent with the observed distribution for giants: a Kolmogorov--Smirnov test yields a $p$-value of 0.7.
    Whether less luminous giants --- assuming they exist --- also form more often in the dense Cosmic Web, remains to be seen when surveys with increased surface brightness sensitivity commence.
    If such giants obey the same radio luminosity--Cosmic Web density relation as other RGs, this will not be the case.
    \item We show that our methodology enables the inference of radio galaxy number densities as a function of Cosmic Web density.
    Whether or not the RG number density is a strong function of Cosmic Web density, depends on the radio luminosity considered.
    In clusters, the number densities of high-luminosity RGs (e.g. $10^{25}$--$10^{27}\ \mathrm{W\ Hz^{-1}}$) are ${\sim}10^2$ times higher than on average; in the same environment type, the number densities of low-luminosity RGs (e.g. $10^{21}$--$10^{23}\ \mathrm{W\ Hz^{-1}}$) are just ${\sim}10^1$ times higher than on average.
    Furthermore, we obtain tentative evidence that the number densities of low-luminosity RGs peak in low-mass clusters, before decreasing again.
    A possible explanation could be that, as cluster mass increases, the underlying population of active galaxies starts favouring the generation of high-luminosity RGs over low-luminosity RGs.
    \item We used the BORG SDSS $T$-web classification to generate probability distributions over four Cosmic Web structure types --- clusters, filaments, sheets, and voids --- for both luminous giants and general RGs.
    Luminous giants inhabit clusters more often ($23\substack{+3\\-3}\%$ versus $16\substack{+1\\-1}\%$).
    Independently, by crossmatching our RGs with a cluster catalogue, we find that the former percentage presumably is an underestimate, while confirming the latter percentage.
    The same crossmatching procedure reveals tentative evidence ($p = 5\%$) that clusters with giant-generating BCGs are more massive than other clusters.
    At the same time, evidence that clusters with RG-generating BCGs are more massive than other clusters is weaker ($p = 7\%$), even though the sample involved is almost twice as big.
\end{enumerate}
Although RG environment characterisations with Bayesian Cosmic Web reconstructions are still far from perfect, we have demonstrated that the current generation of reconstructions already allows one to address open questions in radio galaxy research.
Given the pitfalls of the BORG SDSS, every measured Cosmic Web density should be treated with skepsis, and we warn in particular that the densities around cluster RGs are strongly underestimated.
Nevertheless, given the ever-increasing number of galaxies with spectroscopic redshifts, a growing list of technical refinements, and the widespread availability of computing power, exciting contemporary and near-future opportunities exist to expand and improve upon the results of this work.
It is well possible that Cosmic Web density will soon become a staple quantity that helps to unravel the physics of radio galaxies.

\begin{acknowledgements}
M.S.S.L. Oei, R.J. van Weeren, and A. Botteon acknowledge support from the VIDI research programme with project number 639.042.729, which is financed by the Dutch Research Council (NWO).
A. Botteon acknowledges financial support from the European Union - Next Generation EU.
M.S.S.L. Oei warmly thanks Rafa\"el Mostert, Erik Osinga, and Wendy Williams for helpful discussions and the sharing of data.
LOFAR data products were provided by the LOFAR Surveys Key Science project (LSKSP; \url{https://lofar-surveys.org/}) and were derived from observations with the International LOFAR Telescope (ILT).
LOFAR \citep{vanHaarlem12013} is the Low Frequency Array designed and constructed by ASTRON.
It has observing, data processing, and data storage facilities in several countries, which are owned by various parties (each with their own funding sources), and which are collectively operated by the ILT foundation under a joint scientific policy.
The efforts of the LSKSP have benefited from funding from the European Research Council, NOVA, NWO, CNRS-INSU, the SURF Co-operative, the UK Science and Technology Funding Council and the J\"ulich Supercomputing Centre.
\citet{Oei12022GiantsSample}'s LoTSS DR2 GRG data are avail\-able at the CDS via \url{https://cdsarc.cds.unistra.fr/viz-bin/cat/J/A+A/672/A163}.
The LoTSS DR1 RG data with optical identifications from the value-added catalogue of \citet{Williams12019} are available at \url{https://lofar-surveys.org/dr1_release.html}; those interested in \citet{Hardcastle12019}'s RLAGN subsample should contact Martin J. Hardcastle.
The BORG SDSS data release is publicly available at \url{https://github.com/florent-leclercq/borg_sdss_data_release}.
Individual BORG SDSS HMC Markov chain samples are not available in the public domain; those interested should contact Florent Leclercq.
The cluster catalogue of \citet{Wen12015} is available at the CDS.
\end{acknowledgements}




{\footnotesize\bibliography{cite}}

%
%





\begin{appendix}
\section{Cosmic Web localisation accuracy with spectroscopic and photometric redshifts}
\label{ap:spectroscopicVsPhotometric}
In this appendix, we show that RGs with spectroscopic redshifts can be reliably localised within Cosmic Web reconstructions, while RGs with only photometric redshifts generally cannot.

To determine the accuracy of our radio galaxy localisations in the Cosmic Web, we Monte Carlo simulated a radial comoving distance distribution for each RG and compared its dispersion to the BORG SDSS voxel length.
Each RG has a peculiar velocity $v_\mathrm{p}$ with respect to us.
We treated $v_\mathrm{p}$ as a zero-mean Gaussian random variable (RV): $v_\mathrm{p} \sim \mathcal{N}(0, \sigma_{v_\mathrm{p}})$, and chose a standard deviation $\sigma_{v_\mathrm{p}}$ representative of conditions in low-mass galaxy clusters: $\sigma_{v_\mathrm{p}} = 100\ \mathrm{km\ s^{-1}}$.
Similarly, we treated the measured redshift $z_\mathrm{m}$ as an RV, and again assumed Gaussianity: $z_\mathrm{m} \sim \mathcal{N}\left(\mu_{z_\mathrm{m}}, \sigma_{z_\mathrm{m}}\right)$.
Our catalogue provides the parameters $\mu_{z_\mathrm{m}}$ and $\sigma_{z_\mathrm{m}}$ for each RG.
Using Eqs.~\ref{eq:redshiftCosmological} from left to right, we calculated the relative peculiar velocity RV $\beta_\mathrm{p}$, the peculiar velocity redshift RV $z_\mathrm{p}$, and the cosmological redshift RV $z_\mathrm{c}$:
\begin{align}
    \beta_\mathrm{p} \coloneqq \frac{v_\mathrm{p}}{c};\ \ \ \ z_\mathrm{p} = \sqrt{\frac{1+\beta_\mathrm{p}}{1-\beta_\mathrm{p}}} - 1;\ \ \ \ z_\mathrm{c} = \frac{1 + z_\mathrm{m}}{1 + z_\mathrm{p}} - 1.
\label{eq:redshiftCosmological}
\end{align}
Finally, we calculated the radial comoving distance RV $r = r\left(z_\mathrm{c}\right) = ||\mathbf{r}||_2$.
The distributions of $r$ are approximately Gaussian: for each of the \numberOfGRGsBORGzspUncertainty\ giants in the BORG SDSS volume with a spectroscopic redshift and an associated error, we performed the Shapiro--Wilk test on 1000 Monte Carlo samples and rejected the null hypothesis that $r$ is Gaussian for one at significance level $\alpha = 0.01$.\footnote{If one uses sufficiently many Monte Carlo samples and a fixed significance level, any deviation from Gaussianity will become significant.
Indirectly, the sample size and significance level together specify a degree of non-Gaussianity identifiable with such Shapiro--Wilk tests.}
For these giants, 95\% of the standard deviations $\sigma_r$ are between $1.00\ \mathrm{Mpc}\ h^{-1}$ and $1.12\ \mathrm{Mpc}\ h^{-1}$; the sample median is $1.06\ \mathrm{Mpc}\ h^{-1}$.
By contrast, the 21 giants in the BORG SDSS volume with only a photometric redshift and an associated error have a $\sigma_r$ that ranges from $20\ \mathrm{Mpc}\ h^{-1}$ to $110\ \mathrm{Mpc}\ h^{-1}$; for details, see Table~\ref{tab:GRGsFollowUp}.
To localise a giant in the BORG SDSS, precision up to the scale of a voxel, which are approximately $2.9\ \mathrm{Mpc}\ h^{-1}$ long, is necessary --- but sub-voxel localisation is redundant.
We conclude that giants with a spectroscopic redshift can be tied to an individual voxel, even when taking peculiar velocity into account; on the other hand, giants with only a photometric redshift can be mislocalised more than a typical filament length, and are therefore not subjectable to a precise environment analysis.
Of course, these conclusions extend to any radio galaxy with an associated host, or indeed to any galaxy.\footnote{By jointly inferring the large-scale density field and the radial distances to the galaxies used to constrain the former, it is possible to reduce individual photometric redshift uncertainties \citep{Jasche12012, Tsaprazi12023}, but these reductions are not extensive enough to alter our conclusions.}

As only few (i.e. ${\sim}10^2$) giants are currently known in the volume reconstructed by the BORG SDSS, it is worthwhile to perform host galaxy spectroscopic follow-up for those with photometric redshifts only.
We provide a list with targets in Table~\ref{tab:GRGsFollowUp}.

\begin{table}
\centering
\caption{
Overview of all giants in the constrained BORG SDSS volume without a spectroscopic redshift.\protect\footnotemark}
\label{tab:GRGsFollowUp}
\resizebox{\columnwidth}{!}{%
\begin{tabular}{l l l l l} 
\hline
rank & SDSS DR12 name & photometric & $\sigma_r$ & $l_\mathrm{p}$\\
$\downarrow$ & & redshift $(1)$ & $(\mathrm{Mpc}\ h^{-1})$ & $(\mathrm{Mpc})$\\
\hline
1 & J122009.84+194833.0 & 0.15 $\pm$ 0.04 & 110 & 0.8\\
2 & J155503.00+280430.9 & 0.10 $\pm$ 0.03 & 100 & 1.3\\
3 & J134211.92+565839.3 & 0.11 $\pm$ 0.03 & 70 & 0.7\\
4 & J130444.37+511119.0 & 0.09 $\pm$ 0.02 & 70 & 0.8\\
5 & J122129.95+662644.0 & 0.10 $\pm$ 0.02 & 70 & 1.1\\
6 & J152151.93+570635.0 & 0.07 $\pm$ 0.02 & 50 & 0.8\\
7 & J112033.54+233559.0 & 0.12 $\pm$ 0.02 & 50 & 0.9\\
8 & J140046.38+301900.0 & 0.06 $\pm$ 0.01 & 40 & 0.8\\
9 & J090640.80+142522.9 & 0.13 $\pm$ 0.01 & 40 & 1.2\\
10 & J112248.98+565243.5 & 0.08 $\pm$ 0.01 & 30 & 0.8\\
11 & J085022.99+383547.2 & 0.14 $\pm$ 0.01 & 30 & 0.7\\
12 & J145102.13+301227.0 & 0.16 $\pm$ 0.01 & 30 & 0.8\\
13 & J152229.23+281911.9 & 0.13 $\pm$ 0.01 & 30 & 0.8\\
14 & J140044.26+125219.8 & 0.11 $\pm$ 0.01 & 30 & 0.7\\
15 & J090128.15+145158.1 & 0.14 $\pm$ 0.01 & 30 & 1.2\\
16 & J145827.21+331312.0 & 0.11 $\pm$ 0.01 & 20 & 1.0\\
17 & J152024.13+310557.6 & 0.06 $\pm$ 0.01 & 20 & 0.9\\
18 & J102135.20+420022.3 & 0.13 $\pm$ 0.01 & 20 & 1.5\\
19 & J172715.85+585220.5 & 0.15 $\pm$ 0.01 & 20 & 2.1\\
20 & J141033.40+405932.4 & 0.14 $\pm$ 0.01 & 20 & 1.0\\
21 & J134339.20+195301.7 & 0.15 $\pm$ 0.01 & 20 & 0.9\\
\hline
\end{tabular}%
}
\end{table}
\footnotetext{Roughly speaking, this volume corresponds to the SDSS DR7 MGS footprint up to $z_\mathrm{max} = 0.16$.
The total matter density of the ambient Cosmic Web can only be reliably determined after spectroscopic follow-up.
The giants are sorted (in descending order) by their radial comoving distance standard deviation $\sigma_r$ obtained through Monte Carlo simulation, taking into account photometric redshift errors and a random, zero-centred peculiar velocity component with standard deviation $\sigma_{v_\mathrm{p}} = 100\ \mathrm{km\ s^{-1}}$.
After spectroscopic follow-up, the projected proper length $l_\mathrm{p}$ must be revised.}


\section{Cosmic Web density distribution: the gamma Ansatz}
\label{ap:gammaAnsatz}
In this work, we have modelled several RG Cosmic Web density RVs, such as $1 + \Delta_\mathrm{RG,obs}$, $1 + \Delta_\mathrm{RG}$, and $1 + \Delta_\mathrm{RG}\ \vert\ L_\nu = l_\nu$, as gamma variates.
This choice is largely driven by data (and a preference for simplicity) rather than by theory, although the requirements of continuity and a strictly positive support have a physical basis.
Gamma distributions adequately fit $1 + \Delta_\mathrm{RG,obs}$ (see the bottom row of Fig.~\ref{fig:distributionsDensity}), and their use in modelling $1 + \Delta_\mathrm{RG}\ \vert\ L_\nu = l_\nu$ is justified by the RG prediction of $1 + \Delta_\mathrm{GRG,obs}$, which matches observations (see Fig.~\ref{fig:distributionDensityGRGExplained}).
We post hoc theorised why the gamma distribution arises in the current context.

The late-time density in a voxel is, of course, equal to the late-time mass in the voxel divided by its (fixed) volume of $(2.9\ \mathrm{Mpc}\ h^{-1})^3$.
Therefore, up to a dimensionful constant, the late-time density and the late-time mass have the same probability distribution.
The late-time mass $M$ in the voxel equals the sum of the mass aggregated over cosmic time.
If we are to consider a simplistic treatment of mass aggregation, or structure formation, we should pick the most relevant large-scale structure regime.
The central pie chart of Fig.~\ref{fig:piesTWeb} suggests that most observed RGs have filament environments.
In addition, \citet{Pasini12021} suggest that within filaments, most RGs inhabit galaxy groups, which grow by merging with other groups.
Therefore, we considered a proto-filament --- a massive structure that extends essentially along a single dimension --- on which we considered a Poisson point process with constant spatial rate.
The Poisson points represent peaks in the density field where galaxy groups arise.
As with any homogeneous Poisson process, the distances between the points are exponentially distributed.
It is now helpful to partition the filament in cells, each associated to a single Poisson point.
Assuming that the filament has a constant cross-sectional area and an approximately constant density, the volumes and masses of the cells are simply proportional to the distances between the points, and are thus also exponentially distributed.
Finally, if a fixed fraction of the mass in each cell collapses into the cell's galaxy group, then the group masses within the filament --- at early times --- are, again, exponentially distributed.
A group that exists at late times has built up its mass $M$ by aggregating the masses of early-time groups.
Assuming a constant temporal rate of early-time group aggregation, there exists a typical number $N$ of early-time groups that contribute to the mass of a given late-time group: $M = \sum_{i=1}^N M_i$.
We then invoked the fact that the sum of a fixed number of independent and identically distributed RVs with an exponential distribution is an RV with a gamma distribution.
Thus, as the $M_i$ are exponentially distributed, $M$ is gamma distributed.
This, in turn, implies that the late-time densities of the voxels in which late-time groups fall, are gamma distributed also.
As late-time RGs trace late-time groups, the late-time RG density distribution should be approximately gamma.

In the future, by crossmatching RG catalogues with galaxy group catalogues with accurate masses, the mass distribution of RG-hosting groups in the Local Universe can be sampled and tested against the gamma distribution hypothesis.

\section{Modelling relative baryon density measurement heteroskedasticity}
\label{ap:heteroskedasticity}
We must describe the measured RG relative baryon density given a true RG relative baryon density with a continuous probability distribution that has the positive half-line as its support.
Preferably, one uses a simple analytical prescription.

Motivated by the quality of the fits shown in Fig.~\ref{fig:GRGRGDensityDistributions}, we assumed $1+\Delta_\mathrm{RG,obs}\ \vert\ 1 + \Delta_\mathrm{RG} = 1 + \delta \sim \mathrm{Lognormal}(\mu, \sigma^2)$, and likewise for giants.
The parameters $\mu$ and $\sigma^2$ are determined by assumptions on the RV's mean and variance.
In particular, we assumed that measured RG relative densities are unbiased estimators of the underlying true RG relative densities and that their variances are the same when the underlying true RG relative densities are the same:
\begin{align}
    \mathbb{E}\left[1+\Delta_\mathrm{RG,obs}\ \vert\ 1 + \Delta_\mathrm{RG} = 1 + \delta\right] &= 1 + \delta;\\
    \mathbb{V}\left[1+\Delta_\mathrm{RG,obs}\ \vert\ 1 + \Delta_\mathrm{RG} = 1 + \delta\right] &= g\left(1+\delta\right),
\end{align}
where $g\left(1+\delta\right)$ is some non-negative function characterising the heteroskedasticity.
For the current work, we adopted a two-parameter power-law variance model $g\left(1+\delta\right) \coloneqq a\left(1+\delta\right)^b$.
From basic identities of the lognormal distribution, we found for this choice of $g$
\begin{align}
    \mu &= \ln{\left(1+\delta\right)} - \frac{1}{2}\ln{\left(1 + a\left(1+\delta\right)^{b-2}\right)};\\
    \sigma^2 &= \ln{\left(1 + a\left(1+\delta\right)^{b-2}\right)}.
\end{align}
Using the fact that
\begin{align}
&f_{1+\Delta_\mathrm{RG,obs}}\left(1+\delta\right) =\nonumber\\
&\int_0^\infty f_{1+\Delta_\mathrm{RG,obs}\ \vert\ 1 + \Delta_\mathrm{RG} = x}\left(1+\delta\right) \cdot f_{1 + \Delta_\mathrm{RG}}\left(x\right)\ \mathrm{d}x,
\end{align}
we could deduce the distribution of $1 + \Delta_\mathrm{RG}$ given some parametrisation of it and a procedure like MLE.
We chose $1 + \Delta_\mathrm{RG} \sim \Gamma\left(k, \theta\right)$ on the basis that $1+ \Delta_\mathrm{RG,obs}$ appears to be much better described by a gamma distribution than by a lognormal distribution and that the effect of heteroskedasticity appears minor.
We determined $a$ and $b$ from data.\footnote{
For the fixed voxel method, we found $a = 0.4$ and $b = 1.1$; for the flexible voxel method, we found $a = 0.4$ and $b = 0.9$.
We stuck to a simple prescription here, but there appears to be enough BORG SDSS data to describe the heteroskedasticity with a more accurate (though more complex) model while still avoiding overfitting.}

\section{Radio luminosity--Cosmic Web density relation: Fixed voxel method}
\begin{figure*}[bht!]
    \centering
    \includegraphics[width=\textwidth]{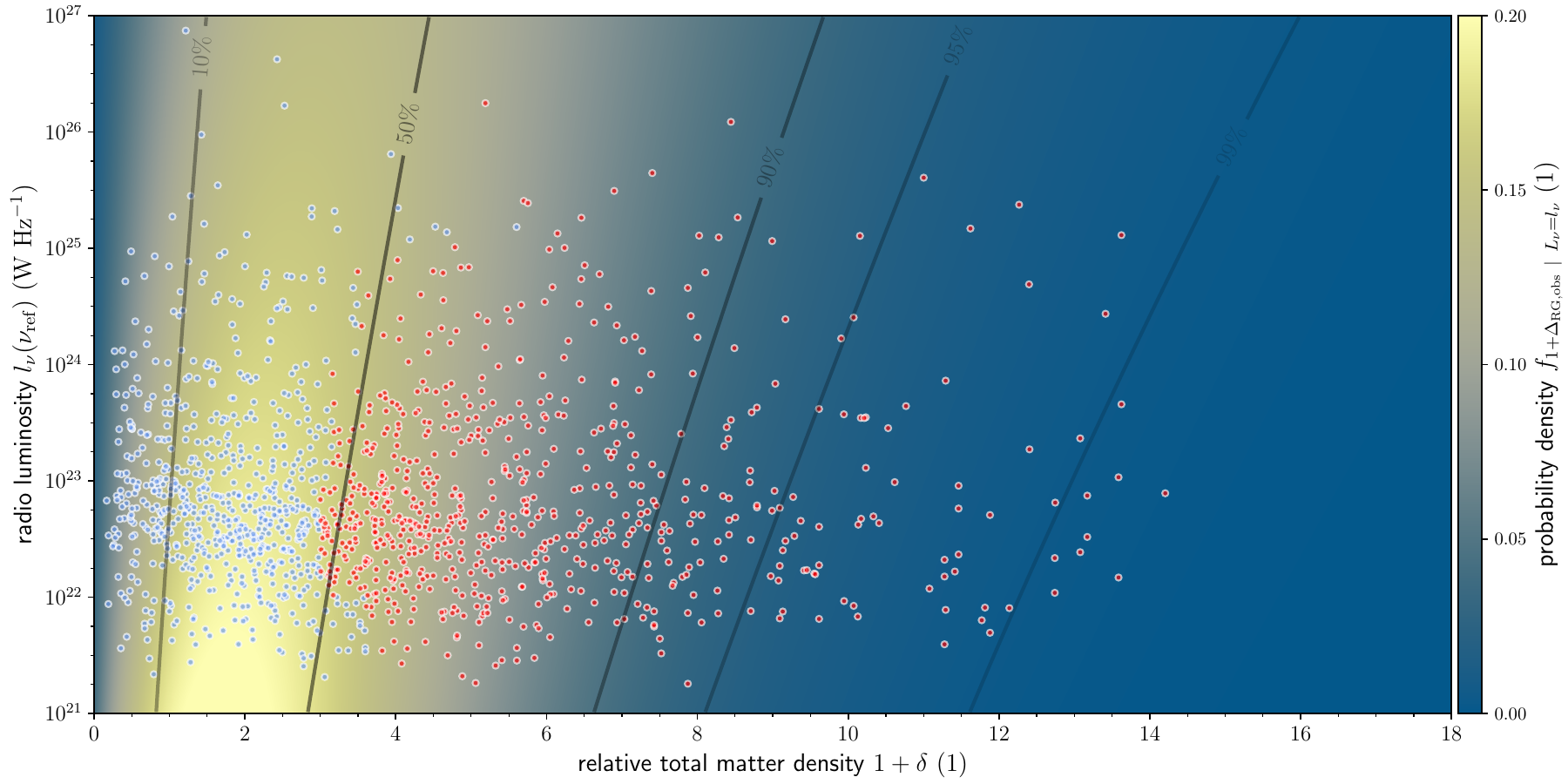}
    \caption{
    PDFs of the observed RG relative total matter density RV given a radio luminosity at $\nu_\mathrm{ref} = 150\ \mathrm{MHz}$, using MLE parameter values for the model described in Sect.~\ref{sec:radioLuminosityCosmicWebDensity}.
    The black contours denote CDF values.
    We overplot all \numberOfRGsBORGzsp\ selected LoTSS DR1 RGs (dots), with those above the empirical median density coloured red, and those below coloured blue.
    We used fixed voxel method densities.
    For flexible voxel method densities, see Fig.~\ref{fig:modelDensityLuminosityScatterFlexible}.
    }
    \label{fig:modelDensityLuminosityScatterFixed}
\end{figure*}\noindent

\begin{figure*}[t!]
    \centering
    \includegraphics[width=.9\textwidth]{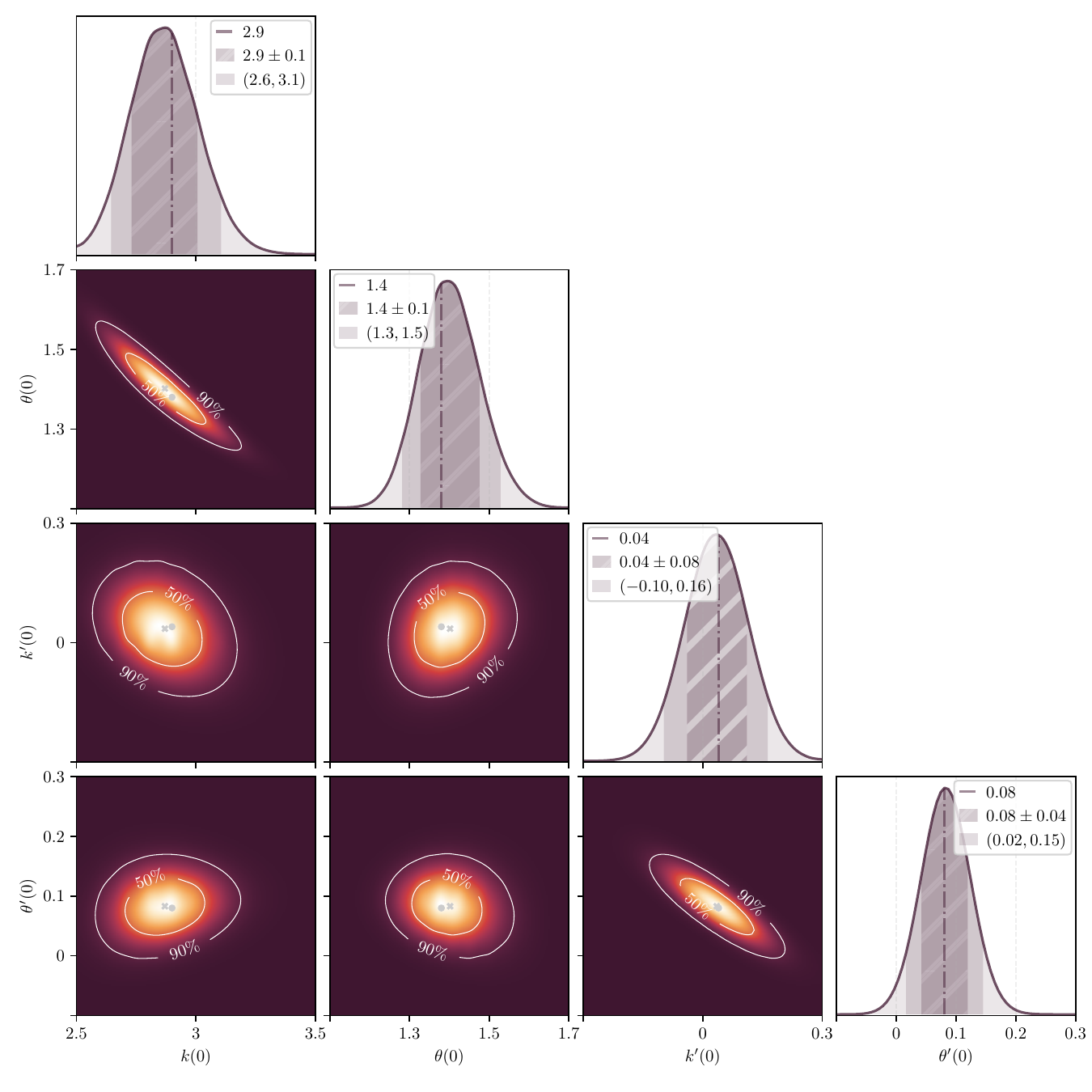}
    \caption{
    Posterior distribution over $k(0)$, $\theta(0)$, $k'(0)$, and $\theta'(0)$, based on selected Local Universe LoTSS DR1 RGs.
    We show all two-parameter marginals of the likelihood function, with contours enclosing $50\%$ and $90\%$ of total probability.
    We mark the MLE (grey dot) and the MAP (grey cross).
    The one-parameter marginals again show the MLE (dash-dotted line), a mean-centred interval of standard deviation--sized half-width (hashed region), and a median-centred $90\%$ credible interval (shaded region).
    We used fixed voxel method densities.
    For flexible voxel method densities, see Fig.~\ref{fig:modelDensityLuminosityCornerFlexible}.
    }
    \label{fig:modelDensityLuminosityCornerFixed}
\end{figure*}\noindent
In Sect.~\ref{sec:radioLuminosityCosmicWebDensity}, we have presented a quantification of the relation between RG radio luminosity and Cosmic Web density.
In particular, we inferred a probability distribution for the $2.9\ \mathrm{Mpc}\ h^{-1}$--scale total matter density around the hosts of Local Universe radio galaxies with a given $150\ \mathrm{MHz}$ radio luminosity --- which can range from $10^{21}$ to $10^{27}\ \mathrm{W\ Hz^{-1}}$.
The MLE model shown in Fig.~\ref{fig:modelDensityLuminosityScatterFlexible} and the one- and two-dimensional posterior marginals shown in Fig.~\ref{fig:modelDensityLuminosityCornerFlexible}, are based on densities measured through Sect.~\ref{sec:flexibleVoxelMethod}'s flexible voxel method.
In this appendix, through Figs.~\ref{fig:modelDensityLuminosityScatterFixed} and \ref{fig:modelDensityLuminosityCornerFixed}, we present the analogous results based on densities measured through Sect.~\ref{sec:fixedVoxelMethod}'s fixed voxel method.
Both methods reveal a positive scaling between RG radio luminosity and Cosmic Web density.

\section{Relative number density derivation}
\label{ap:relativeNumberDensity}
We let the RV $1+\Delta_\mathrm{CW}$ represent the $2.9\ \mathrm{Mpc}\ h^{-1}$--scale relative density at a randomly chosen point in the contemporary Cosmic Web, with $f_{1+\Delta_\mathrm{CW}}$ being its PDF.
Similarly, we let the RV $1+\Delta_\mathrm{RG}$ represent the $2.9\ \mathrm{Mpc}\ h^{-1}$--scale relative density at a randomly chosen RG in the contemporary Cosmic Web, with $f_{1+\Delta_\mathrm{RG}}$ being its PDF.
We considered a cosmologically sized, comoving volume of extent $V$, in which a total of $N_\mathrm{RG}$ radio galaxies exist.
The subvolume in which the relative density is between $1+\delta$ and $1+\delta + \mathrm{d}\delta$ has extent
\begin{align}
    \mathrm{d}V = V f_{1+\Delta_\mathrm{CW}}(1+\delta)\ \mathrm{d}\delta.
\end{align}
The number of RGs in this subvolume is
\begin{align}
    \mathrm{d}N_\mathrm{RG} = n_\mathrm{RG}(1+\delta)\ \mathrm{d}V,
\end{align}
where $n_\mathrm{RG}(1+\delta)$ is the number density of RGs that would arise in an environment of constant density $1+\delta$.
In practice, one never encounters environments of constant density; the average RG number density in the part of the Cosmic Web where the relative density is between $1+\delta_1$ and $1 + \delta_2$ equals
\begin{align}
    \bar{n}_\mathrm{RG}(1+\delta_1, 1+\delta_2) = \frac{\int_{\delta_1}^{\delta_2} n_\mathrm{RG}(1+\delta)f_{1+\Delta_\mathrm{CW}}(1+\delta)\ \mathrm{d}\delta}{\int_{\delta_1}^{\delta_2} f_{1+\Delta_\mathrm{CW}}(1+\delta)\ \mathrm{d}\delta}.
\end{align}
As a result, the cosmic mean RG number density $\bar{n}_\mathrm{RG}$ is given by
\begin{align}
\bar{n}_\mathrm{RG} \coloneqq \int_{-1}^\infty n_\mathrm{RG}(1+\delta) f_{1+\Delta_\mathrm{CW}}(1+\delta)\ \mathrm{d}\delta.
\end{align}
The probability that an RG has a relative density between $1+\delta$ and $1+\delta + \mathrm{d}\delta$, $f_{1+\Delta_\mathrm{RG}}(1+\delta)\ \mathrm{d}\delta$, is
\begin{align}
f_{1+\Delta_\mathrm{RG}}(1+\delta)\ \mathrm{d}\delta = \frac{\mathrm{d}N_\mathrm{RG}}{N_\mathrm{RG}} = \frac{n_\mathrm{RG}(1+\delta) V f_{1+\Delta_\mathrm{CW}}(1+\delta)\ \mathrm{d}\delta}{N_\mathrm{RG}}.
\label{eq:probabilityRelativeDensityRG}
\end{align}
Combining $N_\mathrm{RG} = \bar{n}_\mathrm{RG} V$
with Eq.~\ref{eq:probabilityRelativeDensityRG} leads to
\begin{align}
\frac{n_\mathrm{RG}(1+\delta)}{\bar{n}_\mathrm{RG}} = \frac{f_{1+\Delta_\mathrm{RG}}(1+\delta)}{f_{1+\Delta_\mathrm{CW}}(1+\delta)}.
\label{eq:numberDensityRelativeRG}
\end{align}
In other words, through point-wise division of the PDFs of $1+\Delta_\mathrm{RG}$ and $1+\Delta_\mathrm{CW}$, both of which we determined in this work, we obtained the RG number density that would arise at a given density, up to a constant.
This constant is (the reciprocal of) the cosmic mean RG number density; the LHS of Eq.~\ref{eq:numberDensityRelativeRG} can thus be interpreted as the relative RG number density at $1+\delta$ --- that is, the number density relative to the cosmic mean value.
Fully analogously, the relative GRG number density at $1+\delta$ is
\begin{align}
\frac{n_\mathrm{GRG}(1+\delta)}{\bar{n}_\mathrm{GRG}} = \frac{f_{1+\Delta_\mathrm{GRG}}(1+\delta)}{f_{1+\Delta_\mathrm{CW}}(1+\delta)}.
\label{eq:numberDensityRelativeGRG}
\end{align}
By multiplying both sides of Eq.~\ref{eq:numberDensityRelativeRG} by $f_{L_\nu \vert 1+\Delta_\mathrm{RG}=1+\delta}(l_\nu)$, using the identity $f_{X \vert Y = y}(x)f_Y(y) = f_{Y \vert X = x}(y)f_X(x)$, and dividing both sides by $f_{L_\nu}(l_\nu)$, we obtained
\begin{align}
    \frac{n_\mathrm{RG}(1+\delta) \cdot f_{L_\nu \vert 1+\Delta_\mathrm{RG} = 1+\delta}(l_\nu)}{\bar{n}_\mathrm{RG} \cdot f_{L_\nu}(l_\nu)} = \frac{f_{1+\Delta_\mathrm{RG} \vert L_\nu = l_\nu}(1+\delta)}{f_{1+\Delta_\mathrm{CW}}(1+\delta)}.
\end{align}
The numerator and denominator at the left-hand side are both number density \emph{densities}: they denote radio galaxy numbers per unit of comoving volume and unit of radio luminosity.
As their physical dimensions are distinct from those of ordinary number densities, we denoted them differently.
Calling the numerator $\textphnc{\Anun}_\mathrm{RG}(l_\nu, 1 + \delta)$ and the denominator $\bar{\textphnc{\Anun}}_\mathrm{RG}(l_\nu)$, we wrote\footnote{As all $n$-like symbols in the Roman and Greek alphabets are already in use, we used the Phoenician root $\textphnc{\Anun}$.}
\begin{align}
    \frac{\textphnc{\Anun}_\mathrm{RG}(l_\nu, 1 + \delta)}{\bar{\textphnc{\Anun}}_\mathrm{RG}(l_\nu)} = \frac{f_{1+\Delta_\mathrm{RG} \vert L_\nu = l_\nu}(1+\delta)}{f_{1+\Delta_\mathrm{CW}}(1+\delta)}.
\end{align}

\section{Disparity between mean squared density and squared mean density}
\label{ap:densitySquared}
\begin{figure}[t]
    \centering
    \includegraphics[width=\columnwidth]{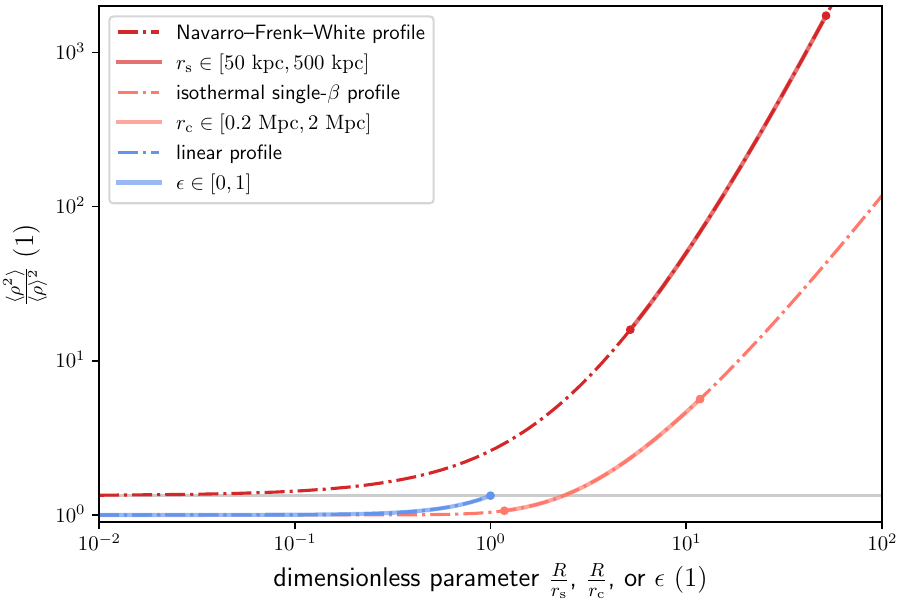}
    \caption{
    Multiplicative discrepancy between the mean squared density $\langle \rho^2 \rangle$ and the squared mean density $\langle \rho \rangle^2$ in voxelised Cosmic Web reconstructions, for density profiles of clusters (red), filaments (orange), and sheets or voids (blue).
    The length scale $R \sim L$, where $L$ is the voxel side length; here, $L = 2.9\ \mathrm{Mpc}\ h^{-1}$.
    The asymptote (grey) corresponds to the ratio $\frac{4}{3}$.
    }
    \label{fig:RGModelSquaredDensityRatios}
\end{figure}\noindent
Several physical quantities that relate to AGN and RGs, such as the bremsstrahlung emissivity around the host galaxy, scale with the square of IGM density $\rho$.
In Sect.~\ref{sec:CosmicWebDensityResolution}, we argue qualitatively that such quantities are likely to be underestimated if calculated using low-resolution densities $\langle \rho \rangle$, such as the ones offered by the BORG SDSS.
In this appendix, we explicitly calculate how much bigger $\langle \rho^2\rangle$ is than $\langle \rho \rangle^2$ --- in the sense of a multiplicative Jensen's gap.
Our calculations demonstrate a strong dependence on Cosmic Web environment.

When the true density field varies over scales much larger than a voxel side length $L$, such as in sheets or voids, the density field variation within a voxel resembles a gradient from $\left(1-\epsilon\right)\langle \rho \rangle$ to $\left(1+\epsilon\right) \langle \rho \rangle$, where $\epsilon \in [0, 1]$.
If the planes of constant density are parallel to two of the voxel's faces, one finds
\begin{align}
    \frac{\langle\rho^2\rangle}{\langle \rho \rangle^2} = \frac{\epsilon^2}{3} + 1,\text{ so that } 1 \leq \frac{\langle\rho^2\rangle}{\langle \rho \rangle^2} \leq \frac{4}{3}.
    \label{eq:JensenSheetVoid}
\end{align}
We show the multiplicative Jensen's gap of Eq.~\ref{eq:JensenSheetVoid} in Fig.~\ref{fig:RGModelSquaredDensityRatios} (blue curve).

For a filament whose baryon density is modelled with an isothermal single-$\beta$ model \citep{Cavaliere11976, Cavaliere11978} with $\beta = \frac{2}{3}$ and core radius $r_\mathrm{c}$, one finds
\begin{align}
    \frac{\langle\rho^2\rangle}{\langle \rho \rangle^2} = \frac{\left(\frac{R}{r_\mathrm{c}}\right)^4}{\left(1+\left(\frac{R}{r_\mathrm{c}}\right)^2\right)\ln^2{\left(1+\left(\frac{R}{r_\mathrm{c}}\right)^2\right)}},
    \label{eq:JensenFilament}
\end{align}
where we considered a cylinder with length $L$ and radius $R$.
Here, $R$ is such that the area of a cylindrical section perpendicular to the axis equals $L^2$, the area of a voxel face.
Thus, $R = \frac{1}{\sqrt{\pi}}L$.
We show the multiplicative Jensen's gap of Eq.~\ref{eq:JensenFilament} in Fig.~\ref{fig:RGModelSquaredDensityRatios} (orange curve).

When the true density field varies over scales much smaller than a voxel side length $L$, such as in clusters, the ratio between the mean of the squared density and the square of the mean density can be much larger than $1$.
In particular, for a Navarro--Frenk--White (NFW) profile \citep{Navarro11996} with scale radius $r_\mathrm{s}$,
\begin{align}
    \frac{\langle\rho^2\rangle}{\langle \rho \rangle^2} = \frac{1 - \left(\frac{r_\mathrm{s}}{r_\mathrm{s}+R}\right)^3}{9\left(\frac{r_\mathrm{s}}{R}\right)^3\left(\ln{\frac{r_\mathrm{s}}{r_\mathrm{s}+R}} + 1 - \frac{r_\mathrm{s}}{r_\mathrm{s}+R}\right)^2}.
    \label{eq:JensenCluster}
\end{align}
where we considered a sphere with radius $R$.
Here, $R$ is such that the volume of the sphere is equal to the volume of the voxel: $R = \left(\frac{3}{4\pi}\right)^{\frac{1}{3}}L$.
We show the multiplicative Jensen's gap of Eq.~\ref{eq:JensenCluster} in Fig.~\ref{fig:RGModelSquaredDensityRatios} (red curve).
In the limit $\frac{r_\mathrm{s}}{R} \to \infty$, $\frac{\langle\rho^2\rangle}{\langle \rho \rangle^2} \to \frac{4}{3}$.

\end{appendix}
\end{document}